\documentclass[a4paper,11pt]{article}
\usepackage{jcappub} 
\usepackage{lineno}
\usepackage{amsmath,amssymb}
\usepackage{hyperref}
\usepackage{comment}

\newcommand{\bs}[1]{\boldsymbol{#1}}
\newcommand{\be}{\begin{equation}}
\newcommand{\ee}{\end{equation}}
\newcommand{\bea}{\begin{eqnarray}}
\newcommand{\eea}{\end{eqnarray}}

\title{Fast Likelihood-free Reconstruction\\of Gravitational Wave Backgrounds} 

\author{Androniki Dimitriou, Daniel G. Figueroa, and Bryan Zaldívar}
\affiliation{Instituto de Física Corpuscular (IFIC), Consejo Superior de Investigaciones Científicas
(CSIC) and Universitat de València 46980, Valencia, Spain}

\emailAdd{androniki.dimitriou@ific.uv.es}
\emailAdd{daniel.figueroa@ific.uv.es}
\emailAdd{bryan.zaldivar@ific.uv.es}

\abstract{
We apply state-of-the-art, likelihood-free statistical 
inference (machine-learning-based) techniques for reconstructing the spectral shape of a gravitational wave background (GWB). We focus on the reconstruction of an arbitrarily shaped signal (approximated by a piecewise power-law in many frequency bins) by the LISA detector, but the method can be easily extended to either template-dependent signals, or to other detectors, as long as a characterisation of the instrumental noise is available. As proof of the technique, we quantify the ability of LISA to reconstruct signals of arbitrary spectral shape ({\it blind} reconstruction), considering a diversity of frequency profiles, and including astrophysical backgrounds in some cases. As a teaser of how the method can reconstruct signals characterised by a parameter-dependent template ({\it template} reconstruction), we present a dedicated study for power-law signals. While our technique has several advantages 
with respect to traditional MCMC methods, we validate it with the latter for concrete cases. This work opens the door for both fast and accurate Bayesian parameter estimation of GWBs, with essentially no computational overhead during the inference step. Our set of tools are integrated into the package {\tt GWBackFinder}, which is publicly available in \href{https://github.com/AndronikiDimitriou/GWBackFinder}{GitHub}.}

\begin{document}
\maketitle
\flushbottom

\section{Introduction}
\label{sec:intro}

The direct detection of gravitational waves (GWs) offers a unique opportunity to probe scales and circumstances of the Universe far beyond the reach of electromagnetic probes. The detection of dozens of transient GW signals by the LIGO/Virgo network since 2015~\cite{Abbott:2016nmj,Abbott:2017vtc,Abbott:2017gyy,Monitor:2017mdv,Abbott:2017oio,TheLIGOScientific:2017qsa,LIGOScientific:2018mvr,LIGOScientific:2021usb,LIGOScientific:2021djp} has begun to reveal the great potential of GW astronomy, yielding outstanding results on the abundance and dynamics of astrophysical compact binaries, while enabling stringent tests of gravity at the same time. Due to well-motivated theoretical reasons, 
the Universe is expected to be filled up with stochastic GW backgrounds (GWBs) of astrophysical and cosmological origin~\cite{Caprini:2018mtu,Maggiore:2018sht,Christensen:2018iqi}. While the LIGO/VIRGO/KAGRA (LVK) collaboration has set stringent upper bounds on the presence of GWBs\footnote{We define the energy density spectrum $\Omega_{\rm GW}(f)$ of a GWB in Sect.~\ref{sec:data}.} around the $\sim$ dHz-kHz frequencies, as $\Omega_{\rm GW} \lesssim 5\times 10^{-9}$~\cite{KAGRA:2021kbb} and $\Omega_{\rm GW} \lesssim 8.6\times 10^{-9}$~\cite{KAGRA:2021kbb,KAGRA:2021mth,Regimbau:2011rp,Stiskalek:2020wbj}, for scale-invariant and astrophysical signals, respectively, pulsar timing array (PTA) collaborations~\cite{NANOGrav:2023gor, Antoniadis:2023ott,Reardon:2023gzh, Xu:2023wog} have just announced the first strong evidence of detection of a GWB, at their natural $\sim$nHz frequency window. The origin of the PTA signal is so far unclear, and while a GWB from a population of supermassive black hole binaries (SMBHBs) is naturally expected at the PTA frequencies~\cite{Kelley:2017lek,NANOGrav:2023hfp,Antoniadis:2023xlr}, cosmological backgrounds also represent a viable explanation~\cite{NANOGrav:2023hvm,Antoniadis:2023xlr,Figueroa:2023zhu}.

Leading astrophysical signals emerge from the incoherent superposition of weak GW signals from compact binaries that cannot be resolved individually by a given experiment\footnote{The amplitude of an astrophysical GWB within a given detector depends therefore on the ability of such detector to resolve individually some of the binaries that give rise to the GWB.}. This is the case, for example, of the stochastic GWB expected from stellar origin black hole binaries (SOBHBs), neutron star binaries (NSBs) at the LVK frequency window~\cite{Sesana:2016ljz,Regimbau:2011rp,Babak:2023lro,Lehoucq:2023zlt}, or extreme mass-ratio inspirals (EMRIs) \cite{PhysRevD.102.103023, Pozzoli:2023kxy}. It is also the case of the background expected from SMBHBs at PTA frequencies~\cite{Sesana:2008mz,Sesana:2012ak}. While astrophysical backgrounds might probe interesting aspects of the environment and formation history of binaries, cosmological backgrounds, on the other hand, carry information on the high energy phenomena that originated them in the early Universe. Theoretically, the Universe might be permeated by a plethora of cosmological backgrounds with a variety of origins, ranging from vacuum 
fluctuations~\cite{Grishchuk:1974ny,Starobinsky:1979ty, Rubakov:1982df,Fabbri:1983us} or particle production during inflation~\cite{Anber:2006xt,Sorbo:2011rz,Pajer:2013fsa,Adshead:2013qp,Adshead:2013nka,Maleknejad:2016qjz,Dimastrogiovanni:2016fuu,Namba:2015gja,Ferreira:2015omg,Peloso:2016gqs,Domcke:2016bkh,Caldwell:2017chz,Guzzetti:2016mkm,Bartolo:2016ami,DAmico:2021zdd,DAmico:2021vka,Fumagalli:2020nvq,Fumagalli:2021mpc}, to post-inflationary phenomena like preheating~\cite{Easther:2006gt,GarciaBellido:2007dg,GarciaBellido:2007af,Dufaux:2007pt,Dufaux:2008dn,Dufaux:2010cf,Bethke:2013aba,Bethke:2013vca,Figueroa:2017vfa,Adshead:2018doq,Adshead:2019lbr,Adshead:2019igv}, kination-domination~\cite{Giovannini:1998bp,Giovannini:1999bh,Boyle:2007zx,Li:2016mmc,Li:2021htg,Figueroa:2018twl,Figueroa:2019paj,Li:2021htg,Gouttenoire:2021wzu,Co:2021lkc,Gouttenoire:2021jhk,Oikonomou:2023qfz}, thermal plasma motions~\cite{Ghiglieri:2015nfa, Ghiglieri:2020mhm, Ringwald:2020ist, Ghiglieri:2022rfp}, oscillon dynamics~\cite{Zhou:2013tsa,Antusch:2016con,Antusch:2017vga,Liu:2017hua,Amin:2018xfe}, first order phase transitions~\cite{Kamionkowski:1993fg,Caprini:2007xq,Huber:2008hg,Hindmarsh:2013xza,Hindmarsh:2015qta,Caprini:2015zlo,Hindmarsh:2017gnf,Cutting:2018tjt,Cutting:2018tjt,Cutting:2019zws,Pol:2019yex,Caprini:2019egz,Cutting:2020nla,Han:2023olf,Ashoorioon:2022raz, Athron:2023mer,Li:2023yaj}, cosmic defect dynamics~\cite{Vachaspati:1984gt,Sakellariadou:1990ne,Damour:2000wa,Damour:2001bk,Damour:2004kw,Figueroa:2012kw,Hiramatsu:2013qaa,Blanco-Pillado:2017oxo,Auclair:2019wcv,Gouttenoire:2019kij,Figueroa:2020lvo,Gorghetto:2021fsn,Chang:2021afa,Yamada:2022aax,Yamada:2022imq,Kitajima:2023cek}, large scalar fluctuations~\cite{Matarrese:1992rp,Matarrese:1993zf, Matarrese:1997ay,Nakamura:2004rm,Ananda:2006af,Baumann:2007zm,Domenech:2021ztg, Dandoy:2023jot}, or others, see~\cite{Caprini:2018mtu} for a comprehensive review. The possibility of detecting a cosmological background opens therefore a new window into the physics of the early Universe, probing energy scales that cannot be accessed otherwise by other terrestrial means. 
Astrophysical backgrounds may act, however, as contaminating foregrounds, diminishing our ability to detect and differentiate potentially measurable cosmological signals.

The evidence at $\sim 3-4\sigma$ of a GWB at the PTA $\sim$nHz window, has marked a new {\it milestone} in GW astronomy. While the evidence for the signal needs to become yet stronger by further data analysis and observations, 
present data can be already used to strongly constrain cosmological signals~\cite{NANOGrav:2023hvm,Antoniadis:2023xlr,Figueroa:2023zhu}, see also~\cite{Franciolini:2023pbf,Vagnozzi:2023lwo,Guo:2023hyp,Kitajima:2023cek,Ellis:2023tsl,Han:2023olf,Cai:2023dls,Madge:2023cak,Bai:2023cqj,Liu:2023ymk,Unal:2023srk,Servant:2023mwt,Ellis:2023oxs,Wang:2023ost}. This marks the dawn of {\it early universe gravitational wave astronomy}, which provides unprecedented opportunities for high-energy physics beyond the standard model (BSM) and early universe cosmology. The holy grail of GW astronomy would be to detect a GWB and demonstrate that it cannot be ascribed to astrophysics, therefore naturally pointing to a potential cosmological origin. This is of course an extremely hard task, no matter the experiment considered. The recent PTA data analysis has reinforced the importance and difficulty of such task~\cite{NANOGrav:2020bcs,EPTA:2023fyk,Antoniadis:2023xlr,NANOGrav:2023gor,NANOGrav:2023hvm,NANOGrav:2023hfp}, stimulating at the same time the search for signals at other frequency ranges. Fortunately, a detection program including a large variety of experiments is emerging, out of the effort to detect GWBs over a wide range of frequencies. These experiments range from increasingly precise cosmic microwave background (CMB) and PTA observations (see e.g.~\cite{Janssen:2014dka,Abazajian:2016yjj}), 
to present and proposed GW direct detection experiments (see e.g.~\cite{Sesana:2019vho,LISA:2017pwj,Ruan:2020smc,Baker:2019pnp,Kawamura:2011zz,Mei:2020lrl,Kuns:2019upi,Reitze:2019iox,Punturo:2010zz}), 
and atom interferometers (see e.g.~\cite{Graham:2017pmn,Bertoldi:2019tck}), 
altogether spanning $\sim 20$ decades in frequency, from $\sim 10^{-17}$ Hz to $\sim 10^3$ Hz. Furthermore, a high-frequency detection program at $\sim$ MHz frequencies and above, has been recently also put forward~\cite{Aggarwal:2020olq}. An evaluation of the ability of many of these experiments to probe new high energy physics is however still in progress. 

This new era of GW cosmology comes, of course, with several important challenges from the data analysis point of view. First of all, there is the need of inferring the instrumental noise of a given experiment. This is certainly not a trivial task for any experiment, 
as characterising all possible noise contributions can be extremely challenging, and might require calibration during operation time. Our ability to identify a potential GWB in the data of any given experiment may depend on our ability to characterize the noise of such experiment. Secondly, and partially due to the previous point, complex simulations and calculations will be required in order to understand/test the modelling of all possible contributions to the data, including cosmological GWBs, astrophysical ``foregrounds'', and  noise contributions. Clearly, sophisticated data analysis techniques are going to be needed in order to optimise the extraction of information from the measured data to come.

The above challenges motivate the consideration of modern statistical methods that are being successfully applied in other domains, such as e.g. particle colliders, neutrino experiments, or large-scale structure observations. Like in the case at hand, studies in those domains that rely on complex Monte Carlo simulations containing ${\cal O}(10^6)$ or more variables, have benefited already from state-of-the-art {\tt simulation-based} (SB) statistical inference techniques~\cite{Cranmer_2020,brehmer2020,lueckmann2021,Miller2022,Tejero-Cantero2020,Cole:2021gwr,Montel:2022fhv,Makinen:2021nly,Dimitriou:2022cvc,Gagnon-Hartman:2023soa,Delaunoy:2020zcu,Karchev:2022xyn,Lin:2022ayr,Alvey:2023naa,Bhardwaj:2023xph}. Even though typically only few of those variables may be of physical interest (in our case the parameters characterising a GWB), traditional MCMC methods, as any likelihood-based approach, require the specification of the likelihood $p(D|\bs{\theta})$, i.e.~the probability of data $D$ given the parameters of interest $\bs{\theta}$. For simulators with internal (latent) stochastic variables, this is, in reality, a ``marginal likelihood'', resulting from marginalising the full likelihood over all latent variables. 
Clearly the marginalisation over such a large number of variables is in general intractable. On the other hand, running MCMC methods over the full (unintegrated) likelihood, while formally possible, is known to suffer from the curse of dimensionality, requiring typically unfeasible computational costs\footnote{While this is true for generic MCMC methods, there are flavors of it, like the Hybrid Monte Carlo (HMC) algorithm, which can handle parameter spaces of the aforementioned dimensionalities or even more, as demonstrated in domains as Lattice Field Theory.}. Modern SB methods rely instead on ``likelihood-free'' inference, i.e.~statistical procedures allowing to make inference on the parameters of interest, without the need to have an explicit form of the likelihood.
~Instead, the likelihood is defined implicitly by the simulator and samples of it can be obtained by running the simulator forward, i.e, $D \sim p(D|\bs{\theta})$. While this idea is not new (it dates back to the 1980's), nowadays it enjoys an accelerated expansion and major qualitative improvements thanks to the exponential development of Machine Learning (ML) methods, which are at the heart of modern SB statistical inference. Another worth-mentioning advantage of these methods is that they typically rely on ``pre-trained'' (a.k.a.~``amortised'') strategies: their computational bottleneck lies mostly at the first stage where a sufficiently exhaustive set of simulated datasets have to be generated. After training the model, an observed (or test) dataset is presented to the fitted model in the final inference stage, with the output corresponding to the posterior distribution of the parameters of interest. This takes essentially no time to evaluate, as opposed to what happens with MCMC methods, which build a costly Markov chain for any given observed dataset. Finally, unlike MCMC, the methods in consideration implement approximate posterior inference. However, note that all the good properties of MCMC are valid as long as the assumed likelihood is the correct one, which as argued above, could be intractable in nowadays's simulation-based studies.

In this work we propose a new method for reconstructing the spectral shape of a GWB, considering a particular flavour of state-of-the-art SB methods, belonging to the family of Neural Posterior Estimation (NPE) \cite{rezende2016}. We adopt the same spirit as Ref.~\cite{Karnesis:2019mph} and Refs.~\cite{Caprini:2019pxz,Flauger:2020qyi}, where a model-independent approach is used to reconstruct the frequency shape of an unknown SGWB (see also the recent proposals~\cite{Baghi:2023qnq,Pozzoli:2023lgz}). Our procedure is based on dividing the frequency range of a detector into sub-intervals (\textit{bins}) within which the signal is approximated by a power law. This corresponds to a Taylor expansion of the signal within each interval, truncated at linear order in a log-log scale. While Refs.~\cite{Caprini:2019pxz, Flauger:2020qyi} start with an arbitrary number of initial bins and subsequently combine them according to information criteria for the sake of model selection (a la Bayesian evidence), our procedure is based on a fixed and sufficiently large number of bins, such as to have a compromise between enough model complexity and robustness to noise (more on this in sec. \ref{subsec:Agnostic}). 
With respect to other related methods, our procedure has the important advantage of straightforwardly delivering samples from the joint posterior $p(\bs{\theta}|D)$, opening the door to both fast and accurate Bayesian parameter estimation of GWBs, with essentially no computational overhead during the inference step.  
We provide a detailed description of our technique in Sects.~\ref{sec:data} \&~\ref{sec:analysis}. Using our method we analyse in detail the ability of the {\it Laser Interferometer Space Antenna} (LISA) experiment~\cite{Audley:2017drz} to detect GWBs and to reconstruct their spectra 
in the milli-Hertz domain. 
In Sect.~\ref{sec:BlindReconstruction} we focus on the ability of LISA to reconstruct a GWB signal with arbitrary spectral shape ({\it blind} reconstruction). We analyse the reconstruction of a diversity of frequency profiles (Sect.~\ref{subsec:arbitraryShape}), propose methods to quantify the  accuracy of reconstruction of power law signals 
and backgrounds with general spectral shapes (Sect.~\ref{subsec:uncertainty}), and compare our technique against MCMC methods (Sect.~\ref{subsec:MCMCcomparison}). In Sect.~\ref{sec:foreground}, for completeness, we consider blind reconstruction of signals on top of
astrophysical foregrounds. Finally, in Sect.~{\ref{sec:template}}, we consider the reconstruction by LISA of signals characterised by a template depending on a set of parameters ({\it template} reconstruction), albeit focusing on power-law profiles only. In Sect.~\ref{sec:conclusions} we summarise our results and discuss future applications of our technique, highlighting that our method can be easily extended to either more complex multi-parameter template-dependent signals, or to other detectors, as long as a characterisation of the instrumental noise is available. 

\section{Signal and noise characterisation}
\label{sec:data}

\subsection{General characterisation of a GWB}

For clarity, we review, very briefly first, the notion of GWs. These are tensor spatial perturbations $h_{ij}$ ($i, j = 1, 2, 3$) that propagate according to the wave equation at the speed of light. They verify the transverse ($\partial_i h_{ij} = 0 $) and traceless ($h_{ii} = 0$) conditions, so $h_{ij}$ is a real symmetric 2-rank tensor (with respect to spatial rotations), with only two independent degrees of freedom, identified as the two polarisations of GWs. The transverse-traceless field $h_{ij}(\mathbf{x}, t)$ in real space can be then decomposed into the two polarisation states $r = +, \times$ in Fourier space as
\be
\label{eq:polarizations}
h_{ij}(\mathbf{x}, t) = \sum_{r = + , \times} \, \int \frac{d^3 \mathbf{k}}{(2 \pi)^{3}} \, h_r(\mathbf{k},t) \,
e^{- i \mathbf{k} \cdot \mathbf{x}} \, e_{ij}^{r}(\mathbf{\hat{k}})\,,
\ee
where the two polarisation tensors $e_{ij}^{r}(\mathbf{\hat{k}})$ can be taken to be real and to satisfy 
$e_{ij}^{r}(\mathbf{- \hat{k}}) = e_{ij}^{r}(\mathbf{\hat{k}})$. The condition for $h_{ij}$ to be real is then $h_r^*(\mathbf{k}, t) = h_r(- \mathbf{k}, t)$. The two polarisation tensors depend only on the unit vector $\mathbf{\hat{k}}$ and are symmetric ($e^r_{ij} = e^r_{ji}$), transverse ($\hat{k}_i\,e^r_{ij} = 0$),  traceless ($e_{ii}^r = 0$), and orthonormal $e_{ij}^{r}(\mathbf{\hat{k}}) \, e_{ij}^{r'}(\mathbf{\hat{k}}) =  \delta_{r r'}$, see e.g.~\cite{Caprini:2018mtu} for further details. 

We characterize the amplitude of a statistically homogeneous and isotropic, unpolarised GWB, via its variance
\bea
\label{eq:GWBvariance}
\langle h_{ij}(\mathbf{x}, t) \, h_{ij}(\mathbf{x}, t) \rangle = 
\int d\log k~\Delta_{\tt h}(k,t)\,, ~~~ \Delta_{\tt h}(k,t) \equiv 2h_c^2(k, t)\,,
\eea
where the factor 2 in front of $h_c^2(k, t)$ is a convention motivated by the fact that the variance involves contributions from two independent polarisations\footnote{If the signal is Gaussian, the spectrum $\Delta_h(k,t)$ fully characterizes the GWB.}. It appears from the above equation that $h_c(k, t)$ represents a (dimensionless) {\it characteristic strain} GW amplitude per logarithmic wave-number interval and per polarization state, at a time $t$. The power spectrum 
of the GWB is defined as 
\bea
\langle h_r(\mathbf{k}, t) \, h^*_{p}(\mathbf{k}', t) \rangle = (2\pi)^3\delta_{r p}\frac{\mathcal{P}_{\tt h}(k,t)}{2} \delta^{(3)}(\mathbf{k}-\mathbf{k^{\prime}})\,, ~~ \mathcal{P}_{\tt h}(k,t) \equiv 
\frac{4\pi^2}{k^3}h_c^2(k,t)\,,
\eea
so that 
$\mathcal{P}_{\tt h}(k,t)/2$ 
represents the variance of each polarization state, at time $t$. In order to make connection with observations, it is necessary to evaluate the GW background today in terms of the present-day physical frequency $f = k / 2 \pi$, with $k$ corresponding to the redshifted  wave-number today. The characteristic GW strain today is then given by
\be
\label{hcf}
h_c(f) \equiv h_c(2\pi f, t_0)\,,
\ee 
where $t_0$ denotes the time today.
A stochastic background is also characterized by its {\it spectral density}
\be
\label{Shf}
S_h(f) = \frac{h_c^2(f)}{2 f} \, ,
\ee
which has dimension $\mathrm{Hz}^{-1}$. This quantity is directly comparable to the noise in a detector, which will be defined later. 

In cosmology it is customary to define the {\it energy density spectrum} $\Omega_{\rm GW}(f)$ of a GWB as the power spectrum of the total energy density $\rho_{\rm GW}^{tot}(t_0)$ encoded in a GWB, normalized to the critical energy density today $\rho_c^{(0)} = 3H_0^2/8\pi G$, as~\cite{Caprini:2018mtu}
\be
\label{eq:GWenergyDensity}
\frac{\rho_{\rm GW}^{tot}(t_0)}{\rho_{c}^{(0)}} \equiv \int d\log f~ 
\Omega_{\rm GW}(f)\,,
\ee
where $G$ is the Newton's constant, $H_0$ today's Hubble rate. Equivalently, the energy density spectrum is defined as:
\bea\label{eq:Omega_GW}
\Omega_{\rm GW}(f) \equiv \frac{1}{\rho_c}\,\frac{d \rho_{\rm GW}}{d \mathrm{log} f} &=& \frac{2 \pi^2}{3 H_0^2} \, f^2 \, h_c^2(f) = \frac{4 \pi^2}{3 H_0^2} \, f^3 \, S_h(f)\,. 
\eea
Each GWB can be therefore characterized (spectrally) by the frequency profile of its energy density spectrum $\Omega_{\rm GW}(f)$. In practice, we will simply refer to $\Omega_{\rm GW}(f)$ as the spectrum of a GWB, understanding implicitly that we refer to its energy density spectrum. 

\subsection{Agnostic parametrisation of a GWB}\label{subsec:Agnostic}

Since one of our aims is to reconstruct the spectral shape $\Omega_{\rm GW}(f)$ for an arbitrary GWB, our strategy for this is to model a  signal with an unknown frequency profile by approximating it in a piece-wise manner as power-law's in intervals (\textit{bins}). This is equivalent to a Taylor expansion of the signal, truncated at linear order in frequency in a log-log scale within each bin. The complexity of the model (and hence its ability to capture accurately arbitrary frequency profiles; in other words, its ``expressivity'') is determined by the number of bins $N_c$, as well as the position of the edges. 
After imposing continuity at each boundary, i.e. at the 
$N_c - 1$ ``break points" $f_k$, the model has $N_c+1$ independent parameters ${\bf s} \equiv\{\Omega_p^{(*)},\gamma_1,...,\gamma_{N_c}\}$, 
given by the amplitude $\Omega_p^{(*)}$ at a pivot frequency $f_p$, plus $N_c$ slopes $\gamma_k$ of the power laws within each bin. From an algorithmic point of view, we start the GW signal construction from the region in the middle of our set of intervals, writing the signal as
\bea\label{eq:parametrisation}
\Omega_{\rm GW}(f; {\bf s})= \Omega_p^{(*)} \cdot 
\begin{cases}
\vdots \\
F_{-2}(f) &\text{ if } f_{-2} \leq f < f_1\\ 
 F_{-1}(f) &\text{ if } f_{-1} \leq f < f_0\\
\hspace*{2.2mm}F_{0}(f) &\text{ if } f_{0} \le f < f_{1}\\
F_{+1}(f) &\text{ if } f_1 \leq f < f_2\\ 
F_{+2}(f) &\text{ if } f_2 \le f < f_3\\ 
 \vdots 
\end{cases}
\label{eq:signal_sim}
\eea
with
\bea
F_{0}(f) &\equiv& \left(\frac{f}{f_0}\right)^ {\gamma_{0}}\,,\\ 
F_{-1}(f) &\equiv& F_{0}(f_0)\left(\frac{f}{f_{0}}\right)^{\gamma_{-1}}\,, ~~ F_{-2}(f) \equiv F_{-1}(f_{-1})\left(\frac{f}{f_{-1}}\right)^{\gamma_{-2}}\,, ~~...\\
F_{+1}(f) &\equiv& F_{0}(f_{1})\left(\frac{f}{f_{1}}\right)^{\gamma_{1}}\,, ~~F_{+2}(f) \equiv F_{+1}(f_{2})\left(\frac{f}{f_{2}}\right)^{\gamma_{2}}\,, ~~...
\eea
where $f_{k}$ are the breakpoints, with $f_{0} = f_p$ the pivot scale so that $\Omega_{\rm GW}(f_p; {\bf s})= \Omega_p^{(*)}$.  
In this work, we consider $N_c = 27$ bins of equal length in log-space for the reconstruction of signals with the LISA experiment (equivalently this amounts to $26$ breaking points, uniformly separated in log-space with fixed locations). The frequencies span within an interval $[f_{\rm min},f_{\rm max}]$, so that $f_{k} \equiv f_p\cdot10^{k\cdot \Delta\log f}$, with $\Delta\log f \equiv \left(\log_{10}(f_{max})-\log_{10}(f_{\rm min})\right)/N_c$, and $k = -13, -12, ...,-1, 0, +1, +2, ..., +13$, so that $f_{-13} = f_{\rm min}$, $f_{0} = f_p$ and $f_{+13} = f_{\rm max}$. These correspond to $N_c = 27$ intervals, $\Delta_f^{(-13)} = [f_{\rm min},f_{-12})$, $\Delta_f^{(-12)} = [f_{-12},f_{-11})$, ..., $\Delta_f^{(0)} = [f_{0},f_{+1})$, ..., $\Delta_f^{(12)} = [f_{12},f_{13})$, $\Delta_f^{(13)} = [f_{13},f_{\rm max}]$. At the end of Sect.~\ref{subsec:uncertainty} we comment on the reasons for our choice of $N_c = 27$ bins, for the LISA frequency window. We note, in any case, that the number and frequency span of bins could be, in principle, adapted to each experiment, and even to specific signal searches, see discussion in Sect.~\ref{sec:conclusions}.

The above is related to the discussion about model complexity. Our choice is based on the heuristic criterion of having as many bins (thus, parameters) as possible, while still being robust against data variability within a single bin. Indeed, the larger the number of bins $N_c$, the smaller the number of data points lying inside one given bin, thus the stronger the degeneracy between signal and noise will be, hindering the inference procedure. While we certainly have not implemented in this analysis a signal-dependent binning optimisation (which we may consider in follow-up studies), our procedure works efficiently for arbitrary signal shapes, as we show in the results. Given that we fix the model complexity, the possibility to have over-fitting may be a concern. However, as we describe in sec. \ref{subsec:methodology} below, 
we implement over-fitting control explicitly during the training procedure.

\subsection{General characterisation of noise}
\label{subsec:noise}

In this paper we present a technique for the detection and spectral reconstruction of a GWB by a given experiment, relying on the availability of noise modeling for such experiment. We assume that one or various interferometric data streams (channels), possibly correlated, are available. We consider a data stream per channel-$\alpha$, $d_\alpha(t)$, that ideally contains only noise, $n_\alpha(t)$, and
a `residual' stochastic signal, $s(t)$, due to the presence of a GWB that we want to infer. We also assume that the data might be split into data streams of finite duration $T$, which we refer to as {\it chunks}. In practice, the data stream of each chunk and channel (we omit for simplicity, for the time being, an index to indicate the chunk), $d_\alpha(t)=s_\alpha(t)+n_\alpha(t)$, is modelled as a real-valued function within the interval $\left[ -T/2, T/2\right]$. Assuming that signal and noise are uncorrelated, we can then treat them similarly, but separately. We assume that both the noise and the GWB signal are {\it stationary}. By defining first the Fourier transform of a single data stream channel as
\begin{eqnarray}\label{eq:s_f}
\tilde{s}_\alpha\left(f\right) =  \int_{-T/2}^{T/2} d t  \;\textrm{e}^{ 2\pi i f t} s_\alpha \left(t\right)\,,
\end{eqnarray}
we can then introduce the ensemble average of Fourier modes as
\begin{eqnarray}\label{eq:spectral_density}
\langle \tilde{s}_\alpha(f) \tilde{s}_\beta^*(f') \rangle \equiv \frac{1}{2} \delta^{(1)}(f-f') S_{\alpha\beta}\left(f\right) \; , 
\end{eqnarray}
where the Greek indices run over the interferometric channels, and the {\it spectral densities} $S_{\alpha\beta}\left(f\right)$ are  Hermitian matrix-valued functions, related to the signal PSD $S_h(f)$ through the detector response functions $\mathcal{R}_{\alpha\beta}$ as 
\begin{eqnarray}
\label{eq:signal_correlators}
S_{\alpha\beta}\left(f\right) =  \mathcal{R}_{\alpha\beta} (f) S_h (f) \; .
\end{eqnarray}
We note that we have included a factor $1/2$ in Eq.~(\ref{eq:spectral_density}) because it is conventional to work with the ``one-side'' spectral densities $S_{\alpha\beta}(f)$, defined only for positive (physical) frequencies.  We highlight that Eq.~(\ref{eq:signal_correlators}) quotes the relationship in the limit of negligible relative motion between detectors (which is only a simplifying assumption for space-based interferometers). In practice, we consider $\mathcal{R}_{\alpha\beta} (f)$ as real and symmetric matrices, though in some peculiar scenarios this property might not hold, see e.g.~Ref.~\cite{Domcke:2019zls}.

Assuming a stationary and real noise $n_\alpha(t)$ in each channel, we also write
\begin{equation}
\label{eq:singlesidednps}
\hspace{-1cm}	\langle \tilde{n}_\alpha (f) \tilde{n}_\beta^*(f') \rangle = \frac{1}{2} \delta^{(1)}(f-f') N_{\alpha\beta}\left(f\right) \;,
\end{equation}
where $N_{\alpha\beta}(f)$ is a Hermitian matrix function of {\it single-sided noise power spectra}. We note that in the literature it is common to refer to the functions $N_{\alpha\beta}(f)$ as $P_{n}^{\alpha\beta}$. Their dimension is $\rm{Hz}^{-1}$ since both $\tilde{n}_\alpha(f)$ and $\delta^{(1)}(f-f')$ have dimension $\rm{Hz}^{-1}$. 

As for cosmological signals, it is common practice to describe the GWB in terms of $\Omega_{\rm GW}(f)$, we can also define an equivalent noise spectrum in ``$\Omega$-units" as
\begin{equation}\label{eq:noiseOmegaUnits}
		\Omega^{\alpha\beta}_{\rm noise} (f) = \frac{4 \pi^2 f^3}{3 H_0^2} \frac{N_{\alpha\beta} (f)}{\mathcal{R}_{\alpha\beta} (f)} \;.
\end{equation}

\subsubsection{LISA noise parametrisation}

We now specialise our discussion on the LISA detector~\cite{Audley:2017drz, Colpi:2024xhw}, for which specific noise contributions are expected~\cite{LISAnoise}. LISA  is a mission planned to be launched in the mid-2030s, and which will probe GWs in the milli-Hertz frequency window. LISA will consist of a constellation of three satellites forming a nearly equilateral triangle with $\sim2.5$ million km length arms, and by monitoring the relative displacements among the satellites, it will perform three correlated interferometric measurements, known as the $X, Y$ and $Z$ data streams, and technically referred to as the time-delay-interferometry (TDI) channels. The latter can be conveniently transformed into three uncorrelated data streams, typically dubbed A, E, and T, which diagonalise the signal and noise covariance matrices~\cite{Hogan:2001jn, Adams:2010vc}. Note that the diagonalisation is a strong simplifying assumption and in this work we will describe the signal and noise in the A, E, and T channel basis, which 
allows to break degeneracies between the signal and noise. 
We also assume, for simplicity, that we are working with `perfect' residuals so that all transient signals (and possible glitches) have been subtracted from the data time stream. Because of periodic operational antenna tasks, LISA data are expected to be split into data streams of finite duration, which for the sake of comparison with~\cite{Caprini:2019pxz,Flauger:2020qyi}, we take as $T=11.5$ days.

While Eqs.~(\ref{eq:s_f})-(\ref{eq:noiseOmegaUnits}) are rather generic for any (multi-channel) GW interferometer, in order to simulate LISA mock data, we need a characterisation of the LISA noise. Our current understanding of the latter is based on the LISA Pathfinder experiment~\cite{Armano:2016bkm} and laboratory tests. The TDI channels of the LISA detector are designed to eliminate the dominant source of noise caused by fluctuations in the central frequency of the laser, as well as the noise caused by displacements of the optical benches. Assuming that is 100\% effective, the residual noise components in each TDI channel are then grouped into two effective contributions coined as ``Interferometry Metrology System" (IMS) and ``acceleration" noise. The IMS and acceleration noise power spectrum densities (PSD) are given by~\cite{LISAnoise}
\begin{eqnarray}
	\begin{aligned}
	P_{\rm IMS}(f, P) &= A_P^2~ \frac{{\rm pm}^2}{\rm Hz}\left[1 + \left(\frac{2\,\textrm{mHz}}{f} \right)^4  \right] \left(\frac{2 \pi f}{c} \right)^2\; , \\
	P_{\rm acc}(f, A) &= A_{\rm acc}^2~ \frac{{\rm fm}^2}{{\rm s}^4\,{\rm Hz}}\left[1 + \left(\frac{0.4\,\textrm{mHz}}{f} \right)^2  \right] \left[1 + \left(\frac{f}{8\,\textrm{mHz}} \right)^4  \right] \left(\frac{1}{2 \pi f } \right)^4 \left(\frac{2 \pi f}{c} \right)^2 \; .
	\end{aligned}
\label{eq:acc_int_noise}
\end{eqnarray}
with $P_{\rm IMS}$ dominating the noise at high-frequencies, $P_{\rm acc}$ at low-frequencies, $f$ is the frequency, $c$ the speed of light in units of $m/s$, and $A_{P},A_{\rm acc}$ are dimensionless noise parameters. Following Ref~\cite{Caprini:2019pxz} we set amplitudes to be $A_{P}=15$ and $A_{\rm acc}= 3$ with $\pm \,$20\% margins. 

The determination of the noise properties is, of course, one of the major challenges of the LISA mission. While ground-based characterisation of the noise is technically difficult, in-flight characterization might be simply not even feasible in realistic conditions, see e.g.~\cite{Muratore:2023gxh}. The precise details of this, however,  go beyond the scope of our work, and for practical purposes we follow Ref~\cite{Flauger:2020qyi}. Therefore we adopt their simplifying assumptions\footnote{These include considering that the noise spectra for all links are identical, stationary, and uncorrelated, that the fluctuations of the masses are isotropic and stationary, the power spectra for all test masses are equal, the fluctuations of the different  masses are uncorrelated, and that the three satellites form an equilateral triangle of length side $L_1=L_2=L_3=L=2.5 \times 10^9\,{\rm m}$.}, which lead to power spectrum densities in the (now diagonal) AET basis as
\begin{align}
 N_{AA}(f,A_{P},A_{\rm acc}) =&~ N_{EE}(f,A_{P},A_{\rm acc})\nonumber \\
 =&~ 8\sin^{2}\left(\frac{2\pi fL}{c}\right)\left\lbrace4\left[1+\cos\left(\frac{2\pi f L}{c}\right)+\cos^{2}\left(\frac{2\pi f L}{c}\right)\right]P_{\rm acc}(f,A_{\rm acc}) \right. \nonumber \\
&\hspace{3cm}+ \left.\left[2+\cos(\frac{2\pi fL}{c})P_{\rm IMS}(f,A_{P})\right]\right\rbrace\,,\vspace{4mm}\\
N_{TT}(f,A_{P},A_{\rm acc}) =&~16\sin^{2}\left(\frac{2\pi fL}{c}\right)\left\lbrace2\left[1-\cos\left(\frac{2\pi f L}{c}\right)\right]^{2}P_{\rm acc}(f,A_{\rm acc})\right. \nonumber \\
&\hspace{3cm}+ \left.\left[1-\cos\left(\frac{2\pi fL}{c}\right)P_{\rm IMS}(f,A_{P})\right]\right\rbrace\,
\end{align}
where $L=2.5 \times 10^9\,{\rm m}$ is the length between every two satellites in the LISA detector, with the three satellites assumed to form an exact equilateral configuration. 

As discussed in detail in Ref.~\cite{Flauger:2020qyi}, the quadratic response functions, $\mathcal{R}_{\alpha\beta}(f) \equiv S_{\alpha\beta}(f)/S_h(f)$, c.f.~Eq.~(\ref{eq:signal_correlators}), can be written as
\begin{equation}
\mathcal{R}_{\alpha\beta}(f) = 16 \sin^2\left(\frac{2 \pi f L}{c}\right) \left(\frac{2 \pi f L}{c} \right)^{2} \tilde{R}_{\alpha\beta}(f) \; , \qquad 
\end{equation}
with $\tilde{R}_{\alpha\beta}$ functions that depends on the geometry of the detector. We take the response functions $\tilde{R}_{\alpha\beta}$ for the different channel cross-spectra from~Ref.~\cite{Flauger:2020qyi}, where a detailed derivation can be found in their Appendix A.3. While the exact form of $\tilde{R}_{\alpha\beta}$ that we use depend on (not very illuminating) expressions given in terms of multidimensional integrals, simple analytic approximations can be obtained nonetheless as~\cite{Cornish:2018dyw,Flauger:2020qyi}
\begin{eqnarray}
	\tilde{R}_{ \text{AA} }(f) = \tilde{R}_{ \text{EE} }(f) \approx \frac{\frac{9}{20}}{1 +0.7 \left(\frac{2 \pi f L}{c}\right)^2} \,,~~ \tilde{R}_{ \text{TT} }(f) = \frac{\frac{9}{20}\left(\frac{2 \pi f L}{c}\right)^6}{1.8 \times 10^3 +0.7 \left(\frac{2 \pi f L}{c}\right)^8} \,. 
\end{eqnarray}

From the power spectrum density and the response functions we can define the strain sensitivities
\begin{equation}
S_{\rm noise}^{\alpha\beta}(f, A_{P},A_{\rm acc}) = \frac{N_{\alpha\beta}(f,A_{P},A_{\rm acc})   }{  \mathcal{R}_{\alpha\beta}(f)  }   = \frac{N_{\alpha\beta}(f,A_{P},A_{\rm acc})   }{   16 \sin^2\left(\frac{2 \pi f L}{c}\right) \left(\frac{2 \pi f L}{c} \right)^{2} \tilde{R}_{\alpha\beta}(f)} \,, 
\end{equation}
and putting everything together, we arrive at the LISA noise model we use for the generation of data realizations and extraction of noise parameters used in our analyses, in units of the energy density parameter
\begin{align}
\Omega_{\rm noise}^{\alpha\beta}(f,A_{P},A_{\rm acc})\equiv \frac{4\pi^{2}f^{3}}{3H_{0}^{2}}S_{\rm noise}^{\alpha\beta}(f,A_{P},A_{\rm acc})\,.
\label{eq:Omeganoise}
\end{align}

\section{Data simulation and statistical analysis}
\label{sec:analysis}

\subsection{Data generation}
\label{subsec:dataGeneration}

To generate mock data we try to imitate as close as possible the real data-taking procedure of LISA, which is expected to take place during a period of $4$ years with $75\%$ efficiency, hence in practice providing an observation time of $3$ years. As discussed before, we assume the data will be divided into data segments (\textit{chunks}) of roughly $11.5$ days each, clean of noise glitches and transient signals. These implies $N_s = 94$ chunks, which we label as $i = 1, 2, ..., N_c$. Since the frequency range for LISA approximately extends from $3 \times 10^{-5}$ Hz to $5 \times 10^{-1}$ Hz, with a frequency spacing of $\Delta f \sim 10^{-6}$ Hz set by the length of each data stream chunk, this amounts to approximately~$\sim 5 \times 10^{5}$ frequency points $\lbrace f_j \rbrace$ per chunk, and thus in total around $\sim 5 \times 10^{7}$ data points.

As seen in Sect.~\ref{subsec:noise}, for a stationary signal (due to a statistically homogeneous and isotropic, unpolarised GWB) and stationary noise (uncorrelated with the signal), the auto- and cross-spectra of different data streams $d_{\alpha}(t) = s_{\alpha}(t) + n_{\alpha}(t)$ read
\begin{equation} 
 \langle \tilde d_\alpha(f) \tilde{d}^*_\beta(f) \rangle' = \mathcal{R}_{\alpha\beta}  S_h (f)   +   N_{\alpha\beta} (f)\,,
\label{eq:data_correlation_app_XYZ}
\end{equation} 
where we have used Eqs.~(\ref{eq:spectral_density})-(\ref{eq:singlesidednps}), and the prime indicates that we have stripped the $\delta^{(1)}$-function and the factor $1/2$. For each pair of channels $\alpha,\beta$ and chunk $i$, we generate the data realizations of $\lbrace \tilde d^{\,i}_{\alpha}(f_j)\rbrace$ directly in the frequency domain, following the procedure described in Ref.~\cite{Caprini:2019pxz}, by defining
\begin{eqnarray}
D_i^{(\alpha\beta)}(f_j) \equiv \frac{4\pi^{2}f^{3}}{3H_{0}^{2}} \frac{d_{\alpha}^i(f_j) d^{i \, *}_{\beta}(f_j)}{ \mathcal{R}_{\alpha\beta}} = \Omega_{{\rm GW},i}(f_j) + \Omega_{{\rm noise},i}^{\alpha\beta}(f_j)\,.
\end{eqnarray}

In practice, for given channels $\alpha,\beta$, we generate for each frequency $f_{j}$ of a time chunk ($i = 1, 2, ..., N_c$), a data point resulting from the sum of the signal and noise contributions
\begin{align}
  D_{i,j}^{\alpha\beta} = S_{i,j} + \mathcal{N}_{i,j}^{\alpha\beta} ,
  \label{eq:data}
\end{align}
where the $S_{i,j}$ and $\mathcal{N}_{i,j}^{\alpha\beta}$ are obtained from the power spectra according to
\begin{align}
  S_{i,j}^{\alpha\beta}=\left|\frac{G_{1}\left(0,\sqrt{\Omega_{\rm GW}^{\alpha\beta}(f_{j})}\right)+iG_{2}\left(0,\sqrt{\Omega_{\rm GW}^{\alpha\beta}(f_{j})}\right)}{\sqrt{2}}\right|^{2}
\end{align}
and
\begin{align}
\mathcal{N}_{i,j}^{\alpha\beta}=\left|\frac{G_{3}\left(0,\sqrt{\Omega_{\rm noise}^{\alpha\beta}(f_{j})}\right)+iG_{4}\left(0,\sqrt{\Omega_{\rm noise}^{\alpha\beta}(f_{j})}\right)}{\sqrt{2}}\right|^{2}.
\end{align}
with $G_{k}(0,\sigma)$ real numbers randomly drawn from a Gaussian distribution with zero mean and standard deviation set by the square root of the GW and noise power spectra (evaluated at $f_j$). For each frequency $f_j$, we obtain the mean across time chunks
\begin{equation}
\bar{ D}_{j}^{\alpha\beta}=\frac{1}{N_c}\sum_{i=1}^{N_c} D_{i,j}^{\alpha\beta} \,.
\end{equation}
As within the LISA frequency range 
of each chunk there are ~$\sim 5 \times 10^{5}$ frequency points $\lbrace f_j \rbrace$, to reduce the computational load we rebin (``coarse grain'') the frequencies from $f = 10^{-3}$ Hz to the maximum frequency $f_{max} = 0.5$~Hz, in 1000 intervals of equal log-spacing, while we keep all the original values from the minimum frequency $f_{min} = 3 \cdot 10^{-5}$ Hz to $f = 10^{-3} $~Hz. The resulting data has a total of 1971 bins per chunk. The frequency $f$ assigned to each bin in the coarse-grained region is then obtained by the mean of the frequencies within the bin. In order to obtain the coarse-grained data, we follow~\cite{Giese:2021dnw} and define the weights of each channel as (for clarity we omit the $\alpha,\beta$ indices):
\bea
w_j \equiv w(f_j, {\bf n}) = \frac{(\Omega_{\rm noise}(f_j, {\bf n}))^{-1}}{ \sum_{l\in{\rm bin}~k} (\Omega_{\rm noise}(f_l, {\bf n}))^{-1} }~,
\label{eq:weights}
\eea
corresponding to the bin $j$-th inside the ``macro-bin'' $k$, and where the parameters of the noise are ${\bf n}\equiv\{A_{P},A_{\rm acc}\}$. With this, we redefine the dataset in the coarse-grained frequency region as
\be\label{eq:weightedFreqs}
f_k \equiv \sum_{j\in{\rm bin}~k} w_j f_j;~~~~ \bar{D}_k \equiv \sum_{j\in{\rm bin}~k} w_j \bar{D}_j~.
\ee
We highlight that the dataset in the coarse-grained frequency region will depend on the values of $A_{P}$ and $A_{\rm acc}$. In practice, as we work in the AET basis, which is diagonal, we simulate data only in the TT, AA, and EE channels:

{\bf TT channel.} This channel is to a very good approximation only sensitive to the instrumental noise for frequencies below 0.02 Hz. We therefore generate TT-channel data using exclusively our noise model as defined in Eq.~(\ref{eq:Omeganoise}) in this frequency region. The values of the parameters $A_{P},A_{\rm acc}$ are drawn from Gaussian priors ${G}(A_{P}|\mu_P,\sigma_P)$, and ${G}(A_{\rm acc}|\mu_{\rm acc},\sigma_{\rm acc})$ respectively, where $\mu_P=15$, $\mu_{\rm acc}=3$, and $\sigma_P/\mu_P = \sigma_{\rm acc}/\mu_{\rm acc} = 0.2$.

{\bf AA, EE channels}. In these channels both noise and signal contributions are present. Consequently, we generate data using, on the one hand, the same noise model Eq.~(\ref{eq:Omeganoise}) as in the TT channel, and on the other hand, on top, a signal according to Eq.~(\ref{eq:signal_sim}). The priors of the signal parameters are taken to be uniform, such that the overall amplitude is drawn from $\log A_s\sim~{\rm unif}[-14,-6]$, while the segment slopes follow $\alpha_n\sim{\rm unif}[-12,12]$. 

\subsection{Simulation-based proposed methodology}
\label{subsec:methodology}

In our present work, we use and adapt the implementation of the methodology given by the {\tt sbi}~\cite{Tejero-Cantero2020} python library. 
In this section, we describe the generic features of our method, while the more technical details can be found in the specialized cited references. We point the reader to appendix \ref{app:A} for a qualitative comparison with other popular methods based on similar techniques. 
Our set of tools are collected all together in our code, {\tt GWBackFinder}, which is publicly available in a {\tt GitHub} repository.\footnote{\href{https://github.com/AndronikiDimitriou/GWBackFinder}{\tt https://github.com/AndronikiDimitriou/GWBackFinder}}

As described in Section~\ref{subsec:dataGeneration}, we have created a simulator that, upon specifying the value of the input parameters $\bs{\theta}\equiv\{{\bf s,n}\}$, produces the data pairs $\{f_j, \bar D_j^{\alpha\beta}\}$ 
for each frequency $f_j$, 
where $\bar D_j^{\alpha\beta}$ is a random realisation (for each channel $\alpha\beta$) of noise $h^2\Omega_{\rm noise}^{\alpha\beta}$ plus signal $h^2\Omega_{\rm GW}$, c.f.~Eq.~(\ref{eq:data}), averaged over chunks. To explain our method next, we ease the notation in the following, dropping the $bar$ and the channel indices from the data variables, $\bar{D}_j^{\alpha\beta} \rightarrow D_j$, understanding implicitly, from now on, that $D_j$ implies the chunk-averaged data amplitude for the frequency $f_j$, for some channel pair $\alpha,\beta$. Our goal is to estimate a posterior probability for the parameters $\bs{\theta}$, $p(\bs{\theta}|{\bf D})$, where ${\bf D}=\{D_j\}$ is an empirical (observed, or simulated test) dataset with as many entries as frequencies, without specifying explicitely the likelihood $p({\bf D}|\bs{\theta})$.

We use a particular flavor of state-of-the-art {\tt simulation-based} (SB) methods, belonging to the family of Neural Posterior Estimation (NPE)~\cite{rezende2016}. The idea of NPE is based on an approximate-inference approach where the desired posterior probability distribution 
is being estimated directly, by using a {\it neural network} model. The approximated posterior $q_{\bs{\phi}}(\bs{\theta}|{\bf D})$ is parameterised in our case by a ``normalizing flow''~\cite{rezende2016, durkan2019}, with {\it optimisable} parameters~$\bs{\phi}$ (see appendix \ref{app:D} for a more technical specification). In order to optimise $q_{\bs{\phi}}$ to be as similar as possible to the true (unknown) posterior $p(\bs{\theta}|{\bf D})$, typically NPE methods minimise the expected value of the forward Kullback-Leibler (FKL) divergence. In particular, the function to minimise is
\bea
\label{eq:fkl}
{\cal L}(\bs{\phi}) =\mathbb{E}_{p({\bf D})}[{\rm FKL}] &=& - \int d{\bf D} d\bs{\theta}~ p({\bf D},\bs{\theta})
\ln q_{\bs{\phi}}(\bs{\theta}|{\bf D}) ~+~{\it const}~,\\
~&\approx& - \frac{1}{N_d} \sum_{s=1}^{N_d} \ln q_{\bs{\phi}}(\bs{\theta}_s|{\bf D}_s)~+~{\it const}~, \nonumber
\eea
where {\it const} in the above equation refers to terms that do not depend on the optimisable  parameters $\bs{\phi}$, and thus can be dropped from the optimisation. Note that the expectation above is taken over $N_d$ datasets. The key observation is that, even though we do not have access to the evidence $p({\bf D})$, in a Monte Carlo approximation of the integral in Eq.~(\ref{eq:fkl}) -- see second line -- we use as samples ${\bf D}_s$ the ones obtained from the simulator, corresponding to the input $\bs{\theta}_s$, the latter being sampled from a given prior $p(\bs{\theta})$. This is where the ``likelihood-free'' (or simulation-based inference) approach comes into play: In this approach, a stochastic simulator is used as an implicit representation of the likelihood, whose numerical evaluation is no longer necessary. In other words, instead of evaluating the likelihood, this method uses simulated datasets ${\bf D}_s$, which are formally samples from the simulator, i.e.~from the implicit likelihood $p(\bs{D|\theta})$. While the number $N_d$ of datasets required by the algorithm depends on the complexity of the data itself, the simulator, as well as on the number of parameters $\bs{\theta}$, in general, these models require a large $N_d$ in order to be trained. This is because, when estimating the posterior for a new dataset ${\bf D}_{N_d+1}$, the closer the latter resembles previously ``seen'' datasets $\lbrace {\bf D}_{s} \rbrace_{s=1, ..., N_d}$, the better the results. This approach is referred, in the specialised literature, as ``amortisation'': since the algorithm has been exposed previously to an exhaustive set of datasets $\lbrace {\bf D}_{s} \rbrace$, making inference on a new dataset ${\bf D}_{N_d+1}$ is amortised, in the sense that no further costly computations are required.  
In our case, we need $N_d\sim \mathcal{O} (10^{6})$.
Having optimised the approximate posterior, $q_{\bs{\phi}_*}(\bs{\theta}|{\bf D})$, where $\bs{\phi}_* ={\rm argmin}{\cal L}(\bs{\phi})$ are the parameters that minimise Eq.(\ref{eq:fkl}) in principle for any possible dataset ${\bf D}$ (due to the expectation in Eq.~(\ref{eq:fkl})), the method readily outputs the posterior probability for any unseen dataset ${\bf D}_o$ that is obtained in the future. On the other hand, it allows by construction to obtain samples from $q_{\bs{\phi}_*}$, as specified in Eq.~(\ref{eq:fkl}).

In minimizing our objective function (Eq. \ref{eq:fkl}), we explicitly implement an over-fitting control procedure by partitioning our set of datasets in ``training'' and ``validation'', as typically done in the machine learning literature\footnote{Note, however, that the presence of the prior in Eq. \ref{eq:fkl} acts as a natural regularizer against over-fitting, since it forces some similarity between the approximate distribution $q$ and the prior distribution. This is well known in the literature of Variational Inference.}.

\subsubsection{Conditional inference}
\label{subsec:conditional}

The statistical methodology just described requires further tuning to our particular problem. As it is already recognised by the GW community, a simultaneous inference of both signal and noise parameters is challenging, given not only the loss of sensitivity for frequencies sufficiently lower or higher than the peak sensitivity (e.g.~$\sim $mHz in the case of LISA), but also given the intrinsic noise of the data themselves.

Fortunately, 
the data acquired by LISA, when expressed 
in the convenient AET diagonal channel basis, is such that the data from the TT channel, ${\bf D}_{\rm TT}$, is mostly sensitive to the noise, with the signal being almost irrelevant. Consequently, we propose to split the inference procedure into two steps: {\it i)} inference of the noise parameters ${\bf n}$ from the TT channel, and {\it ii)} inference of the signal parameters ${\bf s}$ from the AA or EE channel, {\it conditioned} on the previous noise parameters. This strategy was actually already adopted in Ref.~\cite{Flauger:2020qyi}, and is nothing but an implementation of the old idea, common in many other physics domains, of inferring the signal from the residuals obtained after subtracting the noise from the data. As expected, the reconstruction capabilities improve appreciably as compared to an inference of noise and signal parameters all together. 

A note is in order at this point. Given the concrete LISA noise model we use in this paper (which is the same used in~\cite{Flauger:2020qyi}), it turns out that the TT-channel data is mostly 
sensitive to the noise parameter $A_{P}$, and not to 
$A_{\rm acc}$. 
Consequently, our two-step sequential procedure is actually given by: {\it i)} estimating the posterior distribution of $A_{P}$ from the TT-channel data, i.e.~$p(A_{P}|{\bf D}_{\rm TT})$, and {\it ii)} estimating both the noise parameter $A_{\rm acc}$ and the signal parameters ${\bf s}$ from the AA-channel (or alternatively the EE-channel), conditioned on $A_{P}$, i.e.~$p({\bf s},A_{\rm acc}|{\bf D}_{\rm AA}, A_{P})$. Importantly, in step {\it i)} we assume that there is no signal contribution to the TT data, given that 
it is expected that reasonable signals are very subdominant compared to the TT noise, especially at low frequencies where $P_{\rm IMS}$ dominates over any other contribution in the TT channel.


In the context of our technique, we have two approximate posteriors: $q_{\bs{\eta}}(A_{P}|{\bf D}_{\rm TT})$ with optimisable parameters $\bs{\eta}$ from step {\it i)}, and $q_{\bs{\zeta}}({\bf s},A_{\rm acc}|{\bf D}_{\rm AA},A_{P})$ with optimisable parameters $\bs{\zeta}$ from step {\it ii)}. The two-step procedure just described delivers samples $\{{\bf s}_\lambda,{\bf n}_\lambda\}$, with $\lambda = 1, ..., N_s$, where $N_s$ is the number of samples. We choose to present the expected reconstructed signal as $\Omega_{\rm GW}(f; \langle{\bf s}\rangle)$, computed with the mean values of the signal parameters $\langle{\bf s}\rangle$, according to expression~(\ref{eq:signal_sim}). At the same time, we can predict the uncertainty in the spectrum reconstruction from the density regions of the distribution $p(\Omega_{\rm GW}(f;{\bf s}_\lambda))$ inherited from the samples ${\bf s}_\lambda$.

\subsubsection{Sequential procedure}
\label{sec:sequential}

So far, we have described an amortisation procedure which, upon training on a large enough number of datasets, is capable of predicting the posterior distribution corresponding to a given new observation ${\bf D}_o$. 
Sometimes though, the resulting posterior may not be as accurate as the one coming from other approaches that exclusively focus from the beginning on ${\bf D}_o$, like for example MCMC methods. For this reason, some recent strategies have been proposed in the context of simulation-based inference, in order to achieve
more accurate results for particular datasets of interest. This is the case of the multi-round inference, which we {\it also} consider in our work. 

In general, the idea of multi-round inference is to refine the estimate of the posterior $p(\bs{\theta}|{\bf D}_o)$ of the parameters of interest $\bs{\theta}$, focusing on a particular observation ${\bf D}_o$, via a sequential procedure consisting on several rounds. For each round, a relatively small number of datasets is considered in order to update the posterior $p(\bs{\theta}|{\bf D}_o)$, in the same ``amortisation'' spirit as described previously, with the only difference that now the number of datasets is much smaller. The updated posterior is then used in the next round, as the prior to generating the datasets, and so on. In this way, in each round, the attention is more focused on the region that is 
more relevant to ${\bf D}_o$. In the context of the particular implementation of the NPE approach we use in this work, we follow the multi-round inference algorithm presented in \cite{greenberg2019}.

We highlight the benefits of keeping both amortisation and multi-round inference procedures as part of the data analyst's toolbox: while amortisation gives us a very fast prediction for the posterior distribution for a given dataset, the multi-round inference becomes very useful when, as we describe below, we need a more refined prediction on a particular dataset. Furthermore, when compared to some flavors of MCMC (e.g.~\cite{Foreman_Mackey_2013}), it has been shown that multi-round inference is computationally cheaper for certain problems. 

In Sects.~\ref{sec:BlindReconstruction}, \ref{sec:foreground} and \ref{sec:template} we will show the results of three different exercises: first, a blind reconstruction of generic (template-free) signals with a variety of spectral shapes (Sect.~\ref{sec:BlindReconstruction}); secondly, examples of blind reconstruction of signals in the presence of astrophysical foregrounds (Sect.~\ref{sec:foreground}); and thirdly, an example of template-based signal reconstruction, focusing in the simple case of a power law for illustrative purposes (Sect.~\ref{sec:template}).
While we only use the amortisation procedure described above (but no multi-round inference) for our results based on a blind reconstruction of GWB of arbitrary spectra in sections~\ref{sec:BlindReconstruction} and \ref{sec:foreground}, we use as well the multi-round inference procedure in the template reconstruction results presented in  
Sect.~\ref{sec:template}. 

\subsection{Comparison to MCMC}
\label{subsec:mcmc}

Previous studies on GWB spectrum reconstruction rely on MCMC inference, for which a likelihood has to be specified. While certainly the simulated data considered in such studies are formally non-Gaussian, in Ref.~\cite{Flauger:2020qyi} an improved likelihood was used, where a Lognormal contribution is added to the Gaussian one. Such likelihood is inspired by Cosmic Microwave Background (CMB) physics, where the Lognormal contribution accounts for the skewness of the true likelihood which is just slightly non-Gaussian. 
It is however not clear if GWB data as measured by LISA or other experiments enjoy the same properties. 

As commented in the introduction, there are several advantages of our methodology as an alternative to a standard MCMC analysis, like for example: $a)$ being a likelihood-free approach, we do not need to know the correct likelihood, nor to assume any approximation of it, and $b)$ as a pre-trained method, we do not require computationally costly evaluations in order to make inference for a given dataset\footnote{This is strictly true for the amortisation procedure, as in the multi-round inference procedure there is an extra computational cost by training based on a particular dataset. However, typically the number of samples required by multi-round inference is much smaller than those required by typical flavours of MCMC, as Metropolis-Hastings and Nested Sampling.}. 

Still, whenever possible, a comparison with MCMC is desirable, since $1)$ this allows us to compare directly our new methodology with the methodology followed so far in the literature, and $2)$, provided that we know the correct likelihood (or at least, a reliably good approximation of it), it allows for a useful cross-check of any approximate inference method, as in our case.  For our MCMC computation, we adopt the likelihood from~\cite{Giese:2021dnw,Flauger:2020qyi} 
\be
\log {\cal L}_{\rm G+LN}({\bf s},{\bf n}) = \frac{1}{3}\log{\cal L}_{\rm G}({\bf s},{\bf n}) +
\frac{2}{3}\log{\cal L}_{\rm LN}({\bf s},{\bf n})~.
\label{eq:likelihood}
\ee
with Gaussian contribution
\be
\log{\cal L}_{\rm G}({\bf s},{\bf n}) = -\frac{N_c}{2}
\sum_k n_{k} \left(
  \frac{\bar D_k-\Omega_{\rm noise}(f_k;{\bf n})-\Omega_{\rm GW}(f_k;{\bf s})}
  {\Omega_{\rm noise}(f_k;{\bf n})+\Omega_{\rm GW}(f_k;{\bf s})}
\right)^2~,
\label{eq:Gaussian}
\ee
and Log-normal contribution 
\be
\log{\cal L}_{\rm LN}({\bf s},{\bf n}) = -\frac{N_c}{2}
\sum_k n_{k}
\log^2\frac{\Omega_{\rm noise}(f_k;{\bf n})+\Omega_{\rm GW}(f_k;{\bf s})}{\bar{D}_k}~.
\label{eq:lognormal}
\ee
and where $n_{k}$ is the number of frequency values within the $k$-th frequency bin ($n_k>1$ for the coarse-grained frequency region, cf.~Eq.~\ref{eq:weightedFreqs}, and $n_k = 1$ otherwise). 
We note that while we have checked that the likelihoods in Eqs.~(\ref{eq:Gaussian}),(\ref{eq:lognormal}) for the binned data are not formally equivalent to the corresponding unbinned likelihoods, in practice the bias introduced by accounting for the binning in this way turns out to be visibly negligible, at the level of the posterior distributions.


On the other hand, when implementing our MCMC we actually follow the same conditional inference strategy described in Sect.~\ref{subsec:conditional}. Consequently, in practice we infer first the noise parameter $A_P$ by using ${\cal L}_{\rm G+LN}({\bf 0},{A_P})$ in Eq.~(\ref{eq:likelihood}), while using the TT-channel data, $\{D_j\} = {\bf D}_{\rm TT}$. Afterwards, we infer the signal parameters ${\bf s}$, jointly with the noise parameter $A_{\rm acc}$, conditional on a given value of $A_P$, say $A_P=A_P^*$. For this second step we thus use  ${\cal L}_{\rm G+LN}({\bf s},A_P^*,A_{\rm acc})$, using now AA-channel data, $\{D_j\} = {\bf D}_{\rm AA}$,  in order to obtain samples from $p({\bf s},A_{\rm acc}|{\bf D}_{\rm AA})$. In this work, we use the public library {\tt emcee} \cite{Foreman_Mackey_2013}, implementing a popular flavor of MCMC among the cosmology \& astrophysics community, while we have cross-checked some of the results with an independent implementation of the HMC algorithm\footnote{See \cite{neal:2012} for a well-known presentation of it.}, finding full compatibility.


\section{Blind reconstruction}
\label{sec:BlindReconstruction}

In this section, we test the ability of our method to do model-independent signal reconstruction in LISA. Namely, we reconstruct GWB spectra with arbitrary shapes which are a priori unknown to the simulator. We refer to this as {\it blind} reconstruction. We use a variety of signal frequency profiles motivated by astrophysical and cosmological GWBs. Namely, 
%
\renewcommand{\arraystretch}{2}
\begin{table}[tbp]
  \hspace*{-4mm}
  \centering
  \begin{tabular}{|c|c|c|}
    \hline
Type of signal & Functional form of
    $\Omega_{\rm GW}(f)$ & Physical cases\\
    \hline
    \, & \,\vspace*{-10mm}\\
    \hline
1. Single power law &
    $\Omega_p^{(*)}\left(\frac{f}{0.003~{\rm Hz}}\right)^{\gamma}$
    & 
    $\begin{array}{c}
     {\rm SOBHB~background}\vspace*{-4mm}\\ 
     {\rm inflation}\vspace*{-4mm}\\ 
     {\rm kination-domination}\vspace*{-4mm}\\ 
     {\rm cosmic~ defects}
    \end{array}$\\
    \hline
    2. Broken power law &
        $\Omega_p^{(*)} \left[\left(\frac{f}{f_p}\right)^{\gamma_1}\hspace*{-1mm}\Theta(f_p-f) + 
        \left(\frac{f}{f_p}\right)^{\gamma_2}\hspace*{-1mm}\Theta(f-f_p) \right]$ & 
     $\begin{array}{c}
     {\rm 1stOPhT}\vspace*{-4mm}\\ 
     {\rm spinning-axions}\vspace*{-4mm}\\ 
     {\rm KD~within~RD}
    \end{array}$\\        
    \hline
    3. Single bump signal &
       $\Omega_p^{(*)} {\rm Exp}\left\{
        - \frac{[\log_{10}(f/f_b)]^2}{\Delta^2}
    \right\}$ 
    & 
     $\begin{array}{c}
     {\rm Preheating}\vspace*{-4mm}\\ 
     {\rm Oscillons}\vspace*{-4mm}\\
     {\rm axion~fragmentation}\vspace*{-4mm}\\
     {\rm Galactic~background}
     \end{array}$\\
    \hline
   4. Multi-bump signal &
      $\sum_j \Omega_{j}^{(*)} {\rm Exp}\left\{
        - \frac{[\log_{10}(f/f_j)]^2}{\Delta_j^2}
        \right\}$ 
        & 
     $\begin{array}{c}
     {\rm 1stOPhT}\vspace*{-4mm}\\ 
     {\rm Preheating}
    \end{array}$\\
    \hline 
  \end{tabular}
  \caption{Frequency profiles of signals considered for blind reconstruction.}
  \label{tab:mock}
\end{table}

1) {\it Power law} profiles. These can represent, for example, the astrophysical GWB from unresolved binary mergers, see e.g.~\cite{LIGOScientific:2019vic, Sesana:2016ljz,Regimbau:2011rp,Babak:2023lro,Lehoucq:2023zlt}. They can also represent a GWB associated to inflation, for instance a blue tilted signal created by particle production during inflation (see e.g.~\cite{Anber:2006xt,Sorbo:2011rz,Pajer:2013fsa,Adshead:2013qp,Adshead:2013nka,Maleknejad:2016qjz,Dimastrogiovanni:2016fuu,Namba:2015gja,Ferreira:2015omg,Peloso:2016gqs,Domcke:2016bkh,Caldwell:2017chz,Guzzetti:2016mkm,Bartolo:2016ami}), or the high-frequency tail of the inflationary background from vacuum-fluctuations, blue-tilted due to a post-inflationary kination-domination era (see e.g.~\cite{Giovannini:1998bp,Giovannini:1999bh,Boyle:2007zx,Li:2016mmc,Li:2021htg,Figueroa:2018twl,Figueroa:2019paj,Li:2021htg,Gouttenoire:2021wzu,Co:2021lkc,Gouttenoire:2021jhk,Oikonomou:2023qfz}). Furthermore, power laws can mimic as well the high-frequency tail of the GWB expected from cosmic defects, see e.g.~\cite{Vachaspati:1984gt,Sakellariadou:1990ne,Damour:2000wa,Damour:2001bk,Damour:2004kw,Fenu:2009qf,Jones-Smith:2007hib,Figueroa:2012kw,Hiramatsu:2013qaa,Blanco-Pillado:2017oxo,Auclair:2019wcv,Gouttenoire:2019kij,Figueroa:2020lvo,Gorghetto:2021fsn,Chang:2021afa,Yamada:2022aax,Yamada:2022imq}, and in particular~\cite{Auclair:2019wcv} for a review on cosmic string signals potentially detectable by LISA.

2) {\it Broken power-law} profiles. These can represent approximately the GWB spectrum from a strong first order phase transition (1stOPhT), see e.g.~\cite{Kamionkowski:1993fg,Caprini:2007xq,Huber:2008hg,Hindmarsh:2013xza,Hindmarsh:2015qta,Caprini:2015zlo,Hindmarsh:2017gnf,Cutting:2018tjt,Cutting:2018tjt,Cutting:2019zws,Pol:2019yex,Caprini:2019egz,Cutting:2020nla,Han:2023olf,Ashoorioon:2022raz, Athron:2023mer,Li:2023yaj}, and in particular Ref.~\cite{Caprini:2019egz} for a review on the expectation of these signals within the LISA band. They can also describe naturally the spectrum of GWBs expected from spinning-axions or, in general, the high-frequency tail of inflationary or cosmic string GWBs, when considering a kinetic-dominated (KD) period within radiation-domination (RD), see e.g.~\cite{Co:2021lkc,Gouttenoire:2021wzu,Gouttenoire:2021jhk}.

3) {\it Single bump} profiles. These can describe the spectrum of GWBs expected from post-inflationary GW causally-generating mechanisms, like preheating (e.g.~\cite{Easther:2006gt,GarciaBellido:2007dg,GarciaBellido:2007af,Dufaux:2007pt,Dufaux:2008dn,Dufaux:2010cf,Bethke:2013aba,Bethke:2013vca,Enqvist:2012im,Figueroa:2013vif,Figueroa:2014aya,Figueroa:2016ojl,Figueroa:2017vfa,Adshead:2018doq,Adshead:2019lbr,Adshead:2019igv,Figueroa:2022iho,Cosme:2022htl}) and oscillons (e.g.~\cite{Zhou:2013tsa,Antusch:2016con,Antusch:2017vga,Liu:2017hua,Amin:2018xfe}). 
Furthermore, the signal from dark-matter axion-fragmentation scenarios~\cite{Machado:2019xuc,Ratzinger:2020oct} or the (yearly-modulated) signal from unresolved galactic binary mergers, are also characterized by single-peaked GWB spectra (which typically display a low-frequency power-law branch, and a high-frequency exponential fall-off branch, and hence could be 
qualitatively described by either single-bump or broken power-law profiles.).

4) 
{\it Multi-bump} profiles. These capture, at least in essence, the main features of GWBs from phase transitions with multiple terms competing, see e.g~\cite{Caprini:2015zlo, Schwaller:2015tja,Greljo:2019xan,Caprini:2019egz}, or from multi-field preheating scenarios, see e.g.~\cite{Dufaux:2010cf,Figueroa:2022iho}. In particular, Ref.~\cite{Figueroa:2022iho} argues that realistic early universe scenarios that
involve multiple particle species in the generation of GWBs, naturally lead to a sequence of peaks at different
frequencies in the GWB spectrum, known as a {\it stairway} signature. While such multi-peak feature has been shown to occur in preheating scenarios~\cite{Figueroa:2022iho}, it remains as a speculation for 1stOPhT's.

The aim of blind reconstruction is, of course, not to ascribe a given scenario to a reconstructed signal, but rather to test the reconstruction capabilities of our procedure, considering a representative variety of spectral profiles. For ease of comparison with Refs.~\cite{Caprini:2019pxz,Flauger:2020qyi}, we adopt the same parametrisation as those references for the spectrum-profile families $i)-v)$, which are presented in Table~\ref{tab:mock}. 

We note that in each of the following plots we show the power law sensitivity (PLS) curve as an `indicator' of the ability of LISA to measure a GWB with power law spectrum.
We build the PLS curve considering a signal as $\Omega_{GW}(f;\beta)=C_{\beta}f^{\beta}$, so that the value of $C_{\beta}$ is determined to achieve a given signal-to-noise ratio threshold ${\rm SNR}_{\rm thr}$ across the entire LISA frequency band:
\begin{equation}
{\rm SNR}_{\rm thr}=\sqrt{T \int_{f_{min}}^{f_{max}}df \frac{(C_{\beta}f^{\beta})^{2}}{\Omega_s^2(f)} }.
\end{equation}
This computation is repeated for a range of $\beta$ values. The PLS curve is then obtained by associating the maximum value of $C_{\beta}f^{\beta}$ to each frequency $f$. This ensures that any power law signal above this curve will have an SNR exceeding ${\rm SNR}_{\rm thr}$. In the following plots, the LISA PLS is calculated for ${\rm SNR}_{\rm thr}$=10 and an observation time T=3 years \cite{Caprini:2019pxz}.

\subsection{Arbitrary frequency-profile reconstruction}
\label{subsec:arbitraryShape}

\begin{figure}[tbp]
\centering
\includegraphics[width=.47\textwidth]{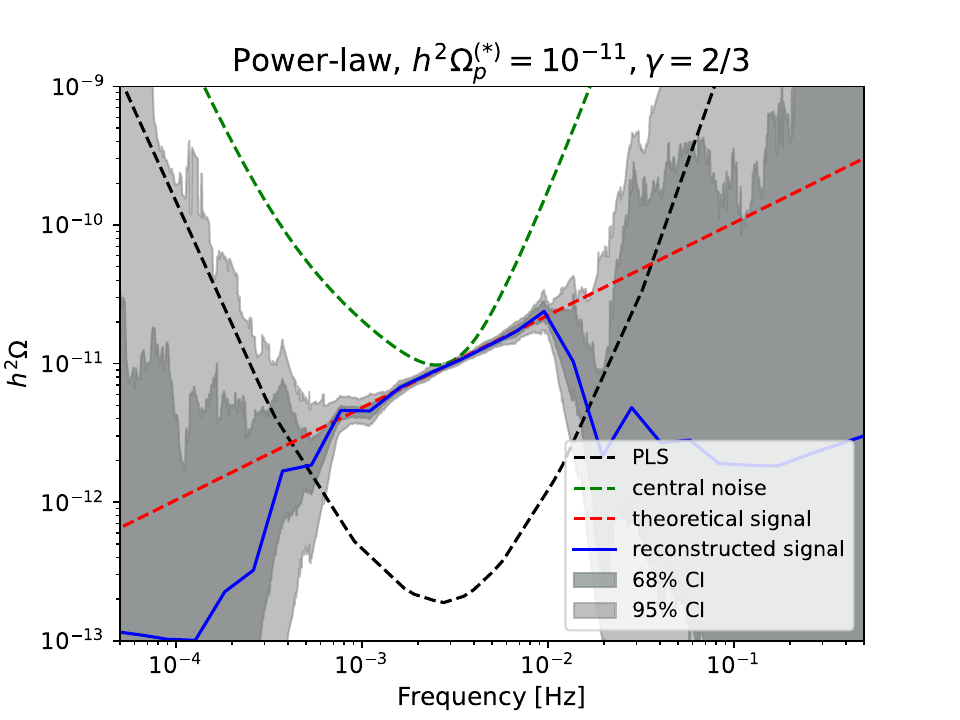}
\qquad
\includegraphics[width=.47\textwidth]{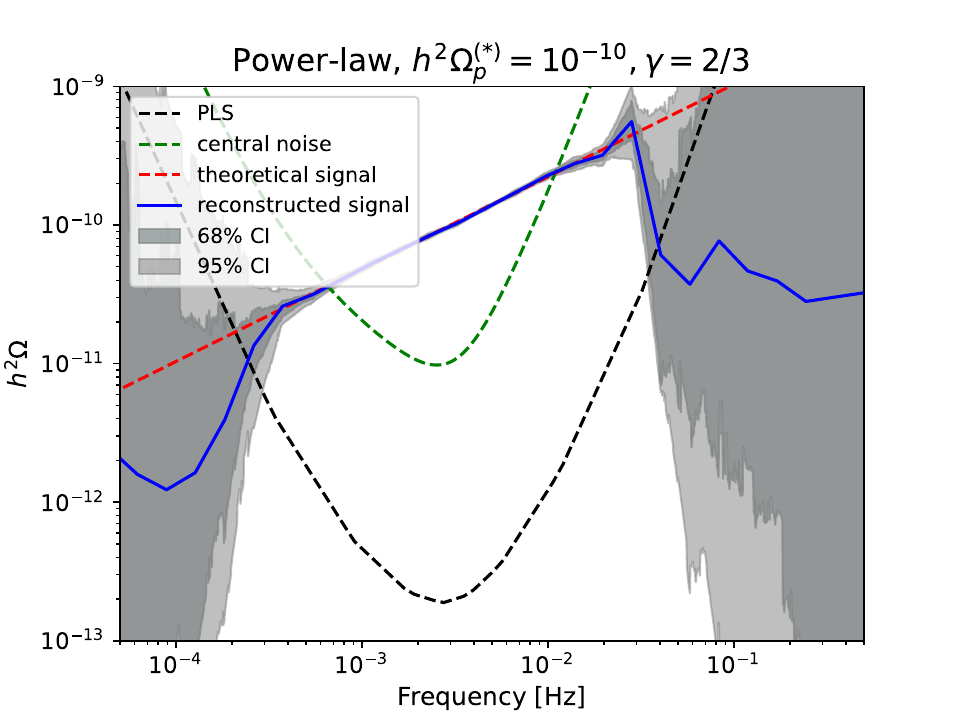}
\qquad
\includegraphics[width=.47\textwidth]{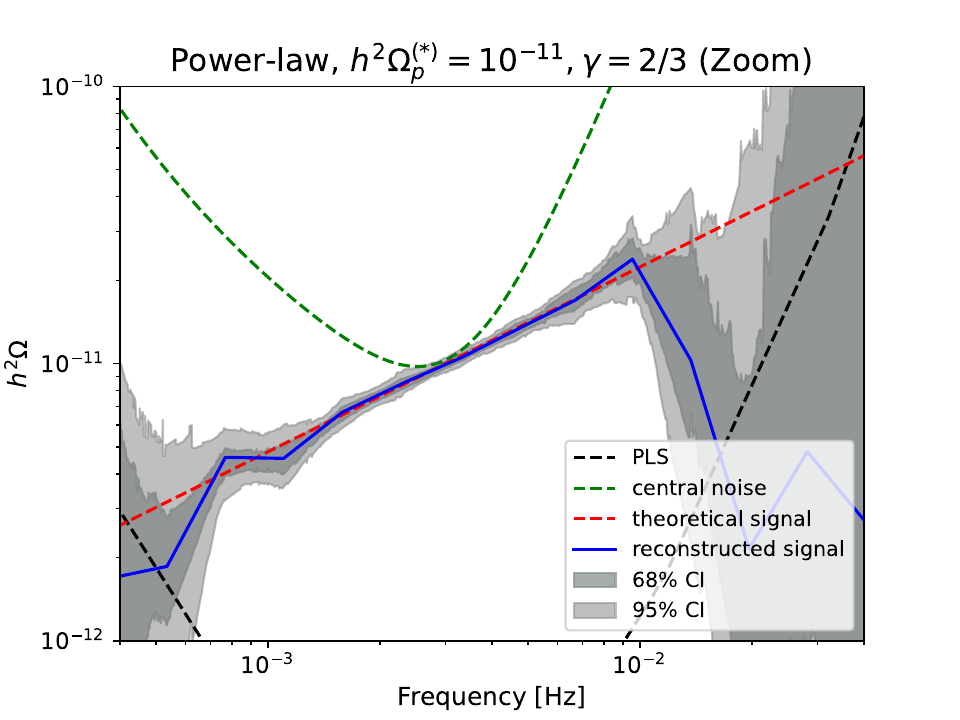}
\qquad
\includegraphics[width=.47\textwidth]{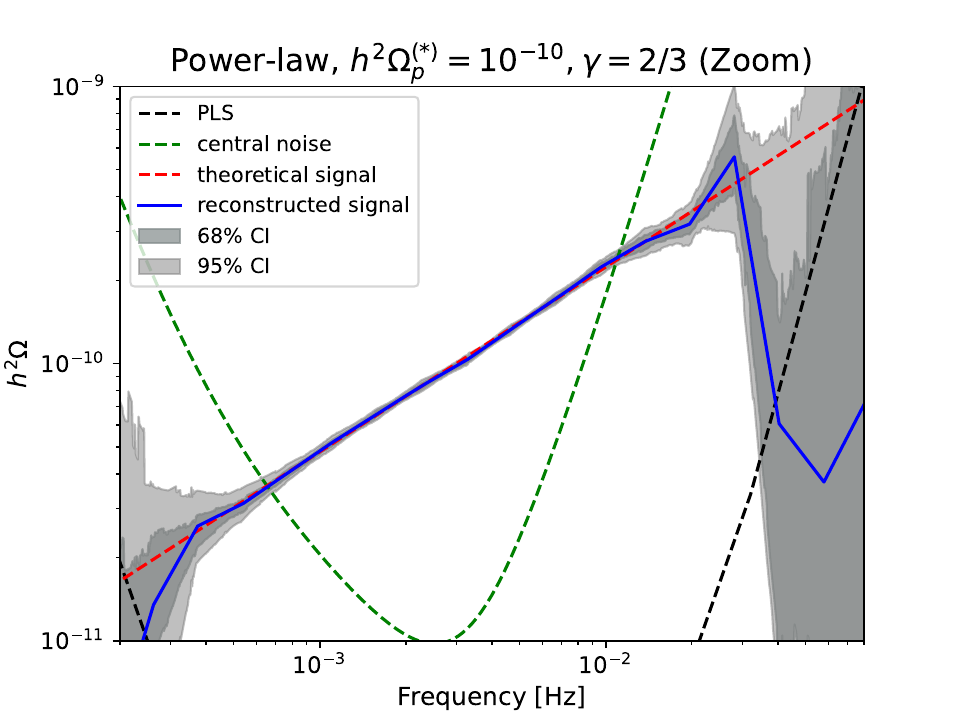}
\caption{Power-law reconstruction. Left) $h^{2}\Omega_{p}^{(*)} = 10^{-11}$, $\gamma = 2/3$, SNR $\approx 528$. Right) $h^{2}\Omega_{p}^{(*)} = 10^{-10}$, $\gamma = 2/3$, SNR $\approx 5281$. Bottom panel: zoom-in of top panel. We see that we are able to reconstruct the theoretical PL shape  in the central frequency region, where LISA is most sensitive. The quality of the reconstruction starts degrading as soon as the true signal approaches and falls below the PLS. The noise (green dashed) and the PLS curve (black dashed) are shown for the AA channel.}
 \label{fig:PL_gen}
\end{figure}

\begin{figure}[tbp]
\centering
\includegraphics[width=.47\textwidth]{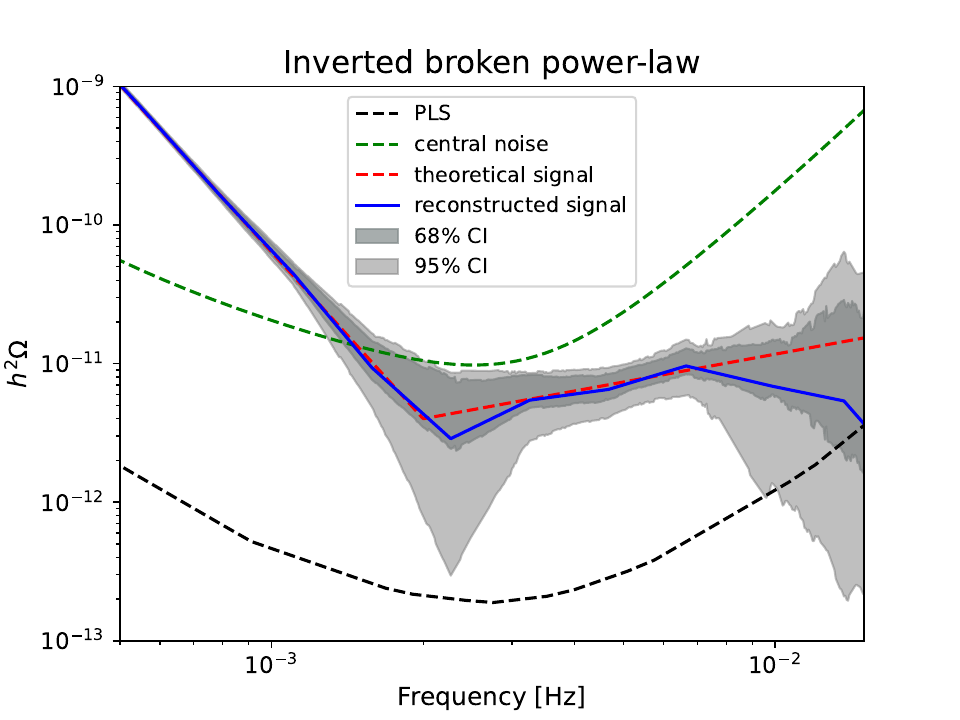}
\qquad
\includegraphics[width=.47\textwidth]{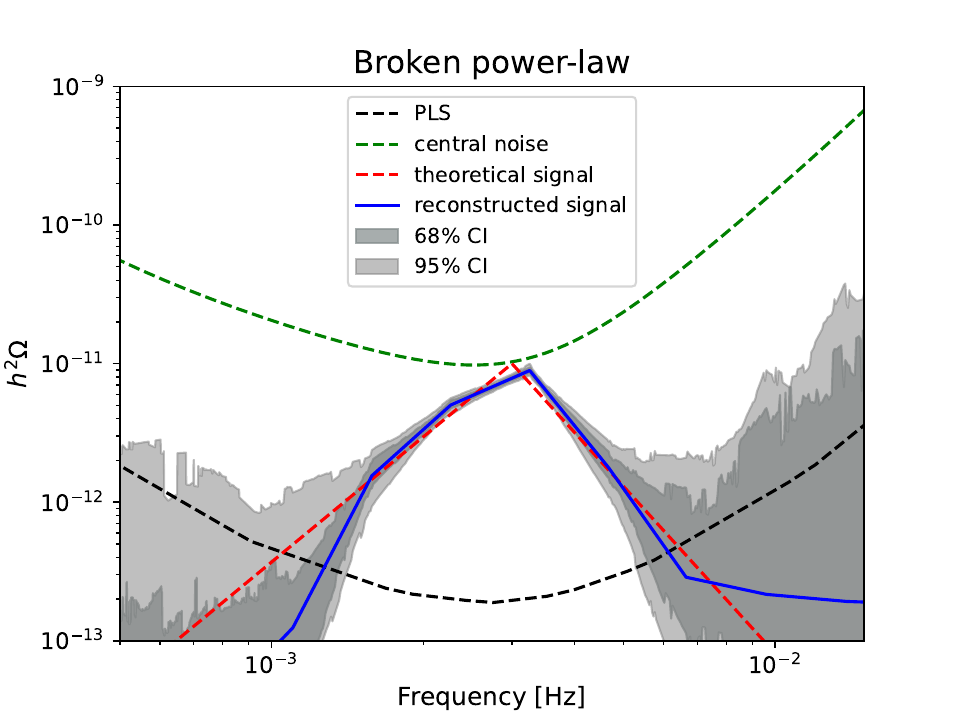}
\includegraphics[width=.47\textwidth]{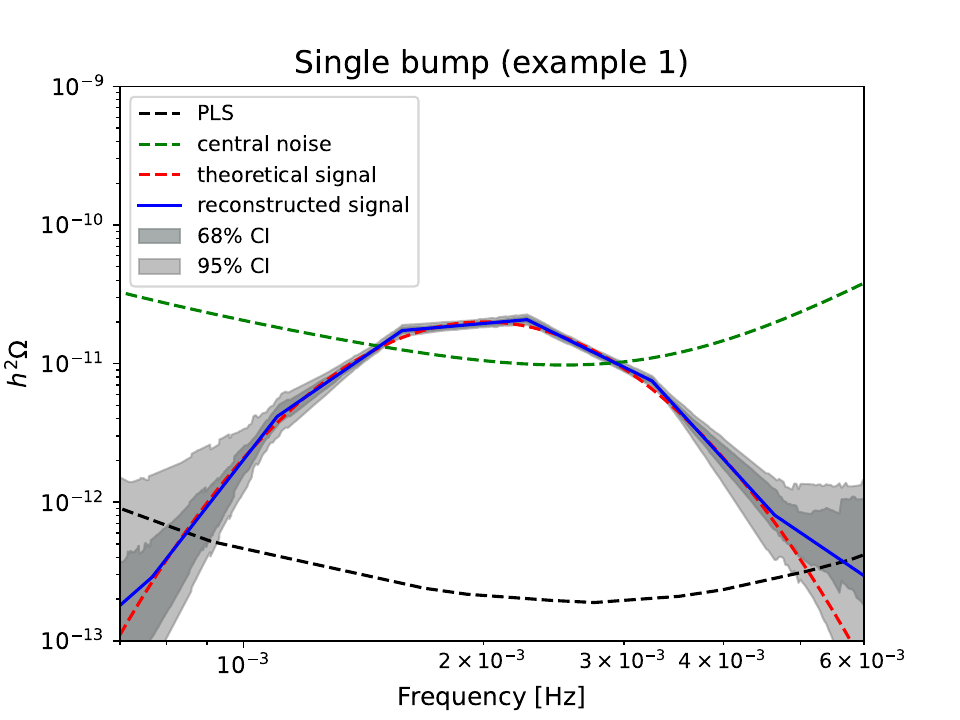}
\qquad
\includegraphics[width=.47\textwidth]{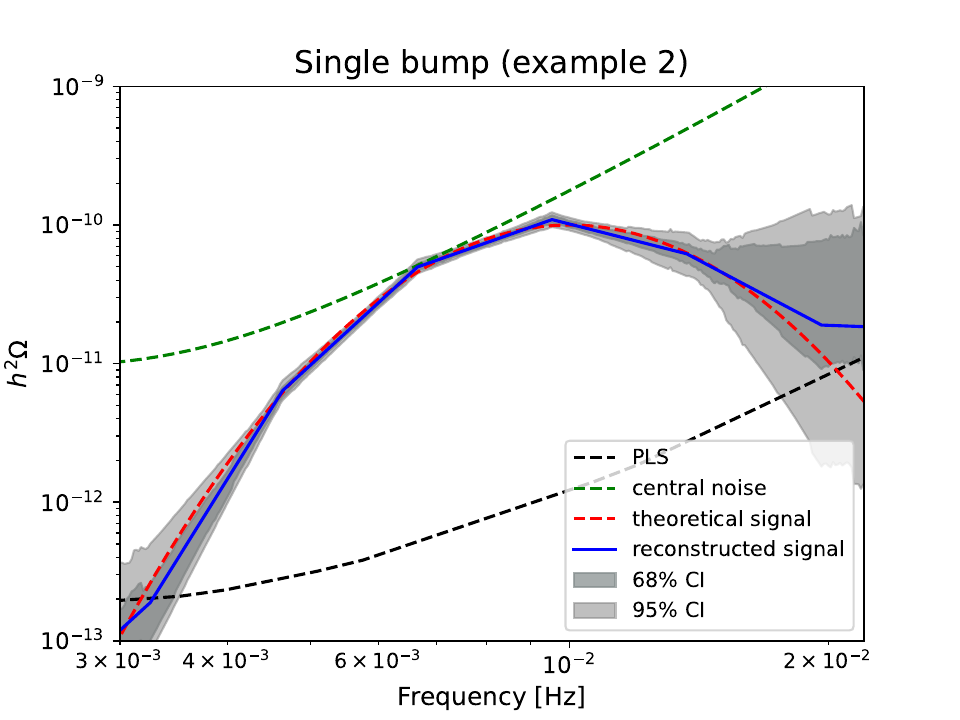}
\caption{Top: Broken power-law reconstruction, with $h^{2}\Omega_{p}^{(*)}=4\times 10^{-12}$, $\gamma_{1}=-4$, $\gamma_{2}=2/3$, $f_{p}=0.002$~Hz, SNR $\approx$ 41728 (Left panel), and $h^{2}\Omega_{p}^{(*)}=4\times 10^{-12}$, $\gamma_{1}=3$, $\gamma_{2}=-4$, $f_{p}=0.002$~Hz, SNR $\approx$ 86 (Right panel). Bottom: Reconstruction of a single peaked signal, with  $h^{2}\Omega_{p}^{(*)}=2\times10^{-11}$, $f_b=0.002$~Hz, $\Delta=0.2$, SNR $\approx$ 626 (left panel), and $h^{2}\Omega_{p}^{(*)}=10^{-10}$, $f_b=0.01$~Hz, $\Delta=0.2$, SNR $\approx$ 585 (right panel).}
 \label{fig:OnePeak_gen}
\end{figure}

\begin{figure}[tbp]
\centering
\includegraphics[width=.47\textwidth]{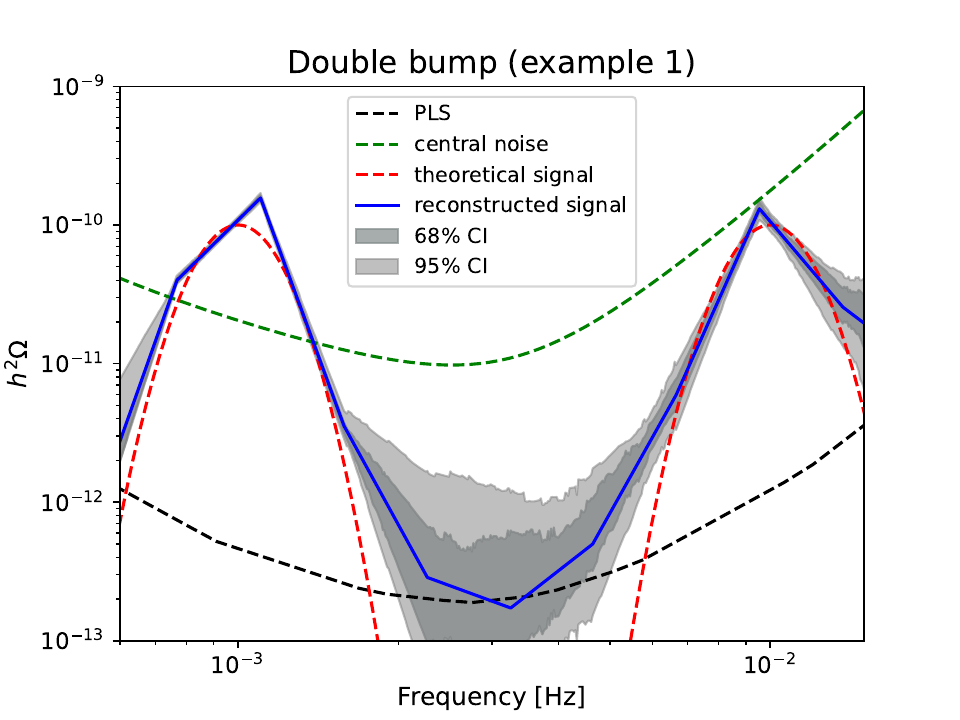}
\qquad
\includegraphics[width=.47\textwidth]{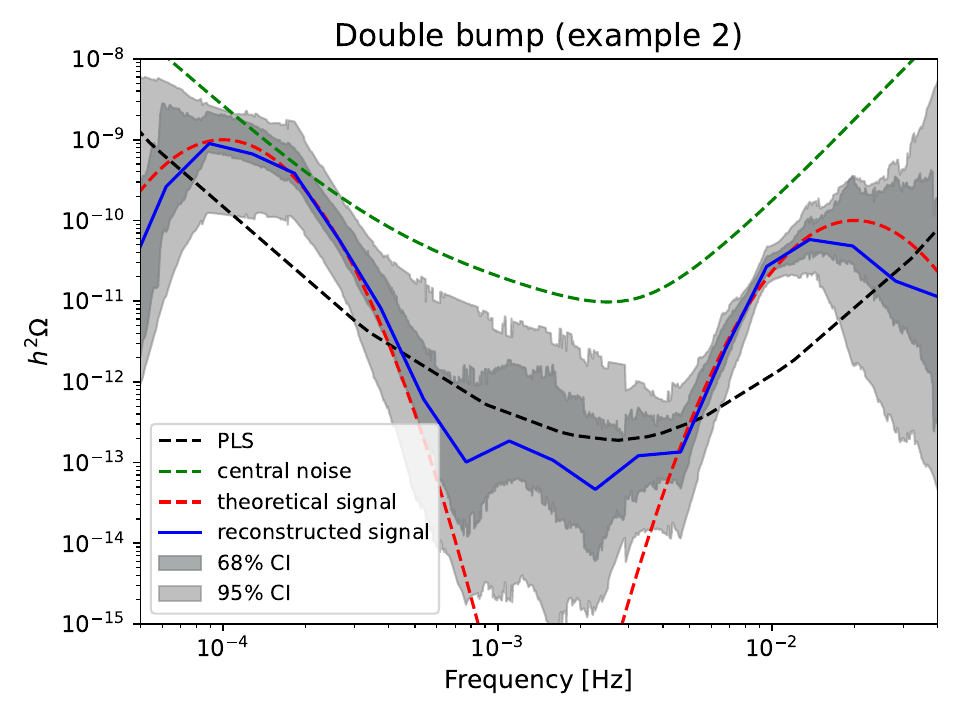}
\includegraphics[width=.47\textwidth]{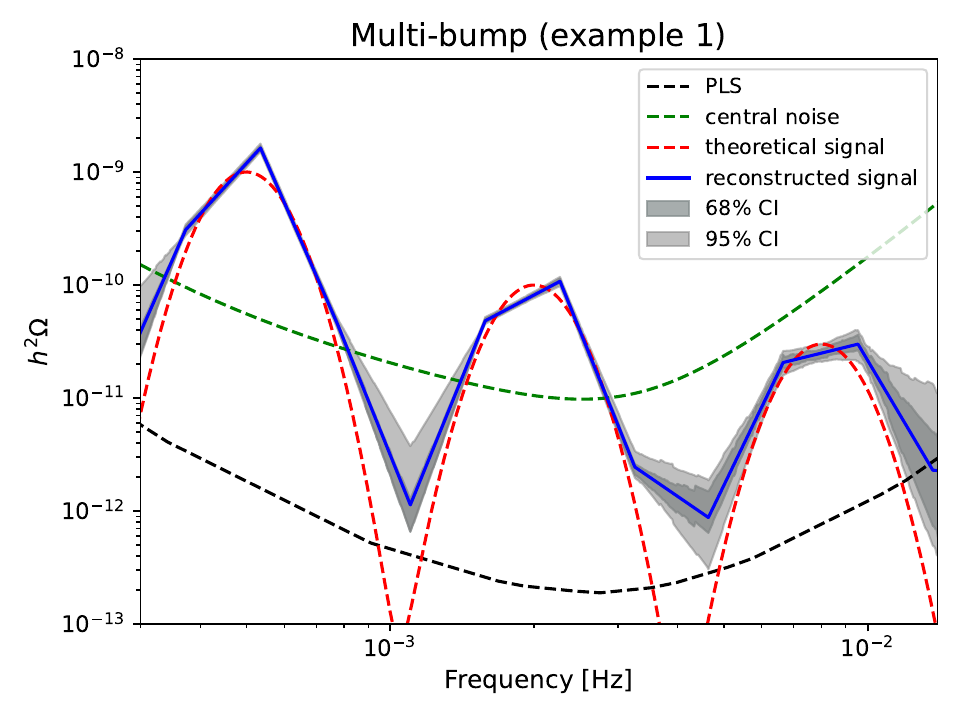}
\qquad
\includegraphics[width=.47\textwidth]{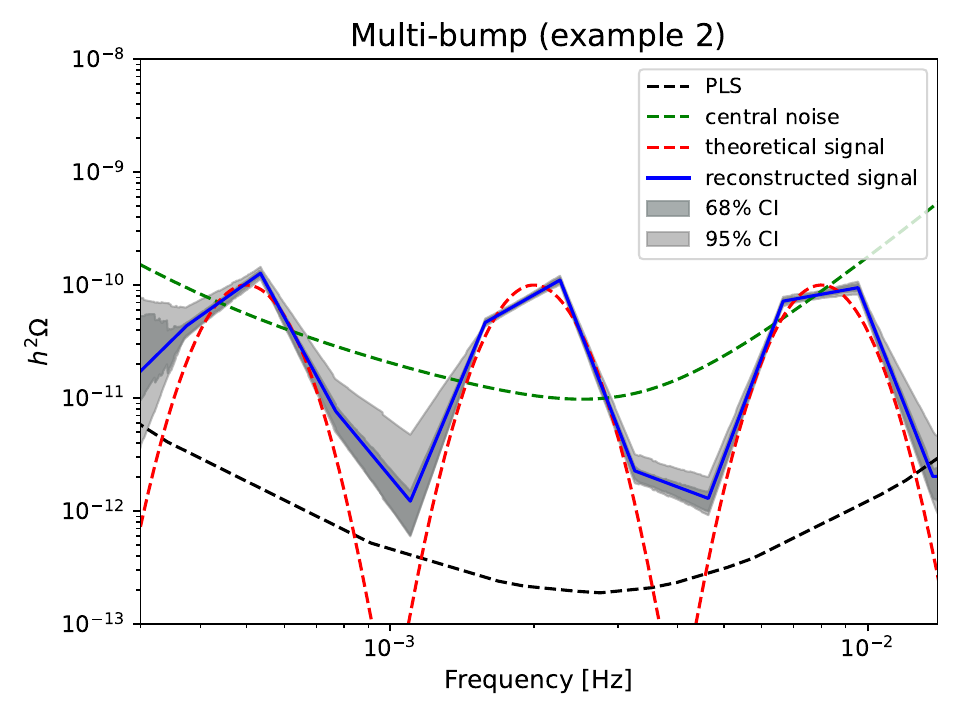}
\caption{Top: Reconstruction of a double-peak signal, with $h^{2}\Omega_{1}^{(*)}= h^{2}\Omega_{2}^{(*)}=10^{-10}$, $f_1=0.001$~Hz, $f_2=0.01$~Hz, $\Delta_1=\Delta_2=0.1$, SNR $\approx$ 895 (left panel), and $h^{2}\Omega_{1}^{(*)} = 10^{-9}, h^{2}\Omega_{2}^{(*)}=10^{-10}$, $f_1=10^{-4}$~Hz, $f_2=0.02$~Hz, $\Delta_1=\Delta_2=0.25$, SNR $\approx$ 148 (right panel). Bottom: Reconstruction of a multipleak signal with $h^{2}\Omega_{1}^{(*)}=10^{-9},  h^{2}\Omega_{2}^{(*)}=10^{-10},  h^{2}\Omega_{3}^{(*)}=10^{-11}$, $f_1= 5\cdot 10^{-4}$~Hz, $f_2=2\cdot 10^{-3}$~Hz, $f_3=8\cdot 10^{-3}$~Hz, $\Delta_1=\Delta_2=\Delta_3=0.1$, SNR $\approx$ 3154 (left panel) and $h^{2}\Omega_{1}^{(*)}=h^{2}\Omega_{2}^{(*)}=  h^{2}\Omega_{3}^{(*)}=10^{-10}$, $f_1= 5\cdot 10^{-4}$~Hz, $f_2=2\cdot 10^{-3}$~Hz, $f_3=8\cdot 10^{-3}$~Hz, $\Delta_1=\Delta_2=\Delta_3=0.1$, SNR $\approx$ 2312 (right panel).}
 \label{fig:Peaks_gen}
\end{figure}

For each of the cases in Table~\ref{tab:mock}, we generate ``test data'' in both TT and AA channels, ${\bf D}^{\rm TT},{\bf D}^{\rm AA}$, from the corresponding signal and noise, and give them as input to the prediction for the joint posterior $q_{\bs{\zeta}^*}({\bf s}, A_{\rm acc}|{\bf D}^{\rm AA},A_P)$, where the noise parameter $A_P$ have been extracted from the corresponding estimated posterior, conditioned on the ${\bf D}^{\rm TT}$ data. We show in Figs.~\ref{fig:PL_gen}-\ref{fig:Peaks_gen} the results of reconstructing signals corresponding to the four cases in Table~\ref{tab:mock}, for different values of the corresponding parameters. In each of these results, we report 
the 68\% and the 95\% intervals of the predictive distribution of $h^2\Omega_{GW}$.

Our first benchmark signals are power law (PL) profiles with amplitudes $h^2\Omega_1^{(*)} = 10^{-11}$ (top-left panel of Fig.~\ref{fig:PL_gen}) and $h^2\Omega_1^{(*)} = 10^{-10}$ (top-right panel of Fig.~\ref{fig:PL_gen}) at the pivot scale $0.003$ Hz, both with slope $\gamma = 2/3$, mimicking the shape of the GWB spectrum from unresolved stellar origin blank hole binaries and neutron star binaries~\cite{Sesana:2016ljz,Regimbau:2011rp,Babak:2023lro,Lehoucq:2023zlt}. As expected, the reconstruction of the theoretical PL shape is more accurate in the central frequency region, where LISA is most sensitive. When injecting a signal with large amplitude, $h^2\Omega_1^{(*)} = 10^{-10}$, the mode of the reconstructed background overlaps roughly within $2$ frequency decades with the injected PL signal, within the range $3\cdot 10^{-4} \lesssim f \lesssim 3\cdot 10^{-2}$, including frequencies where the true signal has dropped below the noise. When injecting a signal with smaller amplitude, $h^2\Omega_1^{(*)} = 10^{-11}$, the reconstructed signal mode overlaps with the theoretical PL over a shorter range, $7\cdot 10^{-4} \lesssim f \lesssim 1\cdot 10^{-2}$, which in any case, it is still larger than a frequency decade. As expected, the quality of the reconstruction starts degrading as soon as the true signal approaches and falls below the PLS. 

Our second benchmark signals are broken power law (BPL) profiles, shown in the top panels of Fig.~\ref{fig:OnePeak_gen}. In the top-left panel we show an {\it inverted} BPL signal, with amplitude $h^2\Omega_{p}^{(*)} = 4\cdot 10^{-12}$ at the pivot scale $0.002$ Hz, with negative left frequency branch slope $\gamma_1 = -4$, and positive right frequency branch $\gamma_2 = 2/3$, mimicking in this way the superposition of the high-frequency fall-off tail of a signal from a phase transition around the TeV scale~\cite{Caprini:2019egz}, and the expected PL signal from unresolved compact binaries~\cite{Sesana:2016ljz,Regimbau:2011rp,Babak:2023lro,Lehoucq:2023zlt}. In the top-right panel we show a {\it straight} BPL signal, with amplitude $h^2\Omega_1^{(*)} = 10^{-11}$ at the pivot scale $0.003$ Hz, but now with positive left frequency branch slope and $\gamma_1 = +3$ and negative right frequency branch slope $\gamma_1 = -4$, mimicking therefore the single peak shape spectrum of a GWB from a TeV scale phase transition~\cite{Caprini:2019egz}. 
In the case of the inverted PBL, the reconstructed mode of the left-branch overlaps pretty well with the injected signal, though the errors are quite large, given the smallness of the signal amplitude. The reconstructed mode of the right-branch overlaps with the true signal only till the latter approaches sufficiently close to the PLS. In the case of the straight BPL, the reconstruction of the signal is pretty decent, with small error bars in the central region, and increasingly larger errors as the branches approach the PLS. 

Our third benchmark signals are single bump (SB) profiles, shown in the bottom panels of Fig.~\ref{fig:OnePeak_gen}. In the bottom-left panel we show a signal with amplitude $h^2\Omega_1^{(*)} = 2\cdot 10^{-11}$ at the pivot scale $0.002$ Hz and width $\Delta = 0.2$, whereas the bottom-right panel shows a signal with amplitude $h^2\Omega_1^{(*)} = 10^{-10}$ at the pivot scale $0.01$ Hz and width $\Delta = 0.2$. As the signal with lower amplitude ($h^2\Omega_1^{(*)} = 2\cdot 10^{-11}$) 
is located only slightly left to the peak of sensitivity of LISA, the reconstruction is quite good, specially for the frequencies around the peak, while the error in reconstruction starts growing as the true signal approaches and crosses the PLS. On the other hand, the reconstruction of the second signal, even though is not bad, it is worse than the first one, despite having a larger amplitude ($h^2\Omega_1^{(*)} = 10^{-10}$). This is because it is peaked at frequencies further away to the right from the LISA sensitivity peak. The mode of the reconstructed signal follows nonetheless pretty well the theoretical injected signal over the range of frequencies where the latter is well above the PLS.


The fourth and final benchmark signals we use to test blind reconstruction are multi-bump (MB) peaked profiles, shown in Fig.~\ref{fig:Peaks_gen}. In the top-left panel we consider a double-peak signal with $h^{2}\Omega_{1}^{(*)}=h^{2}\Omega_{2}^{(*)}=10^{-10}$,  $f_1=0.001$~Hz and $f_2=0.01$~Hz, and $\Delta_1=\Delta_2=0.1$, whereas in the right top-panel we consider and $h^{2}\Omega_{1}^{(*)}=h^{2}\Omega_{2}^{(*)}= 10^{-10}$, $f_1=10^{-4}$~Hz and $f_2=0.02$~Hz, and $\Delta_1=\Delta_2=0.25$. These frequency profiles mimic, at least in spirit, the spectral shape of GWBs from TeV scale strong first-order phase transitions contributed by various sources (e.g.~bubble collisions and sound waves)~\cite{Caprini:2019egz}, as well as possibly multi-peak signatures in preheating signals sourced by various fields~\cite{Dufaux:2010cf}. In the bottom panel we consider a multi-peak signal with $h^{2}\Omega_{1}^{(*)} = 10^{-9}$, $h^{2}\Omega_{1}^{(*)} = 10^{-10}$ and $h^{2}\Omega_{1}^{(*)} = 3\cdot 10^{-11}$,  $f_1=0.0005$~Hz, $f_2=0.002$~Hz and $f_3=0.008$~Hz, with $\Delta_1=\Delta_2=\Delta_3=0.01$, whereas in the right bottom-panel we consider $h^{2}\Omega_{1}^{(*)} = h^{2}\Omega_{2}^{(*)} = h^{2}\Omega_{3}^{(*)} = 10^{-10}$, $f_1=0.0005$ Hz, $f_2=0.002$ Hz and $f_3=0.008$ Hz, and $\Delta_1=\Delta_2=\Delta_3=0.01$. Even though the latter two signals are more speculative than previously used profiles, they illustratively show the capability of LISA to reconstruct complicated multi-peak templates, like e.g.~the stairway signal signatures proposed in~\cite{Figueroa:2022iho}. The accuracy in the reconstruction of all multi-peak signals shown in Fig.~\ref{fig:Peaks_gen} depend essentially on the height of the signal peaks with respect to the PLS at the specific location, with the reconstruction worsening typically as the signal extends towards higher frequencies. 

\subsection{Reconstruction uncertainty}
\label{subsec:uncertainty}

A few comments with respect to all blind reconstructions shown so far are in order at this point:   

$i)$ The uncertainty associated with the reconstruction becomes very large in the frequency regions where the true signal drops below the PLS. 
This is natural, as in the outer frequency regions, the inference of the signal becomes more and more challenging since it becomes increasingly smaller compared to the noise. While a larger 
number of bins for the signal parametrisation [c.f.~Eq.~(\ref{eq:parametrisation})] would produce smaller uncertainties on the sides, this would come with the unwanted sacrifice of the model's expressivity. We note that, in comparison, the expressivity of the {\tt SGWBinner} model~\cite{Caprini:2019pxz} is automatically tuned to the signal at hand, which is an appealing feature we do not have. However, the computational cost of the {\tt SGWBinner} may be inconveniently high. We have found, by  trial an error, that a parametrisation with fixed 27 
bins within the 
LISA band is a reasonable trade-off between model complexity and computational cost. We also note that, even if computationally feasible, a too-large number of bins would not work perfectly, due to the statistical fluctuations of the data becoming more important within the bins. 

$ii)$ The reconstructed signals show a recurrent feature across all cases in consideration: the uncertainties for the higher frequencies are larger than those for the smaller frequencies. Ultimately, this has to do with the asymmetric sensitivity of LISA, as reflected by the PLS curve. For example, in the last (right-most) frequency bin of our signal parametrisation, any real signal is expected to be relatively smaller than the noise, in comparison to the first (left-most) frequency bin. Consequently, even if we increased significantly the slope of a putative signal in the high-frequency end of the LISA band, this would not have an appreciable impact on the total contribution to the data, which would be largely dominated by the noise. This lack of sensitivity is more evident for the last bins in the right side than for the first ones in the left region, see Fig.~\ref{fig:ratios} in Appendix \ref{app:B}, and a more detailed explanation therein. 

In light of the above problems, it seems convenient to define some procedures to quantify the accuracy of the reconstruction of a signal. We propose two different procedures in this section. 
First, a study in Sect.~\ref{subsubsec:PLuncertainty} dedicated to power law profiles, where we characterize the uncertainty of reconstruction of this type of signals by quantifying how well we recover their amplitude $h^2\Omega_{\rm GW}^{(*)}$ and slope $\gamma$. Secondly, in order to provide a more general quantification of the accuracy of the reconstruction of a generic signal, we compute the ``Averaged relative error'' in Sect.~\ref{subsubsec:ARE}, quantifying the degree of agreement between theoretical and reconstructed signals as a function of frequency.

\begin{figure}[tbp]
\centering
\includegraphics[width=.47\textwidth]{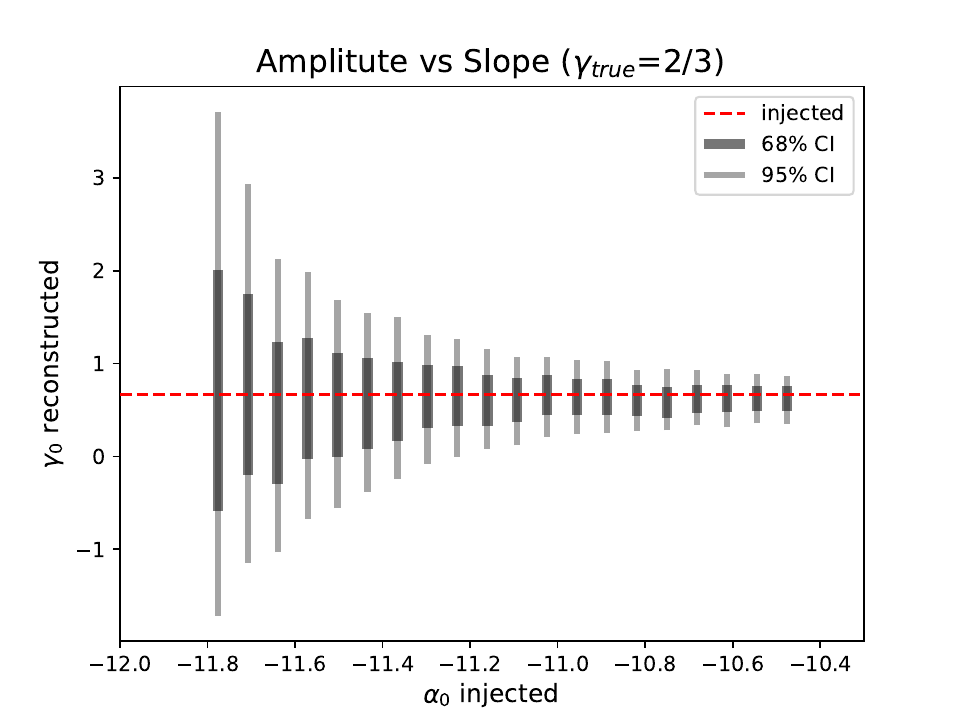}
\qquad
\includegraphics[width=.47\textwidth]{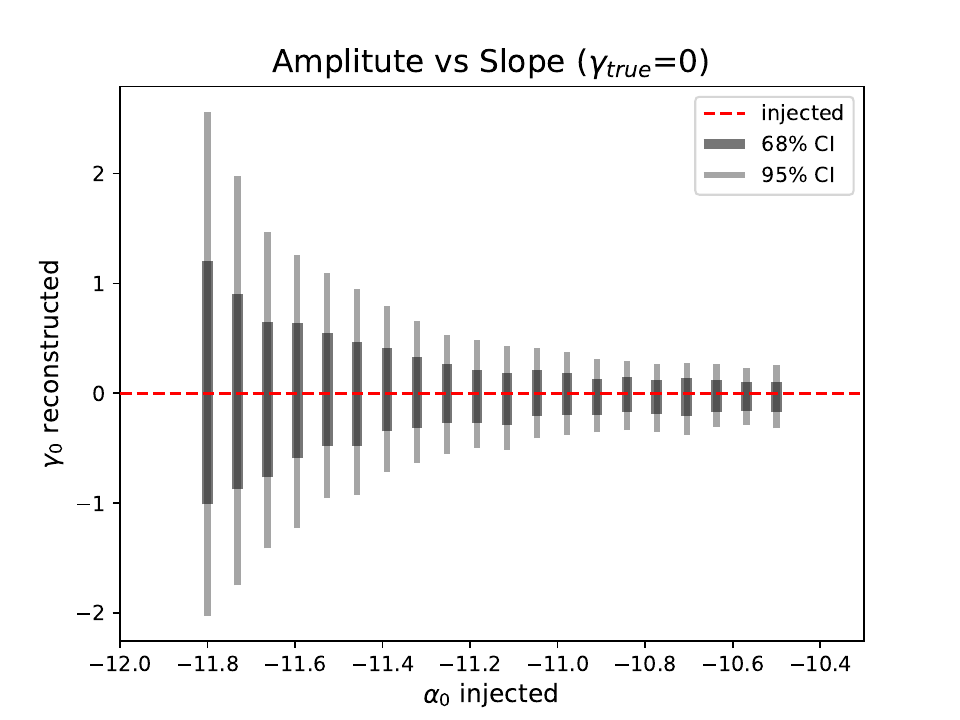}
\qquad
\includegraphics[width=.47\textwidth]{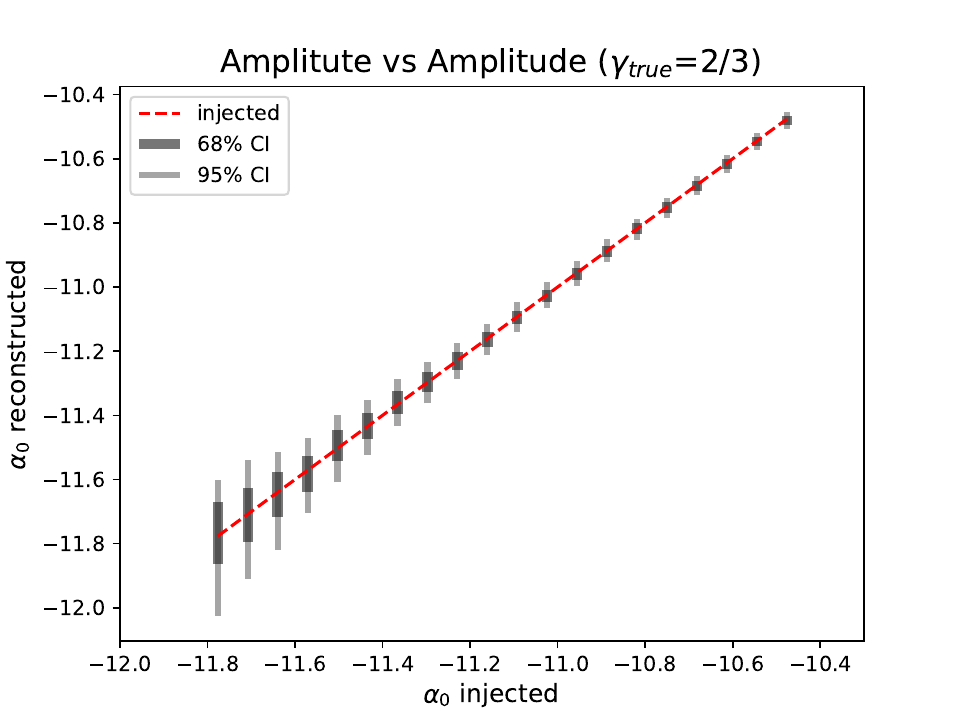}
\qquad
\includegraphics[width=.47\textwidth]{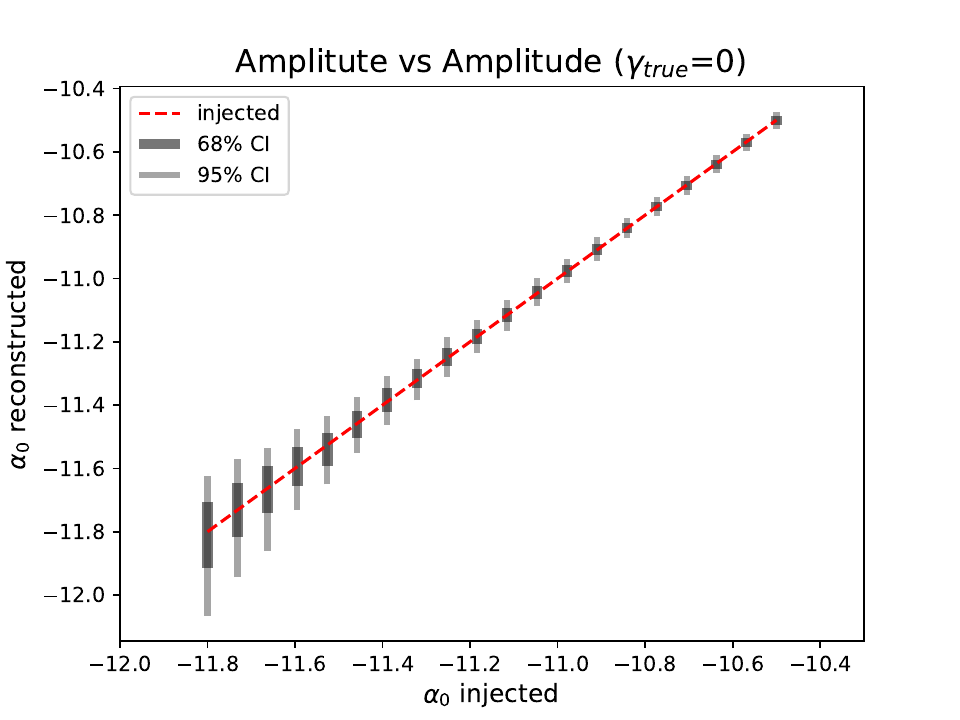}
\caption{Reconstruction of the parameters of a mock power-law signal, as characterised by the central frequency bin of our SGWB parametrisation. The two rows represent reconstructed-slope vs true-amplitude  and reconstructed-amplitude vs true-amplitude (bottom panels). For both panel rows, we fix the true value of the slope to $\gamma_{\rm true}=2/3$ (left panels) and $\gamma_{\rm true}=0$ (right panels).}
 \label{fig:keynote}
\end{figure}

\subsubsection{Power-law case of study}
\label{subsubsec:PLuncertainty}

As an insight into the blind-reconstruction capabilities, we first study the accuracy of the reconstruction of a power law (PL) signal as a function of its amplitude and tilt, which we parametrize now as
\begin{eqnarray}
h^{2}\Omega_{\rm GW}(f) = 10^{\alpha}\left({f\over 0.003\,{\rm Hz}}\right)^{\gamma},
\end{eqnarray}
so that for each $k$-bin (see definition in Sect.~\ref{subsec:Agnostic}) we can reconstruct the tilt $\gamma_k$ and amplitude $\alpha_{k} \equiv \log_{10}(h^{2}\Omega_{\rm GW}(f_k))$. We show the results in Fig. \ref{fig:keynote}, only for the central frequency bin $k=0$ of our parametrisation, for the sake of illustration. In the top panels, the idea is to reconstruct the PL slope $\gamma$ when varying the amplitude values of a true (injected) signal with $\gamma_{\rm true}=2/3$ (top-left) and $\gamma_{\rm true}=0$ (top-right). Moreover, for each pair of slope-amplitude we generate 50 different datasets to eliminate the random fluctuations that are expected to occur around the injected values since the datasets used for inference are inherently noisy. As expected, the precision of the reconstruction worsens visibly as smaller and smaller amplitudes are injected, nonetheless including always the true value (dashed horizontal line). For example, in the case of $\gamma_{\rm true} = 2/3$, the 95\% credible interval on the reconstructed $\gamma_0$ (i.e. the slope of the central frequency bin) goes from $[-1.71, 3.71]$ for $\alpha_0 = -11.78$, down to $[0.082, 1.16]$ for $\alpha_0 = -11.15$, and to $[0.35, 0.87]$ for $\alpha_0 = -10.48$. Analogously, for $\gamma_{\rm true} = 0$, the 95\% credible interval on the reconstructed $\gamma_0$ goes from $[-2.02, 2.56]$ for $\alpha_0 = -11.8$, down to $[-0.50, 0.48]$ for $\alpha_0 = -11.18$, and to $[-0.32, 0.25]$ for $\alpha_0 = -10.5$.

Similarly, the bottom panels show the reconstruction capabilities of the reconstructed amplitude as a function of the true amplitude, for the same two values of $\gamma_{\rm true} = 2/3$ (bottom-left) and $\gamma_{\rm true} = 0$ (bottom-right). In the case of $\gamma_{\rm true} = 2/3$, the 95\% credible interval on the reconstructed $\alpha_0$ goes from $[-12.02, -11.6]$ for $\alpha_0 = -11.78$, down to $[-11.21, -11.16]$ for $\alpha_0 = -11.15$, and to $[-10.51,-10.45]$ for $\alpha_0 = -10.48$. Analogously, for $\gamma_{\rm true} = 0$, the 95\% credible interval on the reconstructed $\alpha_0$ goes from $[-12.07, -11.62]$ for $\alpha_0 = -11.8$, down to $[-11.24, -11.13]$ for $\alpha_0 = -11.18$, and to $[-10.53, -10.47]$ for $\alpha_0 = -10.5$.


. 

\subsubsection{Averaged Relative Error}
\label{subsubsec:ARE}

\begin{figure}[tbp]
\centering
\hspace*{0.2cm}
\includegraphics[width=6.2cm,height=5.1cm]{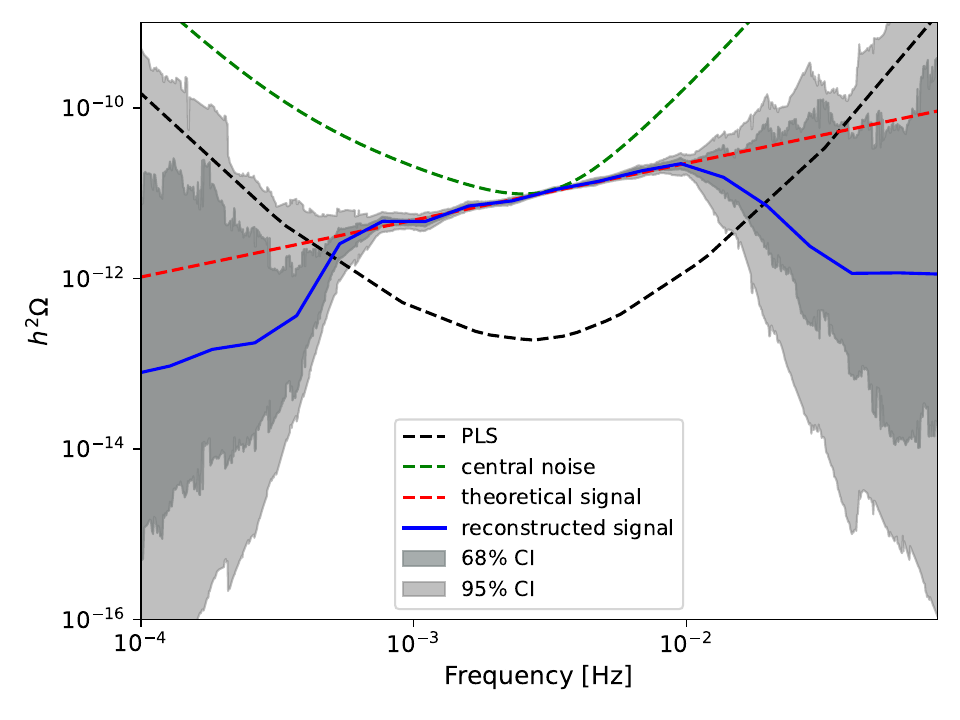}
\hspace{0.0cm}
\includegraphics[width=6.6cm,height=5.1cm]{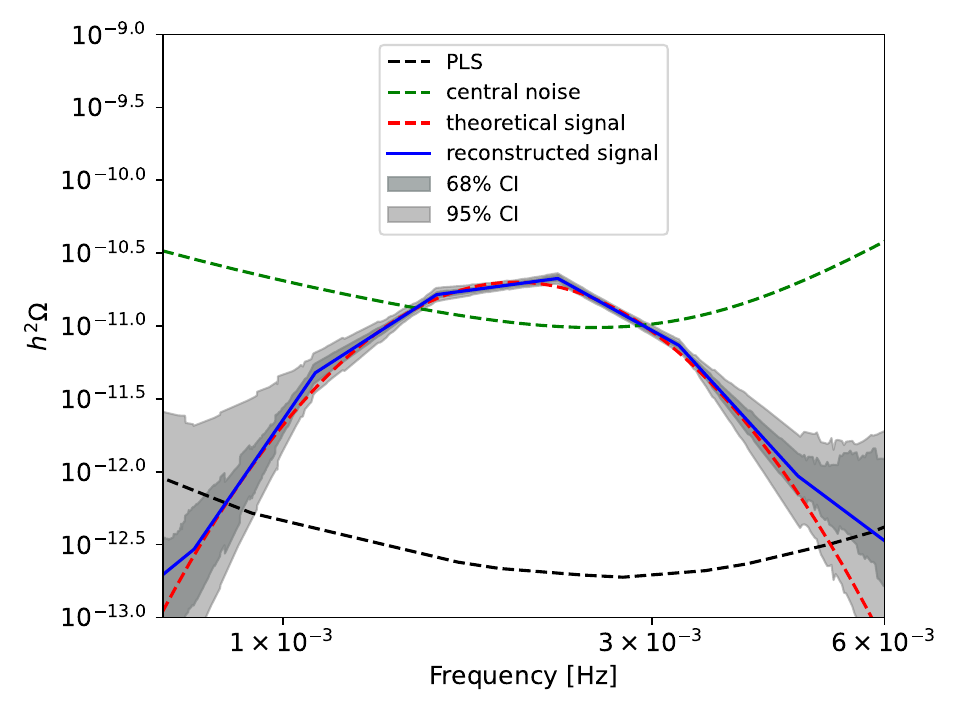}
\qquad
\includegraphics[width=6.2cm,height=5.1cm]{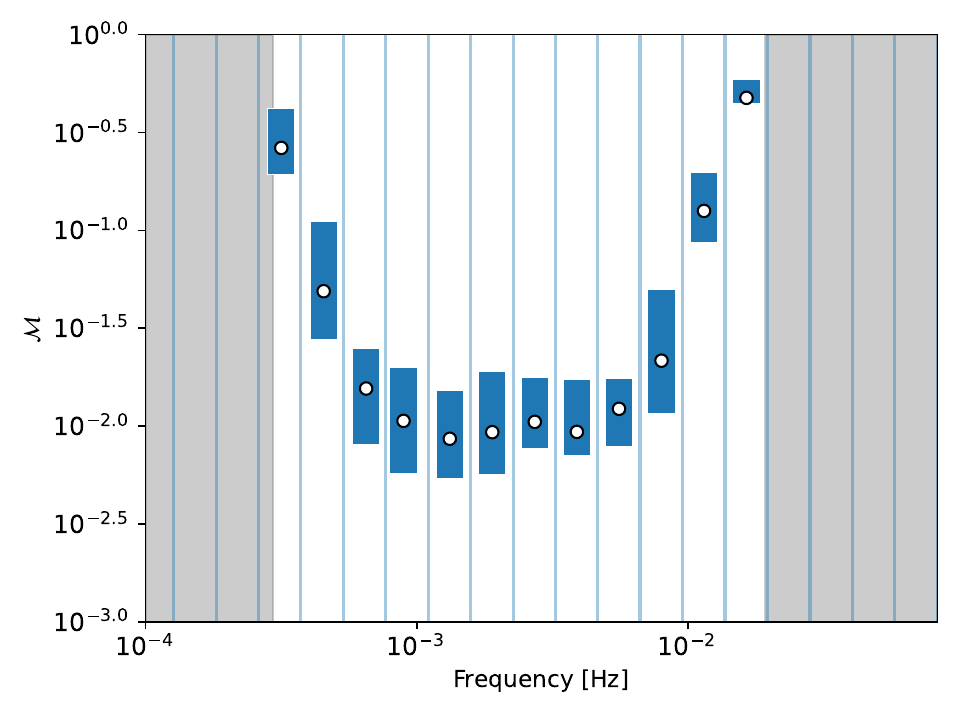}
\hspace{0.2cm}
\includegraphics[width=6.0cm,height=5.1cm]{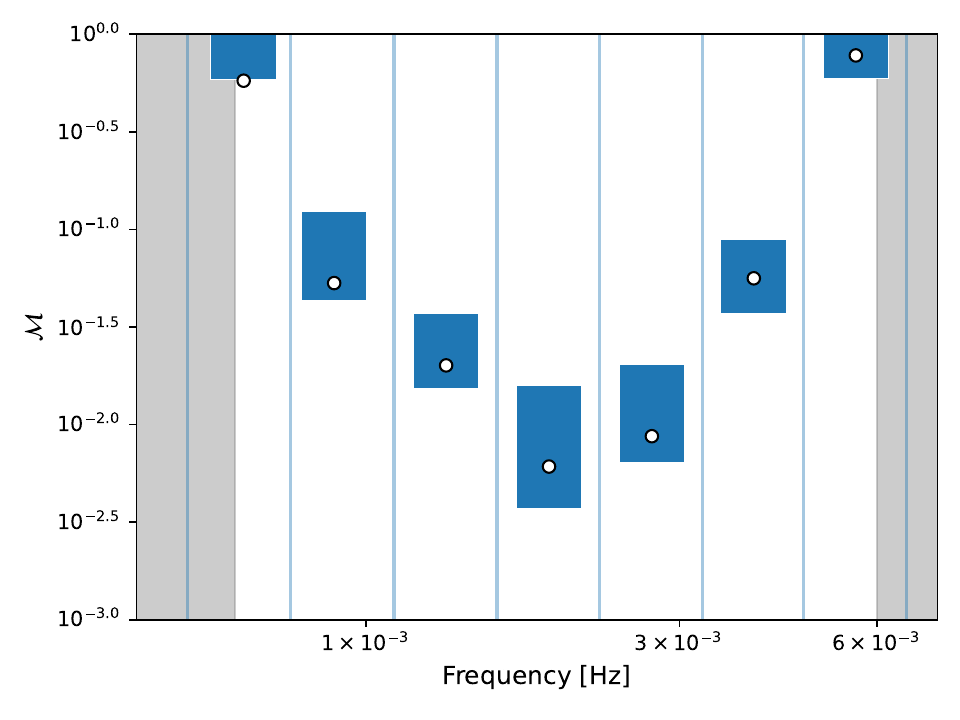}
\caption{Top: Reconstruction of a power-law signal, with $h^{2}\Omega_{p}^{(*)} = 10^{-11}$, $\gamma = 2/3$, SNR $\approx 528$ (left panel), and reconstruction of a single peaked signal, with  $h^{2}\Omega_{p}^{(*)}=2\times10^{-11}$, $f_b=0.002$~Hz, $\Delta=0.2$, SNR $\approx$ 626 (right panel). Bottom: Relative Errors of the two reconstructed signals. The relative errors are presented as ``candle plots'' representing the resulting variability after making parameter inference from $m=50$ simulated datasets corresponding to the same true signal. The mean value over those $m$ simulations is shown as the inner white dot in each candle, while the 68\% containment in represented by the extremes of the solid bars.}
\label{fig:ARE}
\end{figure}

While by visual inspection our method clearly manages to reconstruct generic signal shapes, we apply below a method to quantify the agreement between injected and reconstructed signals. If we had access to a measured value for a given frequency $f$, a possible figure of merit would be to compare it with respect to our prediction, given e.g. by the expected value of the posterior predictive distribution. However, prior to any observation, as we only rely on simulations, a reasonable alternative is to compare, for a given frequency, the true (injected) signal $\Omega_{\rm true}$, with the model prediction $\Omega_{\rm pred}$. We thus construct the Relative Error:
\begin{equation}
  {\cal M}(f) \equiv 
  \left|\frac{\Omega_{\rm true}(f) - \Omega_{\rm pred}(f)}{\Omega_{\rm true}(f)}\right|~,
  \label{eq:ARE}
\end{equation}
where $\Omega_{\rm pred}(f)$ is computed by evaluating our GW parametrisation, Eq.(\ref{eq:parametrisation}), in the mean values of the model parameters according to the resulting posterior distribution. The results are shown in figure \ref{fig:ARE} (bottom panels), for two representative examples: a power-law in the left panel, and a single-bump signal in the right panel, whose correponding reconstructions are shown in the upper panels. Our results (bottom panels) are presented as ``candle plots'', representing the resulting variability after making parameter inference from $m=50$ simulated datasets corresponding to the same true signal. The mean value over those $m$ simulations is shown as the inner white dot in each candle, while the 68\% containment in represented by the extremes of the solid bars. On the other hand, for a given bin of Eq.(\ref{eq:parametrisation}), these $m$ values are computed as the average across all the frequencies $f$ inside the bin $j$.  This is consistent with our aim to quantify the quality of the reconstruction in a per-bin basis\footnote{ and not necessarily in a per-frequency basis, since our model is defined as a piecewise function per bin.}.

Coming back to figure \ref{fig:ARE}, in the bottom panels we show only the values of $\cal M$ for which the signal lies above the PLS curve, while we shade in gray the side bands where this is not the case.  As expected, the figure of merit improves in the central regions, where the sensitivity of LISA is higher. This implies that mean relative errors can only be as low as 1\% in the central bins for the injected signals with an amplitude of $ h^2\Omega_{\rm GW} \sim 10^{-11}$, while the quality of the reconstruction degrades when the signal approaches the PLS from above.

\subsection{{\tt GWBackFinder} vs MCMC}
\label{subsec:MCMCcomparison}

We finish this section by showing a comparison against MCMC results, at the level of the posterior distributions of the parameters of interest. It is worth recalling at this point the features of our SGWB parameterization Eq.~(\ref{eq:signal_sim}), which has 28 independent parameters, given by the signal amplitude $\Omega_p^{(*)}$ of the very central frequency bin, plus 27 exponents corresponding to each of the frequency bins slopes of the piecewise implementation. Consequently, while our modelling is formally a power-law within each bin, their amplitudes are not independent. 

\begin{figure}[tbp]
\centering
\includegraphics[width=.4\textwidth]{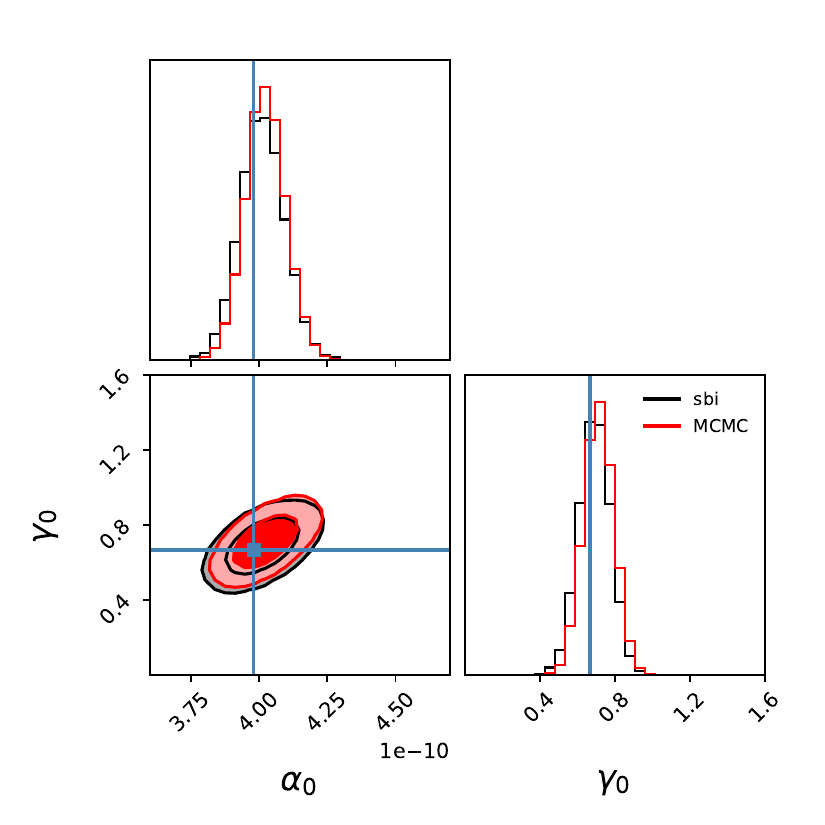}
\qquad
\includegraphics[width=.4\textwidth]{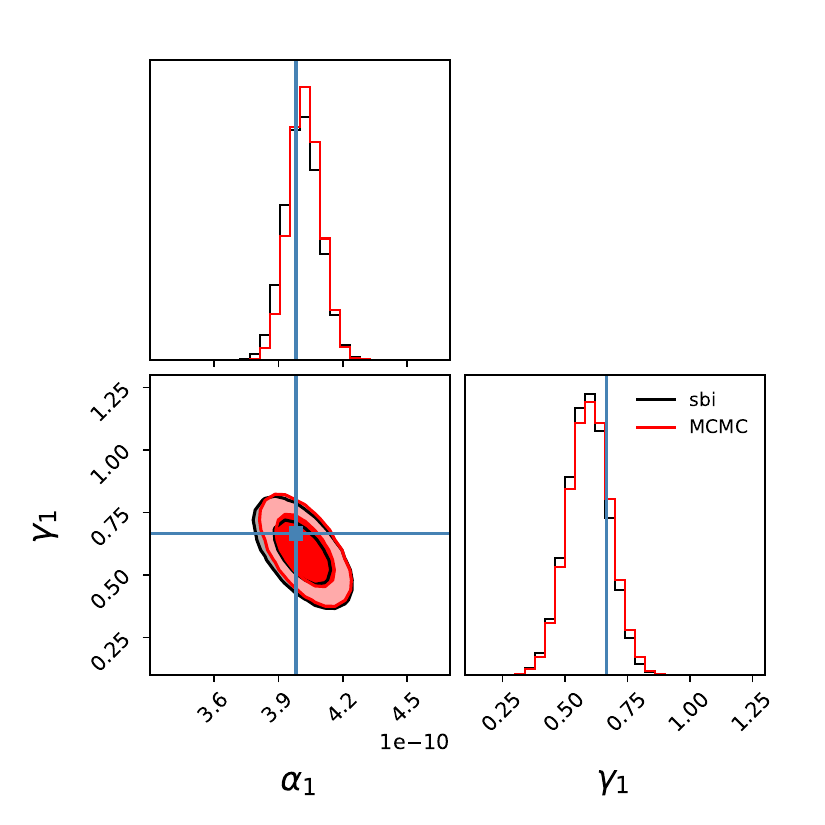}
\qquad
\includegraphics[width=.4\textwidth]{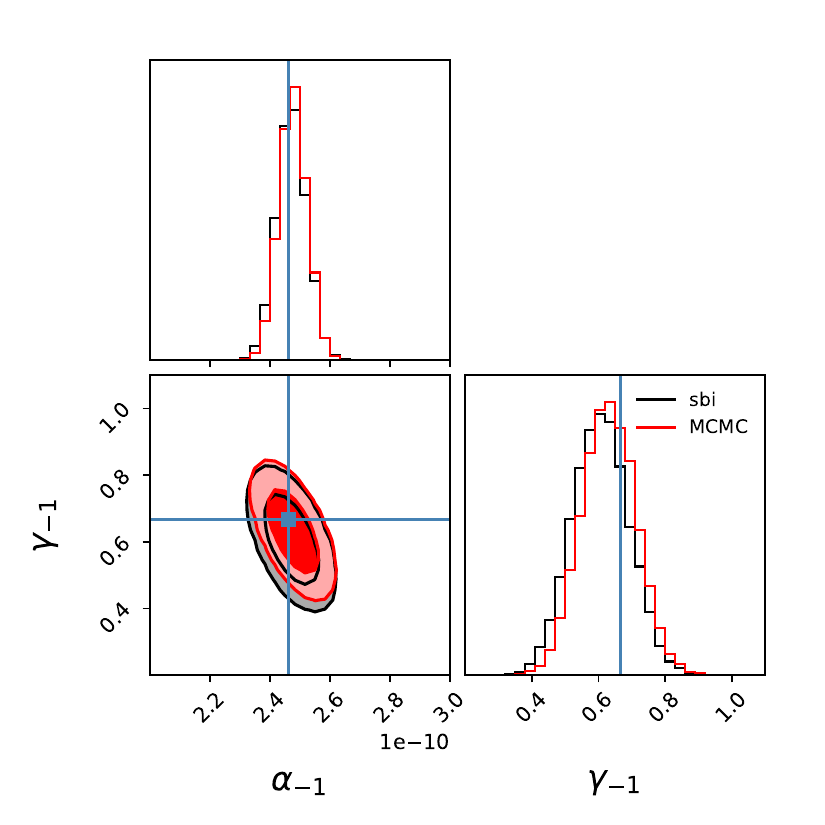}
\qquad
\includegraphics[width=.4\textwidth]{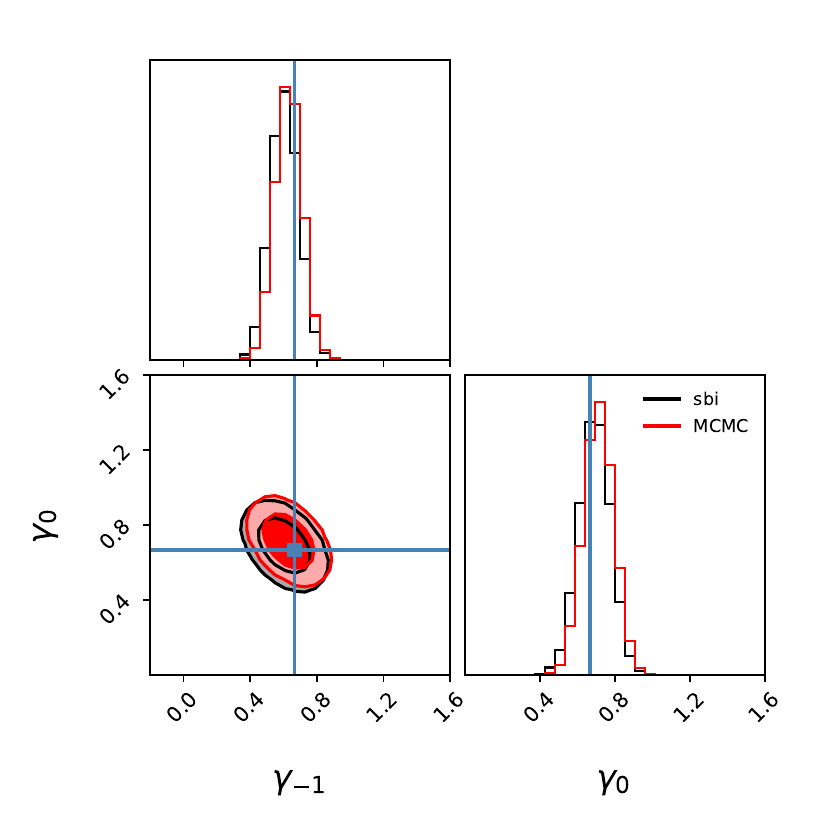}
\caption{Comparison of our results obtained with amortized NPE and MCMC for a power-law injected signal. We focus on three different frequency bins defined in our parameterization, in terms of slope-vs-amplitude corner plots. Top-left panel: central bin, with mean frequency $\bar f\approx 0.003$ Hz. Top-right panel: higher-frequency bin with $\bar f\approx 0.004 $ Hz. Bottom-left panel: lower-frequency bin with $\bar f\approx0.002$ Hz. On the bottom-right panel, we show the correlation between the slopes of the central and the lower-frequency bin.}
 \label{fig:mcmc1}
\end{figure}

\begin{figure}
\centering
\includegraphics[width=.65\textwidth]{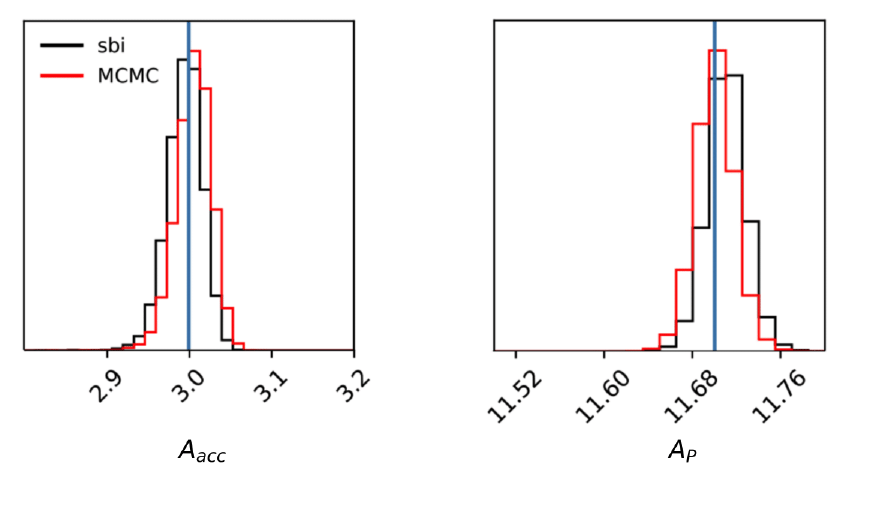}

\caption{Reconstruction of the noise parameters for the same power-law signal and dataset of Fig.(\ref{fig:mcmc1}). The blue lines correspond to the true values of the parameters. NPE agrees with MCMC to very good precision and the $68\%$ intervals of the posteriors capture the correct value of the parameters.}
 \label{fig:mcmc2}
\end{figure}

Since MCMC must sample the full joint posterior, it is slow to converge and it requires many samples - $ \mathcal{O} (10^{5})$ - in order to produce converged contours. Moreover, a different MCMC chain is required to obtain posteriors for different observed datasets. On the other hand, as previously mentioned, NPE can be amortized, meaning that after training the network with many examples (also around $\mathcal{O} (10^{5})$), we have access to the joint posterior of every possible dataset.

For the sake of physics intuition, we present our comparison between amortized NPE and MCMC in terms of slope-vs-amplitude contour plots, obtained for the three most central frequency bins of our parametrisation, which characterise the region of highest LISA sensitivity. In addition, we also show the correlation between the slopes of two of such bins. The injected mock signal we consider corresponds to a power-law with slope $\gamma=0.85$, and amplitude $h^2\Omega_p^{(*)}\approx 1.5\times 10^{-11}$ at a pivot frequency of 0.003 Hz. For the MCMC, we follow the {\tt emcee} documentation. We use $nwalkers = 16$ and we start by initializing the walkers in a tiny Gaussian ball around the maximum likelihood result. Then we run 5000 steps of MCMC. Finally,  we take $nburn = 500$, thin by about half the autocorrelation time and flatten the chain so that we have a flat list of samples.

We show the results in figure \ref{fig:mcmc1}, in terms of corner plots for the parameters corresponding to the central frequency bins, where the 2d posteriors are shown in terms of 68\% (inner) and 95\% (outer) credible intervals.  We see that NPE agrees with the MCMC to very good precision. The $68\%$ intervals of the posteriors capture the true value of the parameters. Note that the accuracy of the reconstruction depends of course on the dataset, and it is thus subject to statistical fluctuations. On the other hand, we show in figure \ref{fig:mcmc2} the posterior distributions of the noise parameters for the same signal and dataset corresponding to figure \ref{fig:mcmc1}, comparing as well with MCMC. Again, we observe that our method is statistically compatible with the latter, showing very precise estimations (namely, narrow posteriors) for both parameters.

\section{Reconstruction over astrophysical foregrounds}
\label{sec:foreground}

Our primary focus in this work has been the development of a model-independent method to reconstruct blindly the spectral shape of an arbitrary GWB, having in mind especially signals of cosmological origin. 
In a realistic situation, however, 
even if a cosmological signal has a very large amplitude to be detectable, astrophysical backgrounds may act as contaminating foregrounds, degrading our ability to detect or differentiate the cosmological signal from the noise. The separation of astrophysical and cosmological backgrounds is expected to become one of the major problems to be addressed in the years to come. Even though we do not aspire to address such a complicate issue here, we expect in any case that our method should be robust enough to reconstruct, under certain circumstances, the GWB spectrum of an unknown signal in the presence of astrophysical foregrounds. 

In the following, we reconstruct a signal in a setup where one astrophysical foreground is also present. For simplicity, we assume the shape of the foreground as known, except for their amplitude. The latter is treated as a parameter whose posterior is to be inferred. In practice, we consider two cases in which the total GWB observed by LISA is the sum of two components: an isotropic astrophysical component (\textit{foreground}) and an unknown cosmological signal, with no signal model. We consider two types of foregrounds, from unresolved merging binaries of extra-galactic origin, and from galactic origin: 
\vspace*{1mm}

{\bf Extra-galactic foreground}: The incoherent superposition of all extra-galactic compact binaries that LISA will be unable to resolve individually, introduces a foreground in the LISA data. The main contributors to this extra-galactic astrophysical signal are expected to be stellar-origin black hole and neutron star binaries~\cite{Sesana:2016ljz,Regimbau:2011rp,Babak:2023lro,Lehoucq:2023zlt}. In the LISA band (also in the LVK band) this foreground takes the form of power-law profile as
\begin{equation}
    \Omega_{\rm EF}(f) = 10^{\alpha_{\rm EF}} \left(\frac{f}{0.001\,{\rm Hz}}\right)^{2/3}~,
\end{equation}
where the slope $\gamma = 2/3$ is fixed by General Relativity for circular binary orbits~\cite{Phinney:2001di}, and following~\cite{Flauger:2020qyi} we consider the (log) amplitude (at the pivot scale $f_*=10^{-3}$Hz) $\alpha_{\rm EF}$ to follow a Gaussian prior, ${\cal N}(\alpha_{\rm EF}^*,\sigma_{\rm EF})$, with $\alpha_{\rm EF}^* = -12.29$ and $\sigma_{\rm EF} = 0.43$. 
\vspace*{1mm}

{\bf Galactic foreground}: The unresolved galactic binary mergers are expected to emit a foreground with a yearly modulation, and hence a proper treatment of such a signal should take into account its variation in each chunk\footnote{As shown in Ref.~\cite{Adams:2013qma}, the periodic modulation can be actually used to partially subtract this foreground.}. For simplicity, we rather consider here the yearly average of the signal, so that it can be described by an isotropic and stationary signal approximately described by~\cite{Cornish:2017vip, Cornish:2018dyw, Schmitz:2020rag}
\begin{equation}\label{eq:gal}
       \Omega_{\rm GF}(f) = 10^{\alpha_{\rm GF}} f^{2/3} {\rm exp}\Big(-f^{a_1}-a_2 f\sin({a_3 f})\Big)\{1+\tanh[a_4(f_k-f)]\}\,. 
\end{equation}
For concreteness we take the values used by Ref.~\cite{Flauger:2020qyi}, which correspond to the values quoted in Table 1 of~Ref.~\cite{Cornish:2018dyw} for the parameters $a_1, \dots, a_4, f_k$ (choosing four years of  observation time). As we assume these parameters to be known, we only fit over the amplitude parametrised by $\alpha_{FG}$, for which we take the Gaussian prior proposed by~\cite{Flauger:2020qyi}, $G(\alpha_{FG}^*,\sigma^2)$, with mean value $\alpha_{FG}^* = -7.95$ and standard deviation $\sigma \simeq 0.28$. We note that in the parametrisation given by Eq.~(\ref{eq:gal}), $10^{\alpha_{\alpha_{\rm GF}}}$ does not represent the peak amplitude of the galactic foreground, so the above mean value choice of $\alpha_{FG}^* = -7.95$ corresponds to a foreground's peak amplitude of order $h^2\Omega_{\rm GF} \sim 10^{-10}$.

\begin{figure}[tbp]
\centering
\includegraphics[width=.47\textwidth]{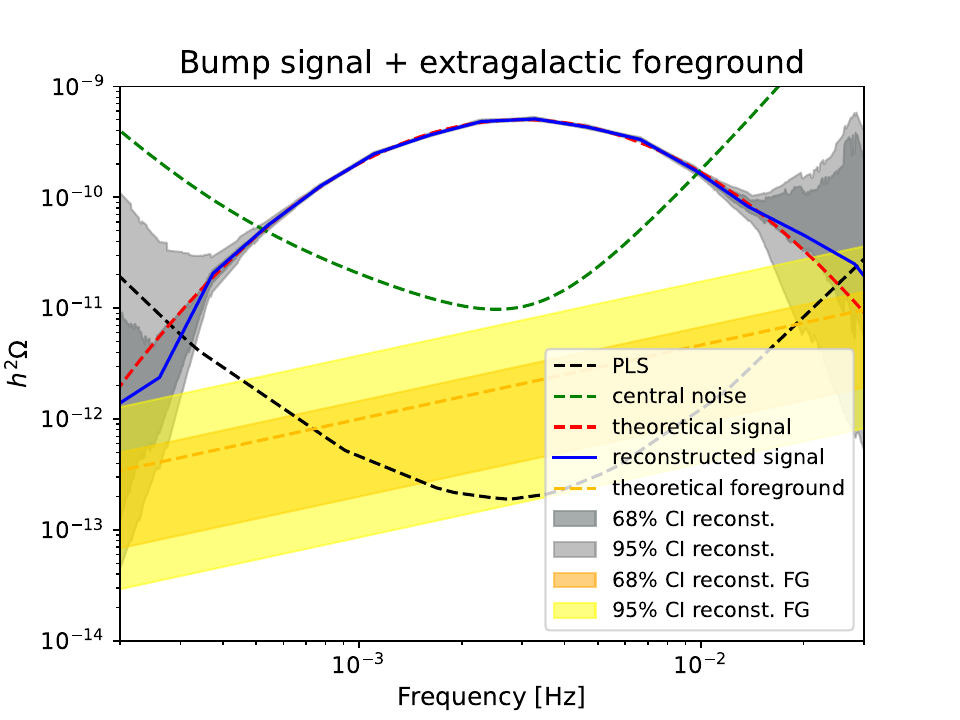}
\qquad
\includegraphics[width=.47\textwidth]{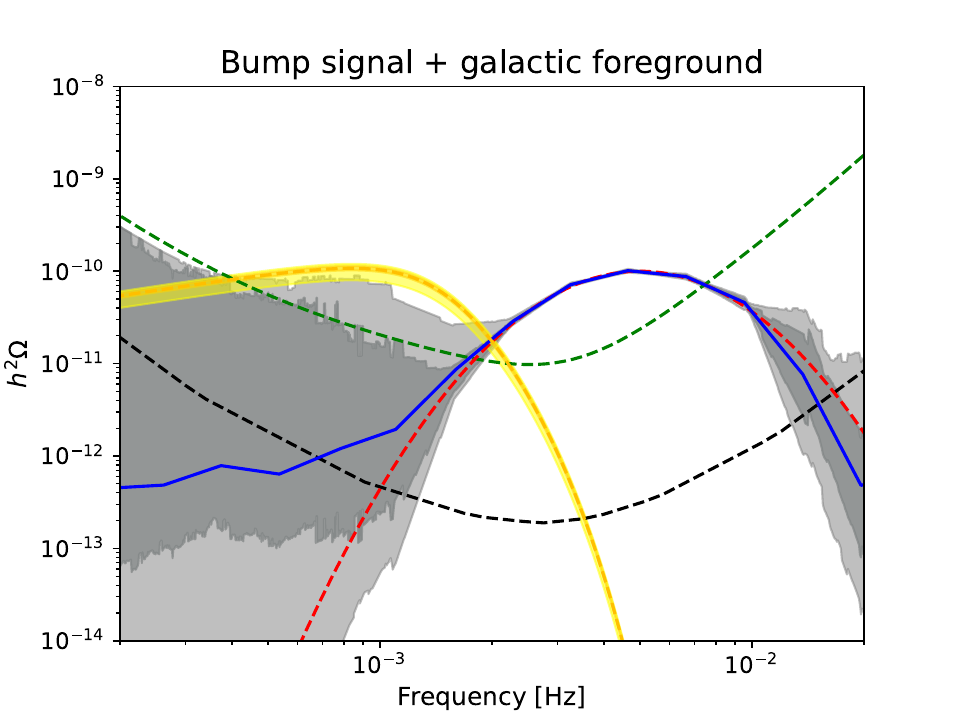 }
\caption{Reconstruction of a single-peaked signal in the presence of a foreground. {Left panel}: Theoretical signal characterised by a single bump profile with $h^{2}\Omega_p^{(*)}=5\cdot 10^{-10}$, $f_p=0.001$ Hz and $\Delta=0.2$, SNR $\sim$ 6492, on top of an extragalactic foreground with (log) amplitude $\alpha_{\rm EF}=-12$. {Right panel}: Theoretical signal characterised by a single bump profile with $h^{2}\Omega_p^{(*)} = 10^{-10}$, $f_p=0.005$ Hz and $\Delta=0.3$, SNR $\sim$ 3053 on top of a galactic foreground with (log) amplitude $\alpha_{\rm GF}=-7.95$.}
 \label{fig:foreground}
\end{figure}

Our results are shown in Fig.~\ref{fig:foreground}. As for the cosmological component, we inject a single bump signal (third entry in Table~\ref{tab:mock}) on top of an extragalactic foreground, and on top of a galactic foreground. As stated before, the foreground model is assumed to be known, so we only reconstruct its amplitude, together with the cosmological component. In the left panel, we reconstruct an injected single bump signal with $\Omega_p^{(*)}=5\cdot 10^{-10}$, $f_p=0.001$Hz and $\Delta=0.2$, on top of an extragalactic foreground with (log) amplitude $\alpha_{\rm EF}=-12$. 
The reconstruction of the foreground's amplitude, while capturing the true value, has an uncertainty compatible with its amplitude's prior, given the very low injected foreground\footnote{Note that we have chosen those values for comparison purposes with the work in ref.\cite{Flauger:2020qyi}. They also recover the foreground's prior, for the same reason commented before.}. 
It would be interesting to see how the reconstruction degrades as we decrease the cosmological signal amplitude with respect to the foreground, but we postpone that exercise for the future, as we believe it will be a more appropriate exercise for a template reconstruction, i.e.~once templates (depending on few parameters) are assumed for both foreground and a cosmological signal.
In the right panel, we reconstruct a single bump signal with $\Omega_p^{(*)} =  10^{-10}$, $f_p=0.005$ Hz and $\Delta=0.3$, on top of a galactic foreground with (log) amplitude $\alpha_{\rm GF}=-7.95$. The reconstruction of the cosmological signal is very precise in the frequency region where the foreground is subdominant. This is expected given the relatively large amplitude of the injected signal. On the other hand, when the
 galactic foreground starts to dominate, the capability of reconstructing the signal rapidly decreases, and the signal reconstruction uncertainties in this frequency range become meaningless. Note that the foreground is very well reconstructed for this example. We postpone for future work further analyses of signal reconstruction on top of astrophysical foregrounds.

\section{Template reconstruction (a teaser)}
\label{sec:template}

We now assume that the spectral shape of the signal is known in advance, and the task is just to infer its parameters. We consider here, for simplicity, the case of a power law signal (cf. first row of table \ref{tab:mock}), characterised by only two parameters: the amplitude $\Omega_p^{(*)}$ and the slope $\gamma$. The same signal template is assumed for both the data generation stage and the inference model (i.e. in this example, the signal model is not given by Eq.(\ref{eq:signal_sim}), but by case 1 of table \ref{tab:mock}, with a single bin covering the whole frequency range). We take $f_p=0.003$ Hz as the reference frequency.

\begin{figure}[tbp]
\centering
\includegraphics[width=.47\textwidth]{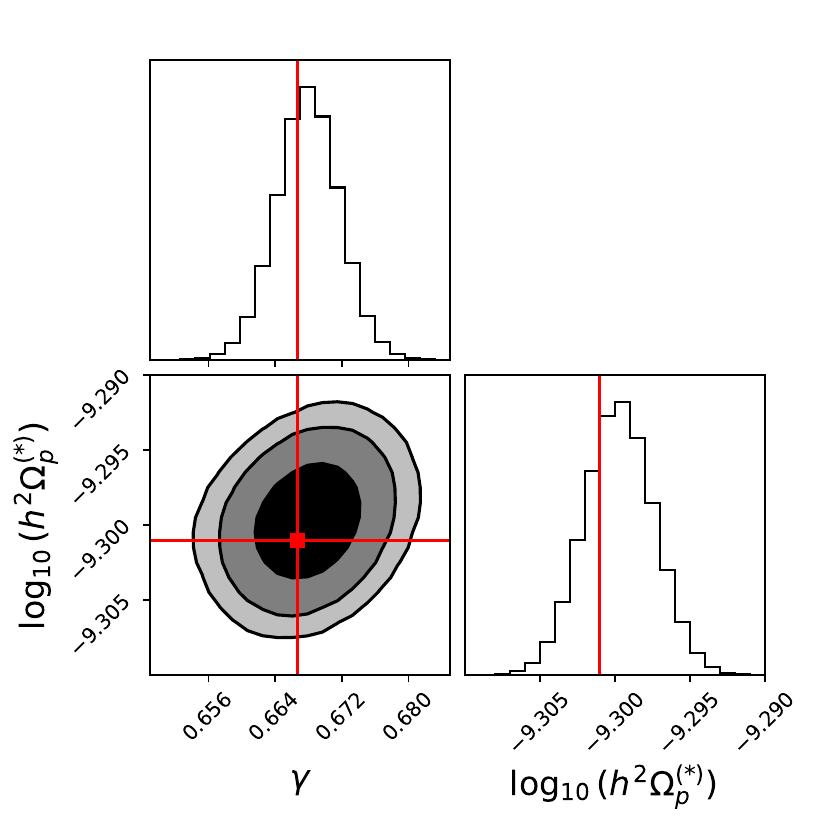}
\qquad
\includegraphics[width=.47\textwidth]{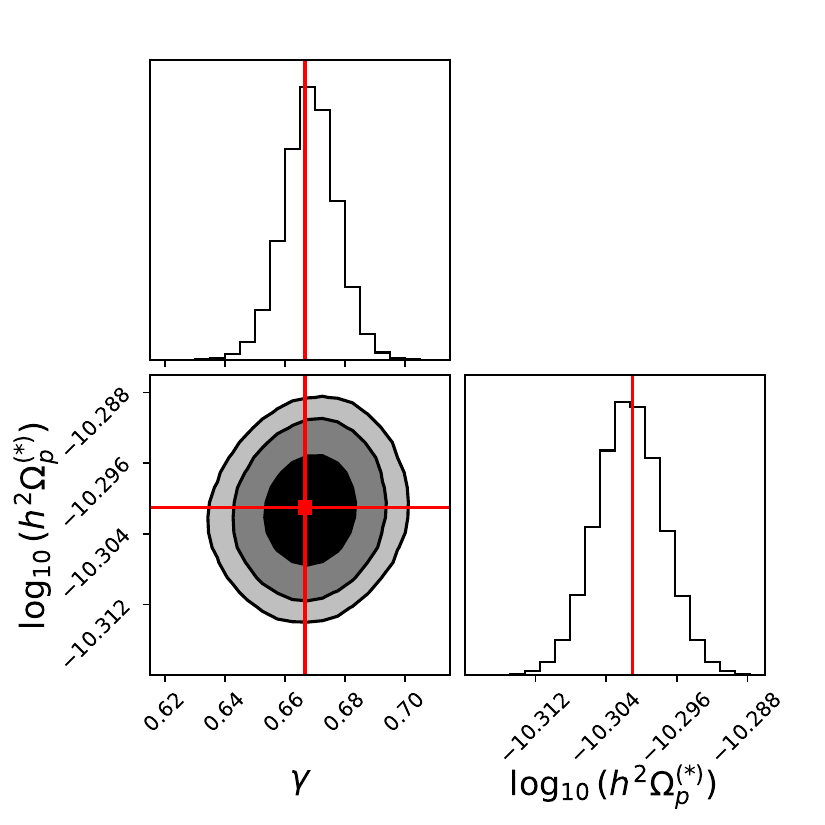}
\qquad
\includegraphics[width=.47\textwidth]{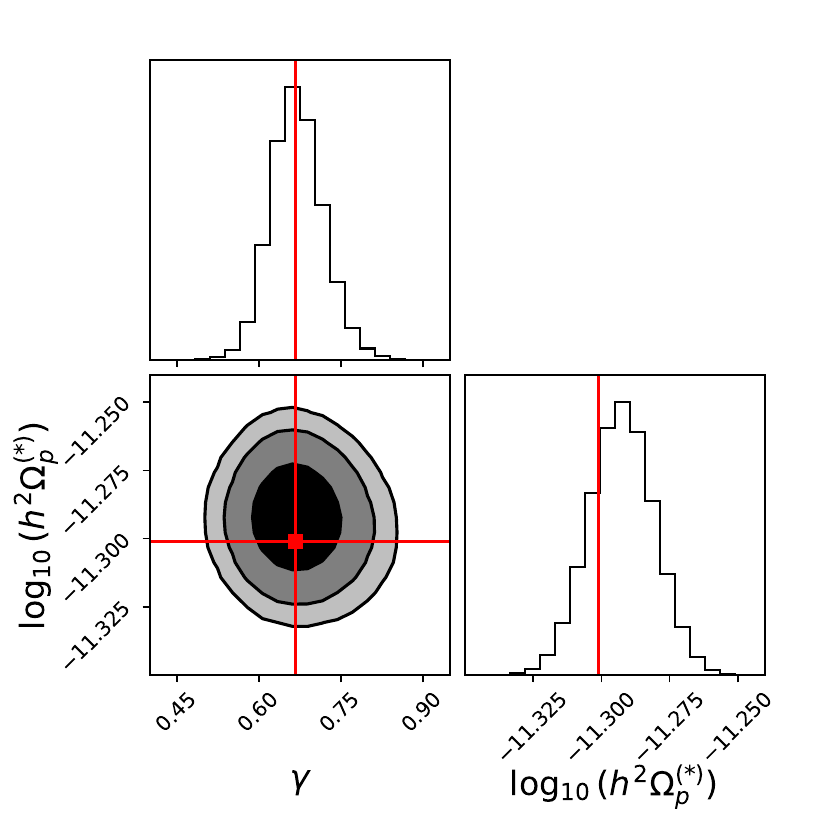}
\quad
\includegraphics[width=.47\textwidth]{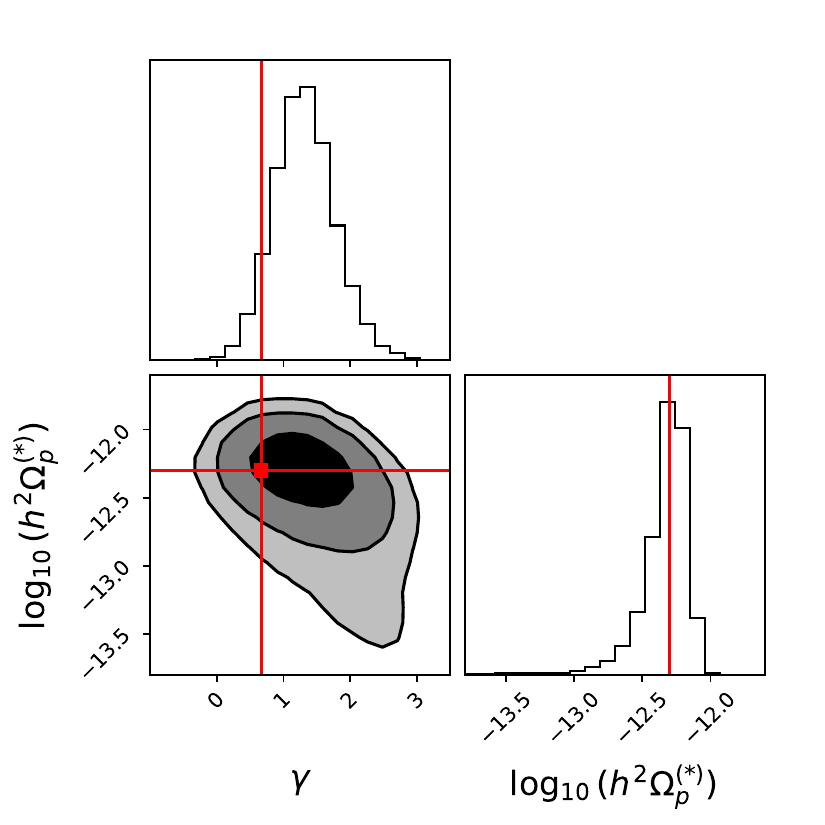}
\caption{Reconstruction of a Power-law signal template with $\gamma = 2/3$. Top-left panel: $h^{2}\Omega_p^{(*)}=5\times 10^{-10}$, SNR $\approx$ 54928, Top-right panel: $h^{2}\Omega_p^{(*)}=5\times 10^{-11}$, SNR $\approx$ 5493, Bottom-left panel: $h^{2}\Omega_p^{(*)}=5\times 10^{-12}$, SNR $\approx$ 549, Bottom-right panel: $h^{2}\Omega_p^{(*)}=5\times 10^{-13}$, SNR $\approx$ 54.9.}
 \label{fig:template}
\end{figure}

The results are shown in figure \ref{fig:template}, for four different cases of injected signal: $h^2\Omega_p^{(*)}=5\times 10^{-10}, 5\times 10^{-11}, 5\times 10^{-12}$, and $5\times 10^{-13}$, for top-left, top-right, bottom-left and bottom-right panels, respectively, all of them with $\gamma=2/3$. As we can see our procedure is able to recover the true parameter values very precisely for relatively large signals (note  the different scales of each panel!). As expected, though, the precision decreases as the amplitude of the injected signal decreases, obtaining roughly an order of magnitude uncertainty for the amplitude reconstruction in the case of the dimmest signal (bottom-right panel). Nonetheless, as a generic feature, we see that the uncertainties are much narrower than for the case of the blind reconstruction. This is expected, as the parametrisation, in this case, is much simpler [2 parameters vs 28 parameters of Eq.(\ref{eq:parametrisation})], besides the fact that the simulation is now performed with the ``correct'' signal model generating the data.

\begin{figure}[tbp]
\centering
\includegraphics[width=.48\textwidth]{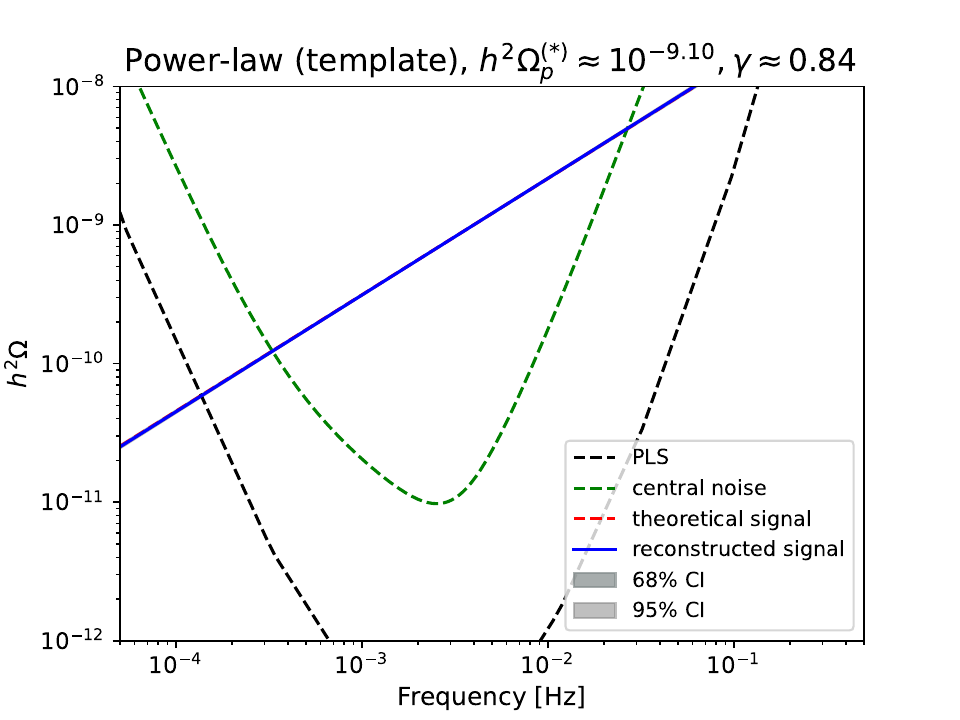}
\qquad
\includegraphics[width=.45\textwidth]{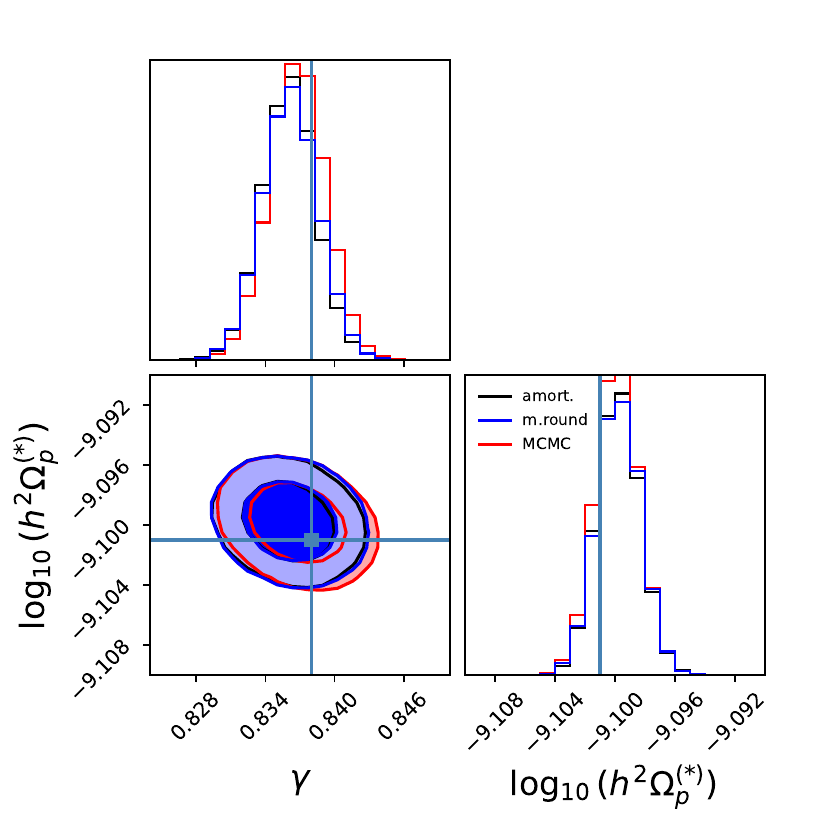}
\caption{Reconstruction of a Power-law signal template with $h^{2}\Omega_p^{(*)}\approx 10^{-9.10}$, $\gamma\approx 0.84$ and SNR $\sim$ 43292. Left panel: The reconstructed signal in the $\Omega$-frequency plane.  Right panel: The corresponding posteriors of the signal parameters, comparing different methods. The blue lines correspond to the true values of the parameters and the contours to the $68\%$ and $95\%$ intervals. All three methods give statistically
compatible results and can cover the true value of the parameters both very accurately
and precisely.  }
 \label{fig:template_mcmc}
\end{figure}

As a way to cross-check our results for this simple case, we also show the comparison from a MCMC analysis, in figure \ref{fig:template_mcmc}. For the MCMC, we follow the {\tt emcee} documentation. We use $nwalkers = 16$ and we start by initializing the walkers in a tiny Gaussian ball around the maximum likelihood result. Then we run 5000 steps of MCMC. Finally,  we take $nburn = 500$, thin by about half the autocorrelation time and flatten the chain so that we have a flat list of samples.
The left panel shows the reconstruction of the signal, and the corresponding posteriors are shown in the right panel in the form of a corner plot. In this case, we perform the MCMC comparison not only with the amortisation procedure followed in all the previous results, but also with the multi-round inference procedure described qualitatively in section \ref{sec:sequential}. While in the case of amortisation we used $\mathcal{O} (10^{5})$ examples for training, for multi-round inference we needed $\mathcal{O} (10^{3})$ and for MCMC $\mathcal{O} (10^{4})$. The difference between the number of samples required for multi-round inference and MCMC becomes more prominent when looking at examples of blind reconstruction with many parameters, as MCMC converges very slowly to the ground truth.  As can be seen in figure \ref{fig:template_mcmc}, all three methods give statistically compatible results, and are able to cover the true value of the parameters very accurately and precisely.
  
\section{Summary and discussion}
\label{sec:conclusions}

In this work, we have developed a new method for fast reconstruction of the spectral shape of a GWB,  based on likelihood-free statistical inference techniques.  
The set of tools we have created is integrated in the package  \href{https://github.com/AndronikiDimitriou/GWBackFinder}{\tt GWBackFinder}, which is a publicly available GitHub repository. Our method can be readily applied to any detector for which a characterization of its instrumental noise is available.  As a demonstration, we have focused in this paper on the reconstruction of an arbitrary shape signal by the LISA detector.

Using our technique,  we quantify the ability of LISA to reconstruct backgrounds of an arbitrary spectral shape ({\it blind} reconstruction) in Sect.~\ref{sec:BlindReconstruction}. This includes an analysis in Sect.~\ref{subsec:arbitraryShape} of the reconstruction of a variety of frequency profiles, for which we obtain, generally, an accurate reconstruction within the central frequency region of LISA. The reconstruction degrades, as expected when the signals approach the power-law sensitivity (PLS) curve. The accuracy of the reconstruction (within each frequency bin), depends of course on the relative height of the signal with respect to the noise or the PLS.  In order to quantify this, we propose two procedures to characterize the accuracy of reconstruction in Sect.~\ref{subsec:uncertainty}, one specialized to power law signals (Sect.~\ref{subsubsec:PLuncertainty}), and another, more general, suitable for signals with an arbitrary spectral shape (Sect.~\ref{subsubsec:ARE}). To validate our technique, we present a comparison of our results against MCMC methods in Sect.~\ref{subsec:MCMCcomparison}. Finally, for completeness, we also consider in Sect.~\ref{sec:foreground} blind reconstruction of representative signals on top of astrophysical foregrounds. This is a proof-of-principle towards more realistic simulations in the spirit of an ultimate global fit.

We highlight that our method can be immediately extended to the reconstruction of multi-parameter template-dependent signals ({\it template} reconstruction).  As a teaser of this,  we show in Sect.~{\ref{sec:template}} a dedicated study to power-law signals. As expected, the procedure is able to reconstruct the signals much more precisely than for the case of the blind reconstruction, mainly due to the fact that the simulations use the correct signal model generating the data.

We note that in this paper, for blind reconstruction, we have fixed the number of frequency sub-intervals within the LISA frequency range $(f_{\rm min},f_{\rm max})$, to $N_c = 27$, choosing equal log-spacing between their extremes. As argued in Sect.~\ref{subsec:uncertainty}, this choice is a compromise between model complexity and computational cost. We note that even though a higher frequency resolution in the reconstruction of a GWB spectrum would be desirable, a larger number of bins would not work perfectly, even if computationally feasible, due to the statistical fluctuations of the data within the bins becoming more important. However, we highlight that our statistical technique does not need to stick to regular bin intervals, and we expect it to work equally well for arbitrary binning configurations. Actually, a future improvement that we are currently developing, consists of allowing break frequencies to be free parameters that are adjusted in the process of signal reconstruction, according to an error minimization criterion. We expect that implementing this 
idea might make a difference in the reconstruction of signals with frequency profiles with many features, especially when these get close enough to the PLS. 

While our technique has several advantages with respect to traditional MCMC methods, we have validated it with the latter for concrete cases, verifying that our results are reliable. Our work opens the door for both fast and accurate Bayesian parameter estimation of GWBs,  with essentially no computational overhead during the inference step. Actually, this paper should be considered the first of a series of works where we intend to use {\tt GWBackFinder} to study the ability of LISA, and other detectors, to reconstruct the parameter space of predicted cosmological and astrophysical GWBs, including many of the signals listed in  Sect.~\ref{sec:intro}.

Last but not least, we comment on how the strategy we have presented in this work can be incorporated into the picture of a ``global fit''~\cite{Cornish:2005qw, Vallisneri:2008ye, MockLISADataChallengeTaskForce:2009wir}. The latter refers to the idea that, when reconstructing GW signals at LISA or similar facilities, we should take into account simultaneously every possible source: not only stochastic backgrounds from cosmological processes on top of astrophysical ones (stochastic foregrounds as the ones considered in this analysis), but also other types of foregrounds as ultra compact binaries, super-massive black holes, extreme mass-ratio inspirals, etc. In our analysis we have incorporated GW foregrounds (galactic or extragalactic) on top of the cosmological backgrounds simply by incorporating the former to the simulation process. Concretely, we have modelled GWB expected signals as a sum of the two contributions: $\Omega_{\rm GW} = \Omega_{\rm cosm.} + \Omega_{\rm foreg.}$, for every channel and frequency (omitted here for the ease of notation), after which we sample the noisy data as specified in sec. \ref{subsec:dataGeneration} as usual. The inference on both contributions is a priori performed simultaneously in the SBI approach, unless a suitable statistical modelling for the dependencies is available, in whose case the latter can be exploited to aid the inference procedure. In the very same way, one can incorporate other parameterised GW contributions directly into the simulator: $\Omega_{\rm GW} = \sum_k \Omega_{\rm k}$, for which the inference procedure, now containing more parameters, remains unaltered.      

\vspace*{0.1cm}
\paragraph{Note added.}  In parallel to this work, a method based on neural ratio estimation (as opposed to neural posterior estimation used in this study) was investigated in \cite{Valerie}, where a single-channel analysis in the presence of transient signals was included, while adopting the same assumptions about the noise modeling as in this work.

\acknowledgments

We are grateful to Peera Simakachorn for help in building the PLS curve; 
Miguel G. Folgado and Aurelio Amerio for helping us with the computational resources; 
Ben Miller for useful discussions on statistical methods, and James Alvey, Uddipta Bhardwaj, Valerie Domcke, Mauro Pieroni, and Christoph Weniger for beneficial discussions, and for having coordinated with us the release of their independent work on a similar subject. We are also grateful for the benefit of our participation in the \href{https://indico.cern.ch/event/1257531/}{\tt TH Institute on GWB data analysis} organized at CERN in July 2023. DGF (ORCID 0000-0002-4005-8915) is supported by a Ram\'on y Cajal contract with Ref.~RYC-2017-23493. AD and BZ akcnowledges
the support from CIDEGENT/2020/055. This work was supported by Generalitat Valenciana grant PROMETEO/2021/083, and by  Spanish Ministerio de Ciencia e Innovacion grant PID2020-113644GB-I00. We also thank the computer resources at IFIC; in particular the {\it SOM cluster}, nowadays funded through the grants EUR2022-134028 and ASFAE/2022/020, as well as Artemisa, funded by the European Union ERDF and Comunitat Valenciana as well as the technical support provided by the
Instituto de F\'isica Corpuscular, IFIC (CSIC-UV), Generalitat
Valenciana, Project “ARTEMISA”, ref. ASFAE/2022/024.

\appendix
\section{Relation to other simulation-based methods}
\label{app:A}

The NPE method adopted in this work is not the only existing one leading the state-of-the-art in the field of simulation-based inference. A second very popular method belongs to the family of Neural Ratio Estimation (NRE). Its similarities and differences with the NPE method adopted in this work are briefly described below, as well as our motivation to adopt the latter.

{\it Similarities}. 1) Both methods consider neural networks in order to infer the posterior of the parameters of interest $\bs{\theta}$, given a dataset; 2) both are likelihood-free, relying on samples from the simulator. 

{\it Main difference}. Instead of optimizing a distribution $q_{\bs{\phi}}$ to approximate the true, intractable posterior, the NRE approach attempts to estimate the likelihood-to-evidence ratio. As it is well known, a neural network classifier is formally an approximation of likelihood ratios; so this idea is at the heart of current implementations ~\cite{Cranmer_2020,brehmer2020,lueckmann2021,Miller2022,Tejero-Cantero2020,Cole:2021gwr,Montel:2022fhv,Makinen:2021nly,Dimitriou:2022cvc,Gagnon-Hartman:2023soa,Delaunoy:2020zcu,Karchev:2022xyn,Lin:2022ayr,Alvey:2023naa,Bhardwaj:2023xph}. Thus, thanks to the Bayes theorem and for a given prior, the NRE becomes in practice an estimate of the posterior.

On the other hand, the NRE method is very convenient when we are interested in few-dimensional marginal posteriors (typically 1d or 2d) as a final object of interest. While estimating high-dimensional posteriors works equally well per se, so far, in practice only 2d posteriors are computed and presented in standard ``corner plots''. This is because NRE struggles, in general, to obtain samples from the estimated high-dimensional posteriors (as usual, sampling generic multivariate distributions is a hard problem). In our case, apart from the posteriors, we are further interested in predictions for $\Omega_{\rm GW}(f;{\bf s})$, which is evaluated with the joint samples, as described above. This is the main reason behind our choice of the NPE over the NRE method: for the former, and thanks to the Normalizing-Flow implementation (cf. Appendix \ref{app:D}), samples are readily available by construction.

\section{Asymmetric reconstruction capabilities}
\label{app:B}

\begin{figure}[htbp]
\centering
\includegraphics[width=.47\textwidth]{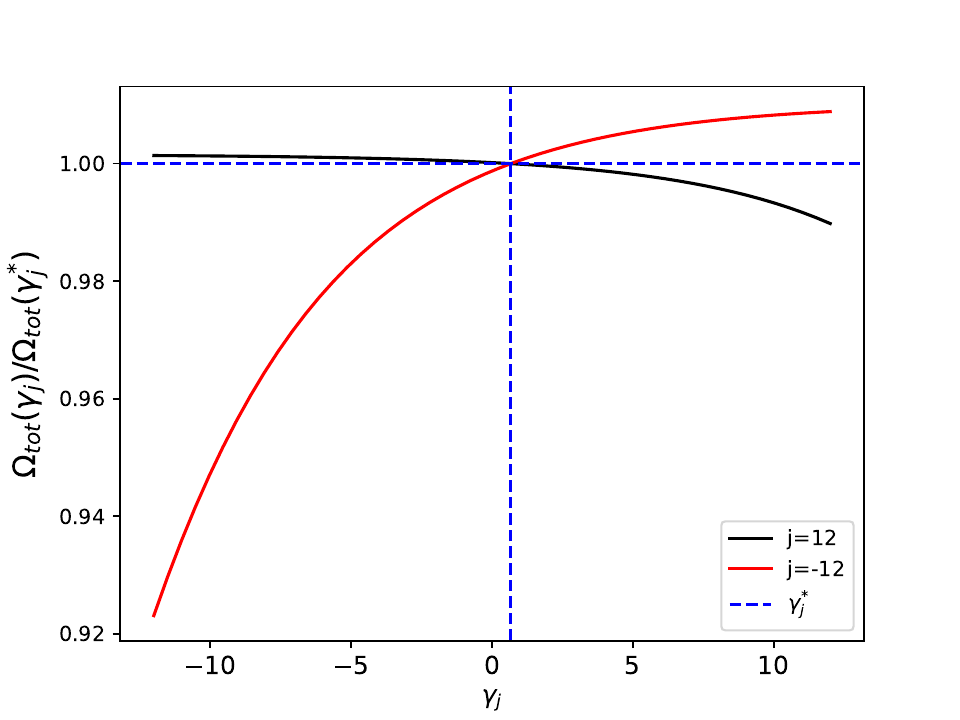}
\qquad
\includegraphics[width=.47\textwidth]{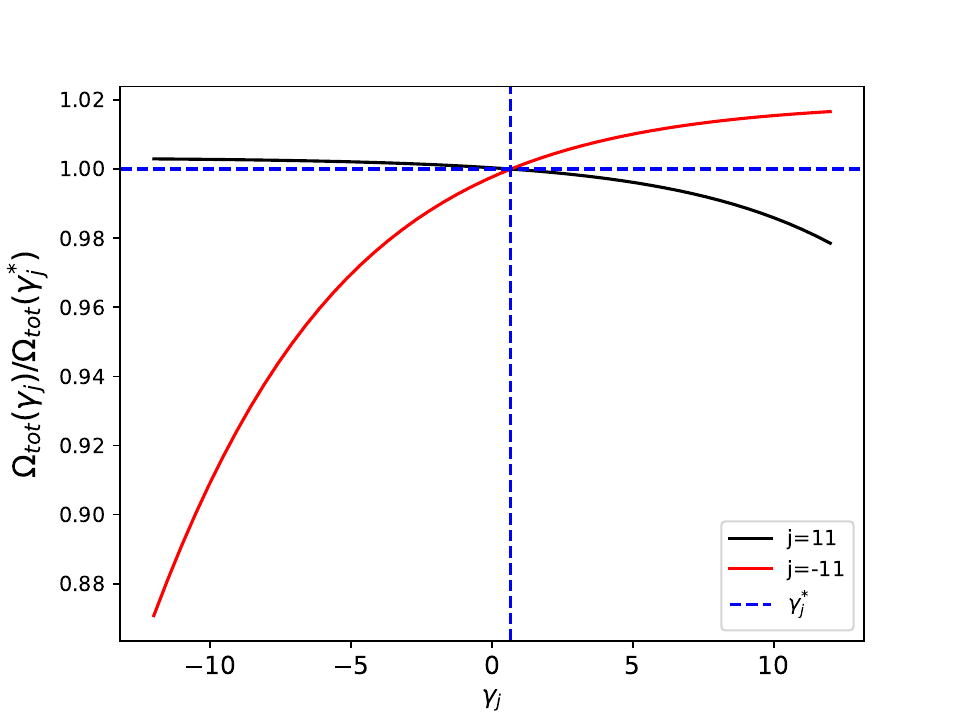}
\qquad
\includegraphics[width=.47\textwidth]{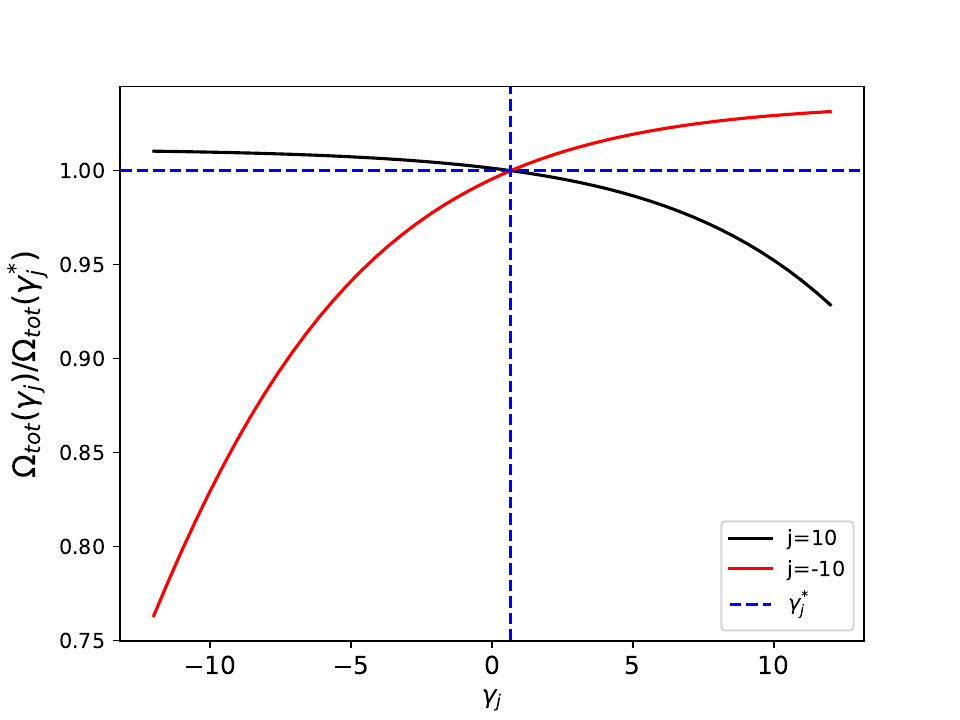}
\qquad
\includegraphics[width=.47\textwidth]{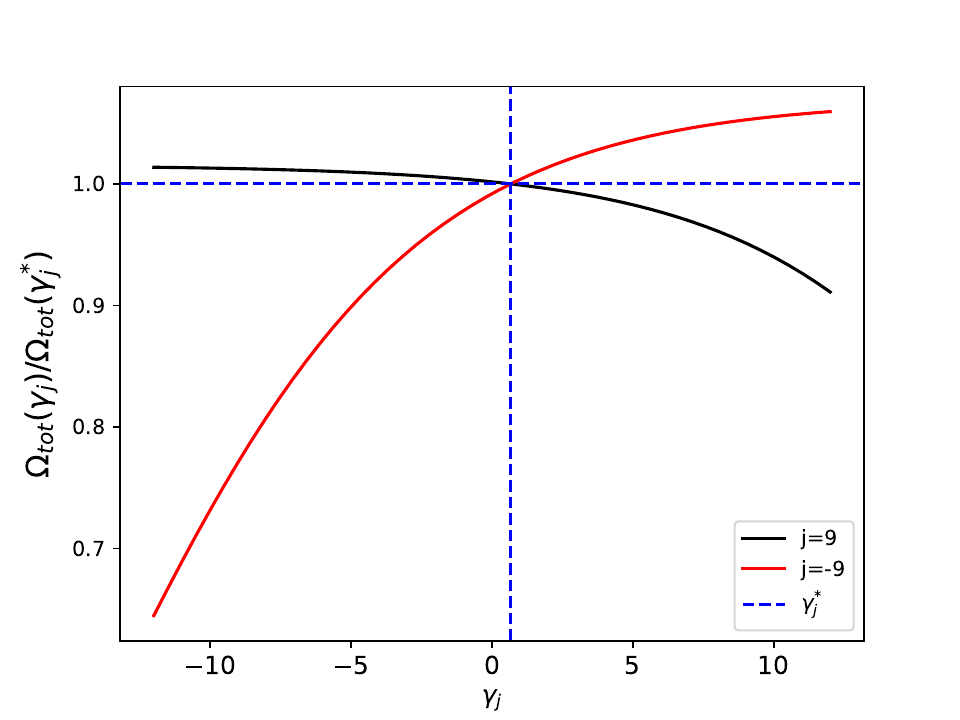}

\caption{Each curve in these panels represent the ratio $\Omega_{\rm tot}(\gamma_j)/\Omega_{\rm tot}(\gamma^*_j)$, as a function of $\gamma_j$, where $j$ is the bin label: $j=12$ vs. $j=-12$ (top-left), $j=11$ vs. $j=-11$ (top-right), $j=-10$ vs. $j=10$ (bottom-left), and $j=-9$ vs. $j=9$ (bottom-right). The vertical line corresponds to $\gamma^*_j$ in each case.}
 \label{fig:ratios}
\end{figure}

In figure \ref{fig:ratios}, we compute for a given frequency region (or bin) $j$ of our signal parametrisation, the ratio of  $\Omega_{\rm tot}(\gamma_j)$ to $\Omega_{\rm tot}(\gamma^*_j)$, as a function of the slope $\gamma_j$ of that bin, with $\gamma^*_j$ is a fixed value. The rest of the parameters of the model (amplitude and all the other slopes) are fixed as well. 

In this figure, the larger the deviation from 1, the larger the sensitivity of the corresponding frequency region to a putative signal, thus the smaller the reconstructed uncertainties. We can see that as the frequency regions in consideration approach the central part, the sensitivity increases, thus decreasing the uncertainties. In all the panels of this figure, we compare pairs of bins on the extreme sides of the frequency range ($j=$-12 vs. $j=$12, $j=$-11 vs. $j=$11, $j=$-10 vs. $j=$10 and $j=$-9 vs. $j=$9, respectively, cf. Eq.\ref{eq:parametrisation}), and in all the cases, the asymmetry is evident: bins in the large frequency region (positive $j$) are less sensitive to changes of the signal slope than the ``mirror'' bins at the low frequency region (negative $j$), and consequently the uncertainties are larger in the former range.

\section{Consistency checks}
\label{app:C}

In this appendix we discuss about ways to check the quality of the approximated posteriors we obtain with our method, independent of MCMC methods\footnote{The methodology followed in this work is an alternative to MCMC methods to start with; cf. sec. \ref{sec:intro} for its motivation. However, in simple cases as the one shown in sec. \ref{subsec:MCMCcomparison}, an MCMC cross check is always desirable.}.

Studying how to check the quality of the posterior inference for simulation-based methods is actually an active area of research \cite{2021arXiv211006581H}. In this study we consider what is arguably one of the most popular procedures, known as the {\it coverage tests}, also known as ``percentile-percentile'' plots\footnote{See e.g. \cite{Cook2006} for an early presentation, much before their use in the context of SBI.}. The idea behind the latter is simple: due to statistical fluctuations, the $x$ credible interval (CI, confidence level $z$) of the estimated posterior distribution of a given parameter, should cover the true value roughly a fraction $x$ of all the cases (EC, empirical coverage, $\hat{z}$), when repeating the inference over different simulated data. See \cite{2021arXiv211006581H} for a more detailed discussion of this idea. Thus, as a function of the CI, a reasonably well gauged posterior would result in a EC curve following closely the diagonal EC = CI.

\begin{figure}[htbp]
\centering
\includegraphics[width=1\textwidth]{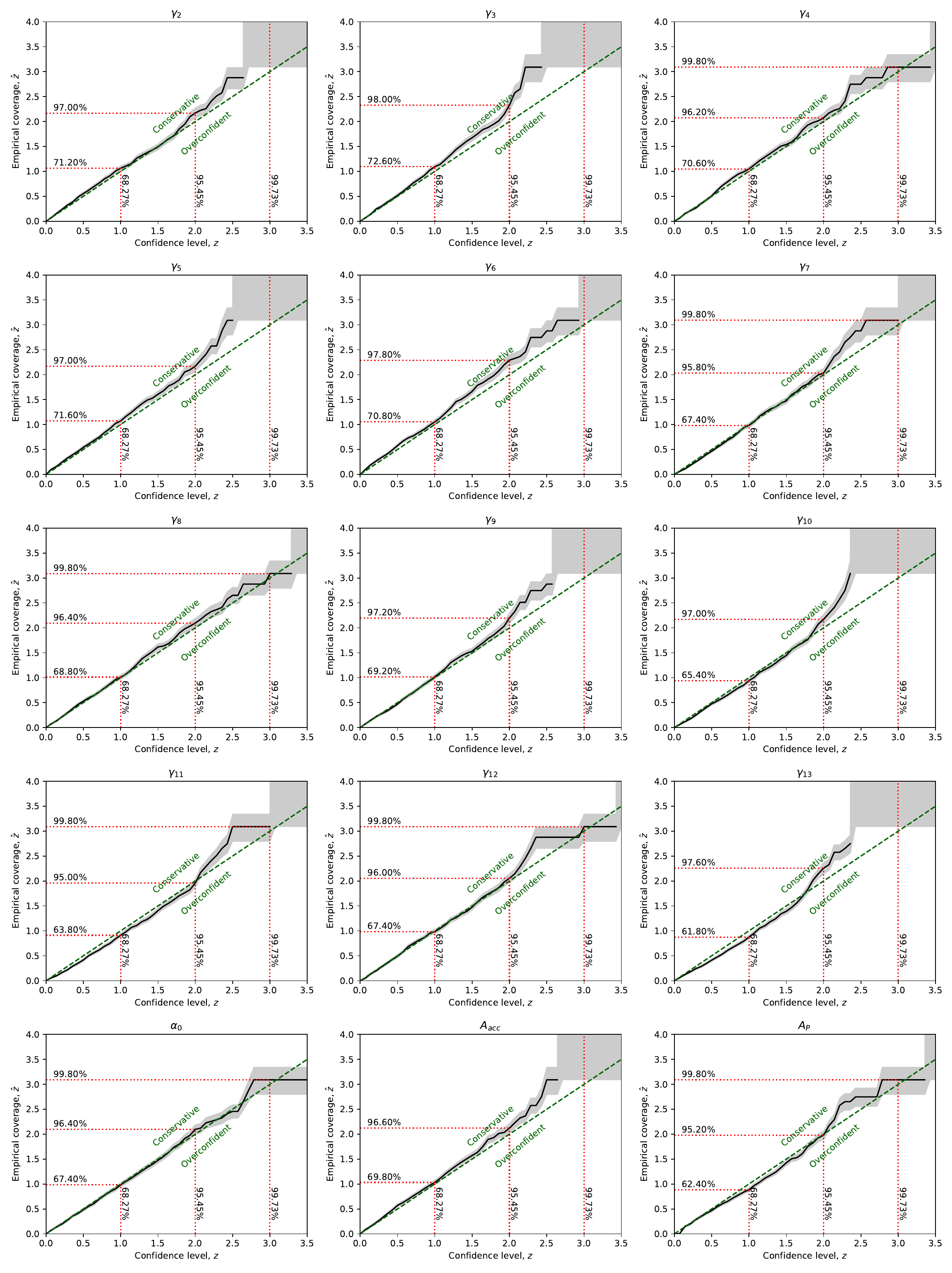}

\caption{Following \cite{2022JCAP...09..004C}, we plot the empirical expected coverage probability as a function of the confidence level, $1- \alpha= 0.6827, 0.9545, 0.9997$, or alternatively of $z=1, 2, 3$, for all the parameters of our signal and noise parametrisation. We see that we have a converged network with good coverage since the empirical expected coverage probability is slightly larger or equal to the confidence levels. The confidence levels $1 - \alpha$ are shown in vertical dashed red lines, while the empirical estimates $1 - \hat{\alpha}$ are shown in the horizontal ones.}
 \label{fig:coverage}
 \vspace{-31.569pt}
\end{figure}

We show in fig. \ref{fig:coverage} the coverage plots for all the parameters. The error bands correspond to the probability-point-function (PPF) range of the EC variable, the latter following a beta distribution\footnote{This is related to the Jeffreys interval, resulting from the fact that the number of times a given CI covers the true value follows naturally a binomial distribution.}. Each of these plots are interpreted as follows: 
curves should ideally be along the diagonal. 
In cases where $\hat{z}> z$, the estimated credible contours of the posteriors are conservative and contain the true value with a frequency higher than nominally expected. On the contrary, when $\hat{z}< z$, the estimated credible contours contain the true value with a frequency lower than expected and hence they are overconfident.
According to \cite{2021arXiv211006581H}, a posterior estimator is considered to be acceptable whenever the empirical expected coverage probability is larger or equal to the confidence level.
We see from this figure that the posteriors of all the parameters are very well calibrated (with very few cases being slightly conservative), since they follow closely the diagonal line.


\begin{figure}[htbp]
\centering
\includegraphics[width=1\textwidth]{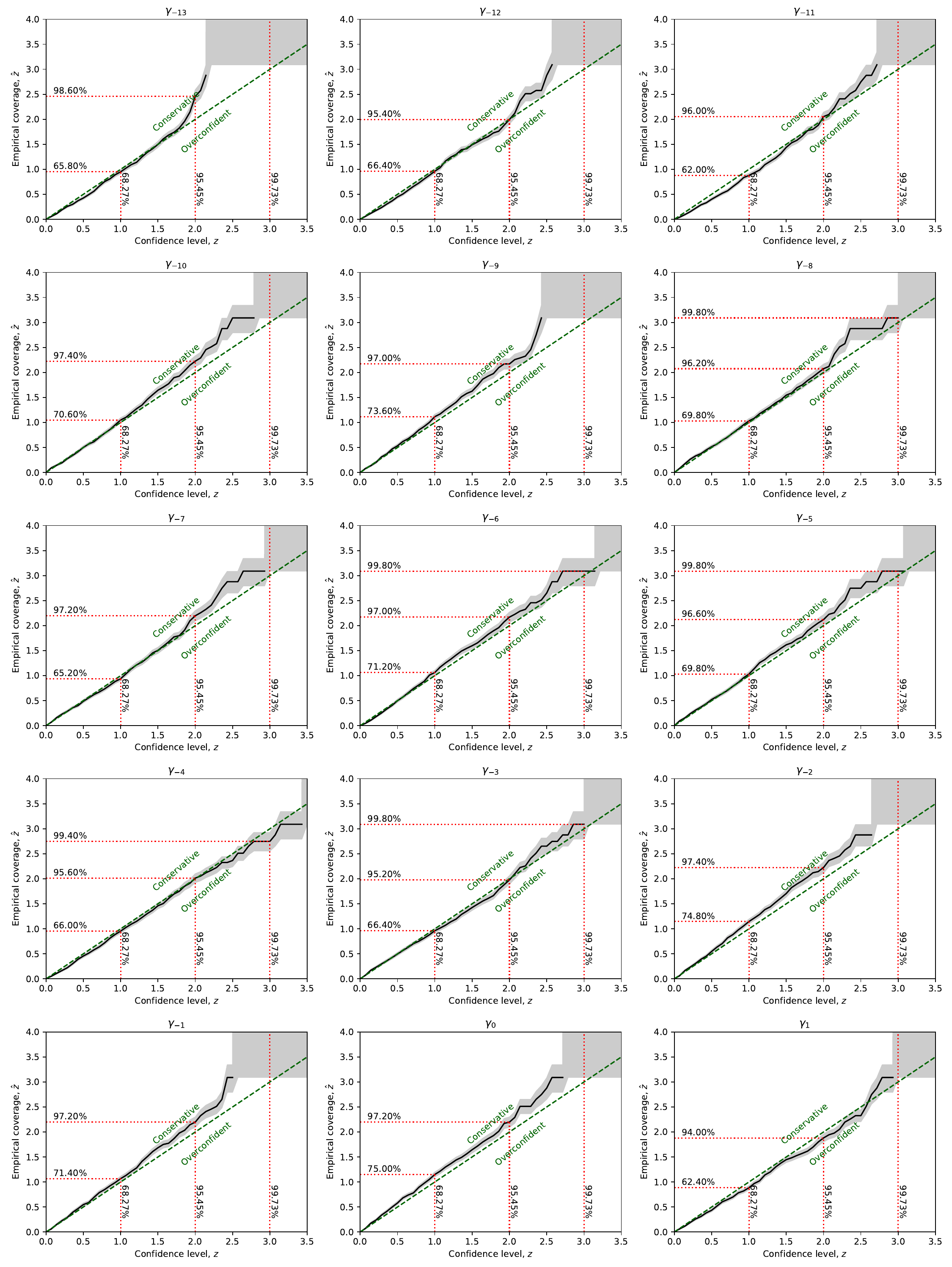}

\end{figure}

\section{Specifying the shape of the approximate posterior}
\label{app:D}

In this appendix, we aim at providing more details on how we model the approximate posterior of the parameters of interest while making an effort to avoid unnecessary technicalities. The interested reader can further refer to the literature cited below.
\newline\newline\noindent
In this work, we parameterise the approximate posterior $q_{\bs{\phi}}(\bs{\theta})$ of the parameters of interest $\bs{\theta}$, with a so-called Normalizing Flow (NF) framework (see \cite{2019arXiv191202762P} for a comprehensive and modern review). The latter is a convenient way to construct flexible probability distributions over continuous random variables. In general they allow us to obtain samples from $\bs{\theta}$ by proposing a transformation $T_{\bs{\phi}}$:

\begin{equation}
  \bs{\theta} = T_{\bs{\phi}}({\bf u}),~~~~~
  {\bf u}\sim p_{\bs{\psi}}({\bf u})~,
\end{equation}
where the parameters $\bs{\phi}$ of the transformation are precisely those of the approximate posterior $q$ we are parameterising, and ${\bf u}$ is a latent random variable, ${\rm dim}({\bf u})={\rm dim}(\bs{\theta})$, but following instead a parameterically easier ``base'' distribution $p_{\bs{\psi}}({\bf u})$ with parameters $\bs{\psi}$; for example, a Gaussian distribution. 

Different flavors of NF differ in the proposal function $T_{\bs{\phi}}$, as well as the base distribution. For sufficiently flexible choices, NFs are able to capture arbitrarily complex distributions. While in our application we are mainly interested in samples of $q_{\bs{\phi}}(\bs{\theta})$, rather than evaluating the distribution itself, note that the latter can be formally obtained simply by inverting the transformation, i.e. obtaining $T^{-1}_{\bs{\phi}}$, and computing the corresponding Jacobian. Therefore, NFs require in general invertible transformations.
\newline\newline\noindent
We use a type of NF called ``neural spline flow" \cite{2019arXiv190604032D}. The idea is that the transformation is composed of splines, i.e. a piecewise function covering the desired range of the variables, whose breakpoints (a.k.a. ``knots'' in this context) are predefined. At each segment, the function consist of a rational-quadratic function, i.e. a ratio of order-2 polynomials, the parameters of which are given by the output of a common neural network. For more details see sec. 3.1 of \cite{2019arXiv190604032D}.

\bibliographystyle{JHEP}
\bibliography{automatic,manual}

\providecommand{\href}[2]{#2}\begingroup\raggedright\begin{thebibliography}{100}

\bibitem{Abbott:2016nmj}
{\scshape LIGO Scientific, Virgo} collaboration, \emph{{GW151226: Observation of Gravitational Waves from a 22-Solar-Mass Binary Black Hole Coalescence}}, \href{https://doi.org/10.1103/PhysRevLett.116.241103}{\emph{Phys. Rev. Lett.} {\bfseries 116} (2016) 241103} [\href{https://arxiv.org/abs/1606.04855}{{\ttfamily 1606.04855}}].

\bibitem{Abbott:2017vtc}
{\scshape LIGO Scientific, VIRGO} collaboration, \emph{{GW170104: Observation of a 50-Solar-Mass Binary Black Hole Coalescence at Redshift 0.2}}, \href{https://doi.org/10.1103/PhysRevLett.118.221101}{\emph{Phys. Rev. Lett.} {\bfseries 118} (2017) 221101} [\href{https://arxiv.org/abs/1706.01812}{{\ttfamily 1706.01812}}].

\bibitem{Abbott:2017gyy}
{\scshape LIGO Scientific, Virgo} collaboration, \emph{{GW170608: Observation of a 19-solar-mass Binary Black Hole Coalescence}}, \href{https://doi.org/10.3847/2041-8213/aa9f0c}{\emph{Astrophys. J. Lett.} {\bfseries 851} (2017) L35} [\href{https://arxiv.org/abs/1711.05578}{{\ttfamily 1711.05578}}].

\bibitem{Monitor:2017mdv}
{\scshape LIGO Scientific, Virgo, Fermi-GBM, INTEGRAL} collaboration, \emph{{Gravitational Waves and Gamma-rays from a Binary Neutron Star Merger: GW170817 and GRB 170817A}}, \href{https://doi.org/10.3847/2041-8213/aa920c}{\emph{Astrophys. J. Lett.} {\bfseries 848} (2017) L13} [\href{https://arxiv.org/abs/1710.05834}{{\ttfamily 1710.05834}}].

\bibitem{Abbott:2017oio}
{\scshape LIGO Scientific, Virgo} collaboration, \emph{{GW170814: A Three-Detector Observation of Gravitational Waves from a Binary Black Hole Coalescence}}, \href{https://doi.org/10.1103/PhysRevLett.119.141101}{\emph{Phys. Rev. Lett.} {\bfseries 119} (2017) 141101} [\href{https://arxiv.org/abs/1709.09660}{{\ttfamily 1709.09660}}].

\bibitem{TheLIGOScientific:2017qsa}
{\scshape LIGO Scientific, Virgo} collaboration, \emph{{GW170817: Observation of Gravitational Waves from a Binary Neutron Star Inspiral}}, \href{https://doi.org/10.1103/PhysRevLett.119.161101}{\emph{Phys. Rev. Lett.} {\bfseries 119} (2017) 161101} [\href{https://arxiv.org/abs/1710.05832}{{\ttfamily 1710.05832}}].

\bibitem{LIGOScientific:2018mvr}
{\scshape LIGO Scientific, Virgo} collaboration, \emph{{GWTC-1: A Gravitational-Wave Transient Catalog of Compact Binary Mergers Observed by LIGO and Virgo during the First and Second Observing Runs}}, \href{https://doi.org/10.1103/PhysRevX.9.031040}{\emph{Phys. Rev. X} {\bfseries 9} (2019) 031040} [\href{https://arxiv.org/abs/1811.12907}{{\ttfamily 1811.12907}}].

\bibitem{LIGOScientific:2021usb}
{\scshape LIGO Scientific, VIRGO} collaboration, \emph{{GWTC-2.1: Deep Extended Catalog of Compact Binary Coalescences Observed by LIGO and Virgo During the First Half of the Third Observing Run}},  \href{https://arxiv.org/abs/2108.01045}{{\ttfamily 2108.01045}}.

\bibitem{LIGOScientific:2021djp}
{\scshape LIGO Scientific, VIRGO, KAGRA} collaboration, \emph{{GWTC-3: Compact Binary Coalescences Observed by LIGO and Virgo During the Second Part of the Third Observing Run}},  \href{https://arxiv.org/abs/2111.03606}{{\ttfamily 2111.03606}}.

\bibitem{Caprini:2018mtu}
C.~Caprini and D.G.~Figueroa, \emph{{Cosmological Backgrounds of Gravitational Waves}}, \href{https://doi.org/10.1088/1361-6382/aac608}{\emph{Class. Quant. Grav.} {\bfseries 35} (2018) 163001} [\href{https://arxiv.org/abs/1801.04268}{{\ttfamily 1801.04268}}].

\bibitem{Maggiore:2018sht}
M.~Maggiore, \emph{{Gravitational Waves. Vol. 2: Astrophysics and Cosmology}}, Oxford University Press (3, 2018).

\bibitem{Christensen:2018iqi}
N.~Christensen, \emph{{Stochastic Gravitational Wave Backgrounds}}, \href{https://doi.org/10.1088/1361-6633/aae6b5}{\emph{Rept. Prog. Phys.} {\bfseries 82} (2019) 016903} [\href{https://arxiv.org/abs/1811.08797}{{\ttfamily 1811.08797}}].

\bibitem{KAGRA:2021kbb}
{\scshape KAGRA, Virgo, LIGO Scientific} collaboration, \emph{{Upper limits on the isotropic gravitational-wave background from Advanced LIGO and Advanced Virgo\textquoteright{}s third observing run}}, \href{https://doi.org/10.1103/PhysRevD.104.022004}{\emph{Phys. Rev. D} {\bfseries 104} (2021) 022004} [\href{https://arxiv.org/abs/2101.12130}{{\ttfamily 2101.12130}}].

\bibitem{KAGRA:2021mth}
{\scshape KAGRA, Virgo, LIGO Scientific} collaboration, \emph{{Search for anisotropic gravitational-wave backgrounds using data from Advanced LIGO and Advanced Virgo\textquoteright{}s first three observing runs}}, \href{https://doi.org/10.1103/PhysRevD.104.022005}{\emph{Phys. Rev. D} {\bfseries 104} (2021) 022005} [\href{https://arxiv.org/abs/2103.08520}{{\ttfamily 2103.08520}}].

\bibitem{Regimbau:2011rp}
T.~Regimbau, \emph{{The astrophysical gravitational wave stochastic background}}, \href{https://doi.org/10.1088/1674-4527/11/4/001}{\emph{Res. Astron. Astrophys.} {\bfseries 11} (2011) 369} [\href{https://arxiv.org/abs/1101.2762}{{\ttfamily 1101.2762}}].

\bibitem{Stiskalek:2020wbj}
R.~Stiskalek, J.~Veitch and C.~Messenger, \emph{{Are stellar--mass binary black hole mergers isotropically distributed?}}, \href{https://doi.org/10.1093/mnras/staa3613}{\emph{Mon. Not. Roy. Astron. Soc.} {\bfseries 501} (2021) 970} [\href{https://arxiv.org/abs/2003.02919}{{\ttfamily 2003.02919}}].

\bibitem{NANOGrav:2023gor}
{\scshape NANOGrav} collaboration, \emph{{The NANOGrav 15 yr Data Set: Evidence for a Gravitational-wave Background}}, \href{https://doi.org/10.3847/2041-8213/acdac6}{\emph{Astrophys. J. Lett.} {\bfseries 951} (2023) L8} [\href{https://arxiv.org/abs/2306.16213}{{\ttfamily 2306.16213}}].

\bibitem{Antoniadis:2023ott}
{\scshape EPTA} collaboration, \emph{{The second data release from the European Pulsar Timing Array III. Search for gravitational wave signals}},  \href{https://arxiv.org/abs/2306.16214}{{\ttfamily 2306.16214}}.

\bibitem{Reardon:2023gzh}
D.J.~Reardon et~al., \emph{{Search for an Isotropic Gravitational-wave Background with the Parkes Pulsar Timing Array}}, \href{https://doi.org/10.3847/2041-8213/acdd02}{\emph{Astrophys. J. Lett.} {\bfseries 951} (2023) L6} [\href{https://arxiv.org/abs/2306.16215}{{\ttfamily 2306.16215}}].

\bibitem{Xu:2023wog}
H.~Xu et~al., \emph{{Searching for the Nano-Hertz Stochastic Gravitational Wave Background with the Chinese Pulsar Timing Array Data Release I}}, \href{https://doi.org/10.1088/1674-4527/acdfa5}{\emph{Res. Astron. Astrophys.} {\bfseries 23} (2023) 075024} [\href{https://arxiv.org/abs/2306.16216}{{\ttfamily 2306.16216}}].

\bibitem{Kelley:2017lek}
L.Z.~Kelley, L.~Blecha, L.~Hernquist, A.~Sesana and S.R.~Taylor, \emph{{The Gravitational Wave Background from Massive Black Hole Binaries in Illustris: spectral features and time to detection with pulsar timing arrays}}, \href{https://doi.org/10.1093/mnras/stx1638}{\emph{Mon. Not. Roy. Astron. Soc.} {\bfseries 471} (2017) 4508} [\href{https://arxiv.org/abs/1702.02180}{{\ttfamily 1702.02180}}].

\bibitem{NANOGrav:2023hfp}
{\scshape NANOGrav} collaboration, \emph{{The NANOGrav 15 yr Data Set: Constraints on Supermassive Black Hole Binaries from the Gravitational-wave Background}}, \href{https://doi.org/10.3847/2041-8213/ace18b}{\emph{Astrophys. J. Lett.} {\bfseries 952} (2023) L37} [\href{https://arxiv.org/abs/2306.16220}{{\ttfamily 2306.16220}}].

\bibitem{Antoniadis:2023xlr}
{\scshape EPTA} collaboration, \emph{{The second data release from the European Pulsar Timing Array: V. Implications for massive black holes, dark matter and the early Universe}},  \href{https://arxiv.org/abs/2306.16227}{{\ttfamily 2306.16227}}.

\bibitem{NANOGrav:2023hvm}
{\scshape NANOGrav} collaboration, \emph{{The NANOGrav 15 yr Data Set: Search for Signals from New Physics}}, \href{https://doi.org/10.3847/2041-8213/acdc91}{\emph{Astrophys. J. Lett.} {\bfseries 951} (2023) L11} [\href{https://arxiv.org/abs/2306.16219}{{\ttfamily 2306.16219}}].

\bibitem{Figueroa:2023zhu}
D.G.~Figueroa, M.~Pieroni, A.~Ricciardone and P.~Simakachorn, \emph{{Cosmological Background Interpretation of Pulsar Timing Array Data}},  \href{https://arxiv.org/abs/2307.02399}{{\ttfamily 2307.02399}}.

\bibitem{Sesana:2016ljz}
A.~Sesana, \emph{{Prospects for Multiband Gravitational-Wave Astronomy after GW150914}}, \href{https://doi.org/10.1103/PhysRevLett.116.231102}{\emph{Phys. Rev. Lett.} {\bfseries 116} (2016) 231102} [\href{https://arxiv.org/abs/1602.06951}{{\ttfamily 1602.06951}}].

\bibitem{Babak:2023lro}
S.~Babak, C.~Caprini, D.G.~Figueroa, N.~Karnesis, P.~Marcoccia, G.~Nardini et~al., \emph{{Stochastic gravitational wave background from stellar origin binary black holes in LISA}}, \href{https://doi.org/10.1088/1475-7516/2023/08/034}{\emph{JCAP} {\bfseries 08} (2023) 034} [\href{https://arxiv.org/abs/2304.06368}{{\ttfamily 2304.06368}}].

\bibitem{Lehoucq:2023zlt}
L.~Lehoucq, I.~Dvorkin, R.~Srinivasan, C.~Pellouin and A.~Lamberts, \emph{{Astrophysical Uncertainties in the Gravitational-Wave Background from Stellar-Mass Compact Binary Mergers}},  \href{https://arxiv.org/abs/2306.09861}{{\ttfamily 2306.09861}}.

\bibitem{PhysRevD.102.103023}
M.~Bonetti and A.~Sesana, \emph{Gravitational wave background from extreme mass ratio inspirals}, \href{https://doi.org/10.1103/PhysRevD.102.103023}{\emph{Phys. Rev. D} {\bfseries 102} (2020) 103023}.

\bibitem{Pozzoli:2023kxy}
F.~Pozzoli, S.~Babak, A.~Sesana, M.~Bonetti and N.~Karnesis, \emph{{Computation of stochastic background from extreme-mass-ratio inspiral populations for LISA}}, \href{https://doi.org/10.1103/PhysRevD.108.103039}{\emph{Phys. Rev. D} {\bfseries 108} (2023) 103039} [\href{https://arxiv.org/abs/2302.07043}{{\ttfamily 2302.07043}}].

\bibitem{Sesana:2008mz}
A.~Sesana, A.~Vecchio and C.N.~Colacino, \emph{{The stochastic gravitational-wave background from massive black hole binary systems: implications for observations with Pulsar Timing Arrays}}, \href{https://doi.org/10.1111/j.1365-2966.2008.13682.x}{\emph{Mon. Not. Roy. Astron. Soc.} {\bfseries 390} (2008) 192} [\href{https://arxiv.org/abs/0804.4476}{{\ttfamily 0804.4476}}].

\bibitem{Sesana:2012ak}
A.~Sesana, \emph{{Systematic investigation of the expected gravitational wave signal from supermassive black hole binaries in the pulsar timing band}}, \href{https://doi.org/10.1093/mnrasl/slt034}{\emph{Mon. Not. Roy. Astron. Soc.} {\bfseries 433} (2013) 1} [\href{https://arxiv.org/abs/1211.5375}{{\ttfamily 1211.5375}}].

\bibitem{Grishchuk:1974ny}
L.P.~Grishchuk, \emph{{Amplification of gravitational waves in an istropic universe}}, {\emph{Zh. Eksp. Teor. Fiz.} {\bfseries 67} (1974) 825}.

\bibitem{Starobinsky:1979ty}
A.A.~Starobinsky, \emph{{Spectrum of relict gravitational radiation and the early state of the universe}}, {\emph{JETP Lett.} {\bfseries 30} (1979) 682}.

\bibitem{Rubakov:1982df}
V.A.~Rubakov, M.V.~Sazhin and A.V.~Veryaskin, \emph{{Graviton Creation in the Inflationary Universe and the Grand Unification Scale}}, \href{https://doi.org/10.1016/0370-2693(82)90641-4}{\emph{Phys. Lett. B} {\bfseries 115} (1982) 189}.

\bibitem{Fabbri:1983us}
R.~Fabbri and M.d.~Pollock, \emph{{The Effect of Primordially Produced Gravitons upon the Anisotropy of the Cosmological Microwave Background Radiation}}, \href{https://doi.org/10.1016/0370-2693(83)91322-9}{\emph{Phys. Lett. B} {\bfseries 125} (1983) 445}.

\bibitem{Anber:2006xt}
M.M.~Anber and L.~Sorbo, \emph{{N-flationary magnetic fields}}, \href{https://doi.org/10.1088/1475-7516/2006/10/018}{\emph{JCAP} {\bfseries 10} (2006) 018} [\href{https://arxiv.org/abs/astro-ph/0606534}{{\ttfamily astro-ph/0606534}}].

\bibitem{Sorbo:2011rz}
L.~Sorbo, \emph{{Parity violation in the Cosmic Microwave Background from a pseudoscalar inflaton}}, \href{https://doi.org/10.1088/1475-7516/2011/06/003}{\emph{JCAP} {\bfseries 06} (2011) 003} [\href{https://arxiv.org/abs/1101.1525}{{\ttfamily 1101.1525}}].

\bibitem{Pajer:2013fsa}
E.~Pajer and M.~Peloso, \emph{{A review of Axion Inflation in the era of Planck}}, \href{https://doi.org/10.1088/0264-9381/30/21/214002}{\emph{Class. Quant. Grav.} {\bfseries 30} (2013) 214002} [\href{https://arxiv.org/abs/1305.3557}{{\ttfamily 1305.3557}}].

\bibitem{Adshead:2013qp}
P.~Adshead, E.~Martinec and M.~Wyman, \emph{{Gauge fields and inflation: Chiral gravitational waves, fluctuations, and the Lyth bound}}, \href{https://doi.org/10.1103/PhysRevD.88.021302}{\emph{Phys. Rev. D} {\bfseries 88} (2013) 021302} [\href{https://arxiv.org/abs/1301.2598}{{\ttfamily 1301.2598}}].

\bibitem{Adshead:2013nka}
P.~Adshead, E.~Martinec and M.~Wyman, \emph{{Perturbations in Chromo-Natural Inflation}}, \href{https://doi.org/10.1007/JHEP09(2013)087}{\emph{JHEP} {\bfseries 09} (2013) 087} [\href{https://arxiv.org/abs/1305.2930}{{\ttfamily 1305.2930}}].

\bibitem{Maleknejad:2016qjz}
A.~Maleknejad, \emph{{Axion Inflation with an SU(2) Gauge Field: Detectable Chiral Gravity Waves}}, \href{https://doi.org/10.1007/JHEP07(2016)104}{\emph{JHEP} {\bfseries 07} (2016) 104} [\href{https://arxiv.org/abs/1604.03327}{{\ttfamily 1604.03327}}].

\bibitem{Dimastrogiovanni:2016fuu}
E.~Dimastrogiovanni, M.~Fasiello and T.~Fujita, \emph{{Primordial Gravitational Waves from Axion-Gauge Fields Dynamics}}, \href{https://doi.org/10.1088/1475-7516/2017/01/019}{\emph{JCAP} {\bfseries 01} (2017) 019} [\href{https://arxiv.org/abs/1608.04216}{{\ttfamily 1608.04216}}].

\bibitem{Namba:2015gja}
R.~Namba, M.~Peloso, M.~Shiraishi, L.~Sorbo and C.~Unal, \emph{{Scale-dependent gravitational waves from a rolling axion}}, \href{https://doi.org/10.1088/1475-7516/2016/01/041}{\emph{JCAP} {\bfseries 01} (2016) 041} [\href{https://arxiv.org/abs/1509.07521}{{\ttfamily 1509.07521}}].

\bibitem{Ferreira:2015omg}
R.Z.~Ferreira, J.~Ganc, J.~Nore\~na and M.S.~Sloth, \emph{{On the validity of the perturbative description of axions during inflation}}, \href{https://doi.org/10.1088/1475-7516/2016/04/039}{\emph{JCAP} {\bfseries 04} (2016) 039} [\href{https://arxiv.org/abs/1512.06116}{{\ttfamily 1512.06116}}].

\bibitem{Peloso:2016gqs}
M.~Peloso, L.~Sorbo and C.~Unal, \emph{{Rolling axions during inflation: perturbativity and signatures}}, \href{https://doi.org/10.1088/1475-7516/2016/09/001}{\emph{JCAP} {\bfseries 09} (2016) 001} [\href{https://arxiv.org/abs/1606.00459}{{\ttfamily 1606.00459}}].

\bibitem{Domcke:2016bkh}
V.~Domcke, M.~Pieroni and P.~Bin\'etruy, \emph{{Primordial gravitational waves for universality classes of pseudoscalar inflation}}, \href{https://doi.org/10.1088/1475-7516/2016/06/031}{\emph{JCAP} {\bfseries 06} (2016) 031} [\href{https://arxiv.org/abs/1603.01287}{{\ttfamily 1603.01287}}].

\bibitem{Caldwell:2017chz}
R.R.~Caldwell and C.~Devulder, \emph{{Axion Gauge Field Inflation and Gravitational Leptogenesis: A Lower Bound on B Modes from the Matter-Antimatter Asymmetry of the Universe}}, \href{https://doi.org/10.1103/PhysRevD.97.023532}{\emph{Phys. Rev. D} {\bfseries 97} (2018) 023532} [\href{https://arxiv.org/abs/1706.03765}{{\ttfamily 1706.03765}}].

\bibitem{Guzzetti:2016mkm}
M.C.~Guzzetti, N.~Bartolo, M.~Liguori and S.~Matarrese, \emph{{Gravitational waves from inflation}}, \href{https://doi.org/10.1393/ncr/i2016-10127-1}{\emph{Riv. Nuovo Cim.} {\bfseries 39} (2016) 399} [\href{https://arxiv.org/abs/1605.01615}{{\ttfamily 1605.01615}}].

\bibitem{Bartolo:2016ami}
N.~Bartolo et~al., \emph{{Science with the space-based interferometer LISA. IV: Probing inflation with gravitational waves}}, \href{https://doi.org/10.1088/1475-7516/2016/12/026}{\emph{JCAP} {\bfseries 12} (2016) 026} [\href{https://arxiv.org/abs/1610.06481}{{\ttfamily 1610.06481}}].

\bibitem{DAmico:2021zdd}
G.~D'Amico, N.~Kaloper and A.~Westphal, \emph{{General double monodromy inflation}}, \href{https://doi.org/10.1103/PhysRevD.105.103527}{\emph{Phys. Rev. D} {\bfseries 105} (2022) 103527} [\href{https://arxiv.org/abs/2112.13861}{{\ttfamily 2112.13861}}].

\bibitem{DAmico:2021vka}
G.~D'Amico, N.~Kaloper and A.~Westphal, \emph{{Double Monodromy Inflation: A Gravity Waves Factory for CMB-S4, LiteBIRD and LISA}}, \href{https://doi.org/10.1103/PhysRevD.104.L081302}{\emph{Phys. Rev. D} {\bfseries 104} (2021) L081302} [\href{https://arxiv.org/abs/2101.05861}{{\ttfamily 2101.05861}}].

\bibitem{Fumagalli:2020nvq}
J.~Fumagalli, S.~Renaux-Petel and L.T.~Witkowski, \emph{{Oscillations in the stochastic gravitational wave background from sharp features and particle production during inflation}}, \href{https://doi.org/10.1088/1475-7516/2021/08/030}{\emph{JCAP} {\bfseries 08} (2021) 030} [\href{https://arxiv.org/abs/2012.02761}{{\ttfamily 2012.02761}}].

\bibitem{Fumagalli:2021mpc}
J.~Fumagalli, G.A.~Palma, S.~Renaux-Petel, S.~Sypsas, L.T.~Witkowski and C.~Zenteno, \emph{{Primordial gravitational waves from excited states}}, \href{https://doi.org/10.1007/JHEP03(2022)196}{\emph{JHEP} {\bfseries 03} (2022) 196} [\href{https://arxiv.org/abs/2111.14664}{{\ttfamily 2111.14664}}].

\bibitem{Easther:2006gt}
R.~Easther and E.A.~Lim, \emph{{Stochastic gravitational wave production after inflation}}, \href{https://doi.org/10.1088/1475-7516/2006/04/010}{\emph{JCAP} {\bfseries 04} (2006) 010} [\href{https://arxiv.org/abs/astro-ph/0601617}{{\ttfamily astro-ph/0601617}}].

\bibitem{GarciaBellido:2007dg}
J.~Garcia-Bellido and D.G.~Figueroa, \emph{{A stochastic background of gravitational waves from hybrid preheating}}, \href{https://doi.org/10.1103/PhysRevLett.98.061302}{\emph{Phys. Rev. Lett.} {\bfseries 98} (2007) 061302} [\href{https://arxiv.org/abs/astro-ph/0701014}{{\ttfamily astro-ph/0701014}}].

\bibitem{GarciaBellido:2007af}
J.~Garcia-Bellido, D.G.~Figueroa and A.~Sastre, \emph{{A Gravitational Wave Background from Reheating after Hybrid Inflation}}, \href{https://doi.org/10.1103/PhysRevD.77.043517}{\emph{Phys. Rev. D} {\bfseries 77} (2008) 043517} [\href{https://arxiv.org/abs/0707.0839}{{\ttfamily 0707.0839}}].

\bibitem{Dufaux:2007pt}
J.F.~Dufaux, A.~Bergman, G.N.~Felder, L.~Kofman and J.-P.~Uzan, \emph{{Theory and Numerics of Gravitational Waves from Preheating after Inflation}}, \href{https://doi.org/10.1103/PhysRevD.76.123517}{\emph{Phys. Rev. D} {\bfseries 76} (2007) 123517} [\href{https://arxiv.org/abs/0707.0875}{{\ttfamily 0707.0875}}].

\bibitem{Dufaux:2008dn}
J.-F.~Dufaux, G.~Felder, L.~Kofman and O.~Navros, \emph{{Gravity Waves from Tachyonic Preheating after Hybrid Inflation}}, \href{https://doi.org/10.1088/1475-7516/2009/03/001}{\emph{JCAP} {\bfseries 03} (2009) 001} [\href{https://arxiv.org/abs/0812.2917}{{\ttfamily 0812.2917}}].

\bibitem{Dufaux:2010cf}
J.-F.~Dufaux, D.G.~Figueroa and J.~Garcia-Bellido, \emph{{Gravitational Waves from Abelian Gauge Fields and Cosmic Strings at Preheating}}, \href{https://doi.org/10.1103/PhysRevD.82.083518}{\emph{Phys. Rev. D} {\bfseries 82} (2010) 083518} [\href{https://arxiv.org/abs/1006.0217}{{\ttfamily 1006.0217}}].

\bibitem{Bethke:2013aba}
L.~Bethke, D.G.~Figueroa and A.~Rajantie, \emph{{Anisotropies in the Gravitational Wave Background from Preheating}}, \href{https://doi.org/10.1103/PhysRevLett.111.011301}{\emph{Phys. Rev. Lett.} {\bfseries 111} (2013) 011301} [\href{https://arxiv.org/abs/1304.2657}{{\ttfamily 1304.2657}}].

\bibitem{Bethke:2013vca}
L.~Bethke, D.G.~Figueroa and A.~Rajantie, \emph{{On the Anisotropy of the Gravitational Wave Background from Massless Preheating}}, \href{https://doi.org/10.1088/1475-7516/2014/06/047}{\emph{JCAP} {\bfseries 06} (2014) 047} [\href{https://arxiv.org/abs/1309.1148}{{\ttfamily 1309.1148}}].

\bibitem{Figueroa:2017vfa}
D.G.~Figueroa and F.~Torrenti, \emph{{Gravitational wave production from preheating: parameter dependence}}, \href{https://doi.org/10.1088/1475-7516/2017/10/057}{\emph{JCAP} {\bfseries 10} (2017) 057} [\href{https://arxiv.org/abs/1707.04533}{{\ttfamily 1707.04533}}].

\bibitem{Adshead:2018doq}
P.~Adshead, J.T.~Giblin and Z.J.~Weiner, \emph{{Gravitational waves from gauge preheating}}, \href{https://doi.org/10.1103/PhysRevD.98.043525}{\emph{Phys. Rev. D} {\bfseries 98} (2018) 043525} [\href{https://arxiv.org/abs/1805.04550}{{\ttfamily 1805.04550}}].

\bibitem{Adshead:2019lbr}
P.~Adshead, J.T.~Giblin, M.~Pieroni and Z.J.~Weiner, \emph{{Constraining axion inflation with gravitational waves from preheating}}, \href{https://doi.org/10.1103/PhysRevD.101.083534}{\emph{Phys. Rev. D} {\bfseries 101} (2020) 083534} [\href{https://arxiv.org/abs/1909.12842}{{\ttfamily 1909.12842}}].

\bibitem{Adshead:2019igv}
P.~Adshead, J.T.~Giblin, M.~Pieroni and Z.J.~Weiner, \emph{{Constraining Axion Inflation with Gravitational Waves across 29 Decades in Frequency}}, \href{https://doi.org/10.1103/PhysRevLett.124.171301}{\emph{Phys. Rev. Lett.} {\bfseries 124} (2020) 171301} [\href{https://arxiv.org/abs/1909.12843}{{\ttfamily 1909.12843}}].

\bibitem{Giovannini:1998bp}
M.~Giovannini, \emph{{Gravitational waves constraints on postinflationary phases stiffer than radiation}}, \href{https://doi.org/10.1103/PhysRevD.58.083504}{\emph{Phys. Rev. D} {\bfseries 58} (1998) 083504} [\href{https://arxiv.org/abs/hep-ph/9806329}{{\ttfamily hep-ph/9806329}}].

\bibitem{Giovannini:1999bh}
M.~Giovannini, \emph{{Production and detection of relic gravitons in quintessential inflationary models}}, \href{https://doi.org/10.1103/PhysRevD.60.123511}{\emph{Phys. Rev. D} {\bfseries 60} (1999) 123511} [\href{https://arxiv.org/abs/astro-ph/9903004}{{\ttfamily astro-ph/9903004}}].

\bibitem{Boyle:2007zx}
L.A.~Boyle and A.~Buonanno, \emph{{Relating gravitational wave constraints from primordial nucleosynthesis, pulsar timing, laser interferometers, and the CMB: Implications for the early Universe}}, \href{https://doi.org/10.1103/PhysRevD.78.043531}{\emph{Phys. Rev. D} {\bfseries 78} (2008) 043531} [\href{https://arxiv.org/abs/0708.2279}{{\ttfamily 0708.2279}}].

\bibitem{Li:2016mmc}
B.~Li, P.R.~Shapiro and T.~Rindler-Daller, \emph{{Bose-Einstein-condensed scalar field dark matter and the gravitational wave background from inflation: new cosmological constraints and its detectability by LIGO}}, \href{https://doi.org/10.1103/PhysRevD.96.063505}{\emph{Phys. Rev. D} {\bfseries 96} (2017) 063505} [\href{https://arxiv.org/abs/1611.07961}{{\ttfamily 1611.07961}}].

\bibitem{Li:2021htg}
B.~Li and P.R.~Shapiro, \emph{{Precision cosmology and the stiff-amplified gravitational-wave background from inflation: NANOGrav, Advanced LIGO-Virgo and the Hubble tension}}, \href{https://doi.org/10.1088/1475-7516/2021/10/024}{\emph{JCAP} {\bfseries 10} (2021) 024} [\href{https://arxiv.org/abs/2107.12229}{{\ttfamily 2107.12229}}].

\bibitem{Figueroa:2018twl}
D.G.~Figueroa and E.H.~Tanin, \emph{{Inconsistency of an inflationary sector coupled only to Einstein gravity}}, \href{https://doi.org/10.1088/1475-7516/2019/10/050}{\emph{JCAP} {\bfseries 10} (2019) 050} [\href{https://arxiv.org/abs/1811.04093}{{\ttfamily 1811.04093}}].

\bibitem{Figueroa:2019paj}
D.G.~Figueroa and E.H.~Tanin, \emph{{Ability of LIGO and LISA to probe the equation of state of the early Universe}}, \href{https://doi.org/10.1088/1475-7516/2019/08/011}{\emph{JCAP} {\bfseries 08} (2019) 011} [\href{https://arxiv.org/abs/1905.11960}{{\ttfamily 1905.11960}}].

\bibitem{Gouttenoire:2021wzu}
Y.~Gouttenoire, G.~Servant and P.~Simakachorn, \emph{{Revealing the Primordial Irreducible Inflationary Gravitational-Wave Background with a Spinning Peccei-Quinn Axion}},  \href{https://arxiv.org/abs/2108.10328}{{\ttfamily 2108.10328}}.

\bibitem{Co:2021lkc}
R.T.~Co, D.~Dunsky, N.~Fernandez, A.~Ghalsasi, L.J.~Hall, K.~Harigaya et~al., \emph{{Gravitational wave and CMB probes of axion kination}}, \href{https://doi.org/10.1007/JHEP09(2022)116}{\emph{JHEP} {\bfseries 09} (2022) 116} [\href{https://arxiv.org/abs/2108.09299}{{\ttfamily 2108.09299}}].

\bibitem{Gouttenoire:2021jhk}
Y.~Gouttenoire, G.~Servant and P.~Simakachorn, \emph{{Kination cosmology from scalar fields and gravitational-wave signatures}},  \href{https://arxiv.org/abs/2111.01150}{{\ttfamily 2111.01150}}.

\bibitem{Oikonomou:2023qfz}
V.K.~Oikonomou, \emph{{Flat energy spectrum of primordial gravitational waves versus peaks and the NANOGrav 2023 observation}}, \href{https://doi.org/10.1103/PhysRevD.108.043516}{\emph{Phys. Rev. D} {\bfseries 108} (2023) 043516} [\href{https://arxiv.org/abs/2306.17351}{{\ttfamily 2306.17351}}].

\bibitem{Ghiglieri:2015nfa}
J.~Ghiglieri and M.~Laine, \emph{{Gravitational wave background from Standard Model physics: Qualitative features}}, \href{https://doi.org/10.1088/1475-7516/2015/07/022}{\emph{JCAP} {\bfseries 07} (2015) 022} [\href{https://arxiv.org/abs/1504.02569}{{\ttfamily 1504.02569}}].

\bibitem{Ghiglieri:2020mhm}
J.~Ghiglieri, G.~Jackson, M.~Laine and Y.~Zhu, \emph{{Gravitational wave background from Standard Model physics: Complete leading order}}, \href{https://doi.org/10.1007/JHEP07(2020)092}{\emph{JHEP} {\bfseries 07} (2020) 092} [\href{https://arxiv.org/abs/2004.11392}{{\ttfamily 2004.11392}}].

\bibitem{Ringwald:2020ist}
A.~Ringwald, J.~Sch\"utte-Engel and C.~Tamarit, \emph{{Gravitational Waves as a Big Bang Thermometer}}, \href{https://doi.org/10.1088/1475-7516/2021/03/054}{\emph{JCAP} {\bfseries 03} (2021) 054} [\href{https://arxiv.org/abs/2011.04731}{{\ttfamily 2011.04731}}].

\bibitem{Ghiglieri:2022rfp}
J.~Ghiglieri, J.~Sch\"utte-Engel and E.~Speranza, \emph{{Freezing-In Gravitational Waves}},  \href{https://arxiv.org/abs/2211.16513}{{\ttfamily 2211.16513}}.

\bibitem{Zhou:2013tsa}
S.-Y.~Zhou, E.J.~Copeland, R.~Easther, H.~Finkel, Z.-G.~Mou and P.M.~Saffin, \emph{{Gravitational Waves from Oscillon Preheating}}, \href{https://doi.org/10.1007/JHEP10(2013)026}{\emph{JHEP} {\bfseries 10} (2013) 026} [\href{https://arxiv.org/abs/1304.6094}{{\ttfamily 1304.6094}}].

\bibitem{Antusch:2016con}
S.~Antusch, F.~Cefala and S.~Orani, \emph{{Gravitational waves from oscillons after inflation}}, \href{https://doi.org/10.1103/PhysRevLett.118.011303}{\emph{Phys. Rev. Lett.} {\bfseries 118} (2017) 011303} [\href{https://arxiv.org/abs/1607.01314}{{\ttfamily 1607.01314}}].

\bibitem{Antusch:2017vga}
S.~Antusch, F.~Cefala and S.~Orani, \emph{{What can we learn from the stochastic gravitational wave background produced by oscillons?}}, \href{https://doi.org/10.1088/1475-7516/2018/03/032}{\emph{JCAP} {\bfseries 03} (2018) 032} [\href{https://arxiv.org/abs/1712.03231}{{\ttfamily 1712.03231}}].

\bibitem{Liu:2017hua}
J.~Liu, Z.-K.~Guo, R.-G.~Cai and G.~Shiu, \emph{{Gravitational Waves from Oscillons with Cuspy Potentials}}, \href{https://doi.org/10.1103/PhysRevLett.120.031301}{\emph{Phys. Rev. Lett.} {\bfseries 120} (2018) 031301} [\href{https://arxiv.org/abs/1707.09841}{{\ttfamily 1707.09841}}].

\bibitem{Amin:2018xfe}
M.A.~Amin, J.~Braden, E.J.~Copeland, J.T.~Giblin, C.~Solorio, Z.J.~Weiner et~al., \emph{{Gravitational waves from asymmetric oscillon dynamics?}}, \href{https://doi.org/10.1103/PhysRevD.98.024040}{\emph{Phys. Rev. D} {\bfseries 98} (2018) 024040} [\href{https://arxiv.org/abs/1803.08047}{{\ttfamily 1803.08047}}].

\bibitem{Kamionkowski:1993fg}
M.~Kamionkowski, A.~Kosowsky and M.S.~Turner, \emph{{Gravitational radiation from first order phase transitions}}, \href{https://doi.org/10.1103/PhysRevD.49.2837}{\emph{Phys. Rev. D} {\bfseries 49} (1994) 2837} [\href{https://arxiv.org/abs/astro-ph/9310044}{{\ttfamily astro-ph/9310044}}].

\bibitem{Caprini:2007xq}
C.~Caprini, R.~Durrer and G.~Servant, \emph{{Gravitational wave generation from bubble collisions in first-order phase transitions: An analytic approach}}, \href{https://doi.org/10.1103/PhysRevD.77.124015}{\emph{Phys. Rev. D} {\bfseries 77} (2008) 124015} [\href{https://arxiv.org/abs/0711.2593}{{\ttfamily 0711.2593}}].

\bibitem{Huber:2008hg}
S.J.~Huber and T.~Konstandin, \emph{{Gravitational Wave Production by Collisions: More Bubbles}}, \href{https://doi.org/10.1088/1475-7516/2008/09/022}{\emph{JCAP} {\bfseries 09} (2008) 022} [\href{https://arxiv.org/abs/0806.1828}{{\ttfamily 0806.1828}}].

\bibitem{Hindmarsh:2013xza}
M.~Hindmarsh, S.J.~Huber, K.~Rummukainen and D.J.~Weir, \emph{{Gravitational waves from the sound of a first order phase transition}}, \href{https://doi.org/10.1103/PhysRevLett.112.041301}{\emph{Phys. Rev. Lett.} {\bfseries 112} (2014) 041301} [\href{https://arxiv.org/abs/1304.2433}{{\ttfamily 1304.2433}}].

\bibitem{Hindmarsh:2015qta}
M.~Hindmarsh, S.J.~Huber, K.~Rummukainen and D.J.~Weir, \emph{{Numerical simulations of acoustically generated gravitational waves at a first order phase transition}}, \href{https://doi.org/10.1103/PhysRevD.92.123009}{\emph{Phys. Rev. D} {\bfseries 92} (2015) 123009} [\href{https://arxiv.org/abs/1504.03291}{{\ttfamily 1504.03291}}].

\bibitem{Caprini:2015zlo}
C.~Caprini et~al., \emph{{Science with the space-based interferometer eLISA. II: Gravitational waves from cosmological phase transitions}}, \href{https://doi.org/10.1088/1475-7516/2016/04/001}{\emph{JCAP} {\bfseries 04} (2016) 001} [\href{https://arxiv.org/abs/1512.06239}{{\ttfamily 1512.06239}}].

\bibitem{Hindmarsh:2017gnf}
M.~Hindmarsh, S.J.~Huber, K.~Rummukainen and D.J.~Weir, \emph{{Shape of the acoustic gravitational wave power spectrum from a first order phase transition}}, \href{https://doi.org/10.1103/PhysRevD.96.103520}{\emph{Phys. Rev. D} {\bfseries 96} (2017) 103520} [\href{https://arxiv.org/abs/1704.05871}{{\ttfamily 1704.05871}}].

\bibitem{Cutting:2018tjt}
D.~Cutting, M.~Hindmarsh and D.J.~Weir, \emph{{Gravitational waves from vacuum first-order phase transitions: from the envelope to the lattice}}, \href{https://doi.org/10.1103/PhysRevD.97.123513}{\emph{Phys. Rev. D} {\bfseries 97} (2018) 123513} [\href{https://arxiv.org/abs/1802.05712}{{\ttfamily 1802.05712}}].

\bibitem{Cutting:2019zws}
D.~Cutting, M.~Hindmarsh and D.J.~Weir, \emph{{Vorticity, kinetic energy, and suppressed gravitational wave production in strong first order phase transitions}}, \href{https://doi.org/10.1103/PhysRevLett.125.021302}{\emph{Phys. Rev. Lett.} {\bfseries 125} (2020) 021302} [\href{https://arxiv.org/abs/1906.00480}{{\ttfamily 1906.00480}}].

\bibitem{Pol:2019yex}
A.~Roper~Pol, S.~Mandal, A.~Brandenburg, T.~Kahniashvili and A.~Kosowsky, \emph{{Numerical simulations of gravitational waves from early-universe turbulence}}, \href{https://doi.org/10.1103/PhysRevD.102.083512}{\emph{Phys. Rev. D} {\bfseries 102} (2020) 083512} [\href{https://arxiv.org/abs/1903.08585}{{\ttfamily 1903.08585}}].

\bibitem{Caprini:2019egz}
C.~Caprini et~al., \emph{{Detecting gravitational waves from cosmological phase transitions with LISA: an update}}, \href{https://doi.org/10.1088/1475-7516/2020/03/024}{\emph{JCAP} {\bfseries 03} (2020) 024} [\href{https://arxiv.org/abs/1910.13125}{{\ttfamily 1910.13125}}].

\bibitem{Cutting:2020nla}
D.~Cutting, E.G.~Escartin, M.~Hindmarsh and D.J.~Weir, \emph{{Gravitational waves from vacuum first order phase transitions II: from thin to thick walls}}, \href{https://doi.org/10.1103/PhysRevD.103.023531}{\emph{Phys. Rev. D} {\bfseries 103} (2021) 023531} [\href{https://arxiv.org/abs/2005.13537}{{\ttfamily 2005.13537}}].

\bibitem{Han:2023olf}
C.~Han, K.-P.~Xie, J.M.~Yang and M.~Zhang, \emph{{Self-interacting dark matter implied by nano-Hertz gravitational waves}},  \href{https://arxiv.org/abs/2306.16966}{{\ttfamily 2306.16966}}.

\bibitem{Ashoorioon:2022raz}
A.~Ashoorioon, K.~Rezazadeh and A.~Rostami, \emph{{NANOGrav signal from the end of inflation and the LIGO mass and heavier primordial black holes}}, \href{https://doi.org/10.1016/j.physletb.2022.137542}{\emph{Phys. Lett. B} {\bfseries 835} (2022) 137542} [\href{https://arxiv.org/abs/2202.01131}{{\ttfamily 2202.01131}}].

\bibitem{Athron:2023mer}
P.~Athron, A.~Fowlie, C.-T.~Lu, L.~Morris, L.~Wu, Y.~Wu et~al., \emph{{Can supercooled phase transitions explain the gravitational wave background observed by pulsar timing arrays?}},  \href{https://arxiv.org/abs/2306.17239}{{\ttfamily 2306.17239}}.

\bibitem{Li:2023yaj}
Y.-Y.~Li, C.~Zhang, Z.~Wang, M.-Y.~Cui, Y.-L.S.~Tsai, Q.~Yuan et~al., \emph{{Primordial magnetic field as a common solution of nanohertz gravitational waves and Hubble tension}},  \href{https://arxiv.org/abs/2306.17124}{{\ttfamily 2306.17124}}.

\bibitem{Vachaspati:1984gt}
T.~Vachaspati and A.~Vilenkin, \emph{{Gravitational Radiation from Cosmic Strings}}, \href{https://doi.org/10.1103/PhysRevD.31.3052}{\emph{Phys. Rev. D} {\bfseries 31} (1985) 3052}.

\bibitem{Sakellariadou:1990ne}
M.~Sakellariadou, \emph{{Gravitational waves emitted from infinite strings}}, \href{https://doi.org/10.1103/PhysRevD.42.354}{\emph{Phys. Rev. D} {\bfseries 42} (1990) 354}.

\bibitem{Damour:2000wa}
T.~Damour and A.~Vilenkin, \emph{{Gravitational wave bursts from cosmic strings}}, \href{https://doi.org/10.1103/PhysRevLett.85.3761}{\emph{Phys. Rev. Lett.} {\bfseries 85} (2000) 3761} [\href{https://arxiv.org/abs/gr-qc/0004075}{{\ttfamily gr-qc/0004075}}].

\bibitem{Damour:2001bk}
T.~Damour and A.~Vilenkin, \emph{{Gravitational wave bursts from cusps and kinks on cosmic strings}}, \href{https://doi.org/10.1103/PhysRevD.64.064008}{\emph{Phys. Rev. D} {\bfseries 64} (2001) 064008} [\href{https://arxiv.org/abs/gr-qc/0104026}{{\ttfamily gr-qc/0104026}}].

\bibitem{Damour:2004kw}
T.~Damour and A.~Vilenkin, \emph{{Gravitational radiation from cosmic (super)strings: Bursts, stochastic background, and observational windows}}, \href{https://doi.org/10.1103/PhysRevD.71.063510}{\emph{Phys. Rev. D} {\bfseries 71} (2005) 063510} [\href{https://arxiv.org/abs/hep-th/0410222}{{\ttfamily hep-th/0410222}}].

\bibitem{Figueroa:2012kw}
D.G.~Figueroa, M.~Hindmarsh and J.~Urrestilla, \emph{{Exact Scale-Invariant Background of Gravitational Waves from Cosmic Defects}}, \href{https://doi.org/10.1103/PhysRevLett.110.101302}{\emph{Phys. Rev. Lett.} {\bfseries 110} (2013) 101302} [\href{https://arxiv.org/abs/1212.5458}{{\ttfamily 1212.5458}}].

\bibitem{Hiramatsu:2013qaa}
T.~Hiramatsu, M.~Kawasaki and K.~Saikawa, \emph{{On the estimation of gravitational wave spectrum from cosmic domain walls}}, \href{https://doi.org/10.1088/1475-7516/2014/02/031}{\emph{JCAP} {\bfseries 02} (2014) 031} [\href{https://arxiv.org/abs/1309.5001}{{\ttfamily 1309.5001}}].

\bibitem{Blanco-Pillado:2017oxo}
J.J.~Blanco-Pillado and K.D.~Olum, \emph{{Stochastic gravitational wave background from smoothed cosmic string loops}}, \href{https://doi.org/10.1103/PhysRevD.96.104046}{\emph{Phys. Rev. D} {\bfseries 96} (2017) 104046} [\href{https://arxiv.org/abs/1709.02693}{{\ttfamily 1709.02693}}].

\bibitem{Auclair:2019wcv}
P.~Auclair et~al., \emph{{Probing the gravitational wave background from cosmic strings with LISA}}, \href{https://doi.org/10.1088/1475-7516/2020/04/034}{\emph{JCAP} {\bfseries 04} (2020) 034} [\href{https://arxiv.org/abs/1909.00819}{{\ttfamily 1909.00819}}].

\bibitem{Gouttenoire:2019kij}
Y.~Gouttenoire, G.~Servant and P.~Simakachorn, \emph{{Beyond the Standard Models with Cosmic Strings}}, \href{https://doi.org/10.1088/1475-7516/2020/07/032}{\emph{JCAP} {\bfseries 07} (2020) 032} [\href{https://arxiv.org/abs/1912.02569}{{\ttfamily 1912.02569}}].

\bibitem{Figueroa:2020lvo}
D.G.~Figueroa, M.~Hindmarsh, J.~Lizarraga and J.~Urrestilla, \emph{{Irreducible background of gravitational waves from a cosmic defect network: update and comparison of numerical techniques}}, \href{https://doi.org/10.1103/PhysRevD.102.103516}{\emph{Phys. Rev. D} {\bfseries 102} (2020) 103516} [\href{https://arxiv.org/abs/2007.03337}{{\ttfamily 2007.03337}}].

\bibitem{Gorghetto:2021fsn}
M.~Gorghetto, E.~Hardy and H.~Nicolaescu, \emph{{Observing invisible axions with gravitational waves}}, \href{https://doi.org/10.1088/1475-7516/2021/06/034}{\emph{JCAP} {\bfseries 06} (2021) 034} [\href{https://arxiv.org/abs/2101.11007}{{\ttfamily 2101.11007}}].

\bibitem{Chang:2021afa}
C.-F.~Chang and Y.~Cui, \emph{{Gravitational waves from global cosmic strings and cosmic archaeology}}, \href{https://doi.org/10.1007/JHEP03(2022)114}{\emph{JHEP} {\bfseries 03} (2022) 114} [\href{https://arxiv.org/abs/2106.09746}{{\ttfamily 2106.09746}}].

\bibitem{Yamada:2022aax}
M.~Yamada and K.~Yonekura, \emph{{Cosmic F- and D-strings from pure Yang\textendash{}Mills theory}}, \href{https://doi.org/10.1016/j.physletb.2023.137724}{\emph{Phys. Lett. B} {\bfseries 838} (2023) 137724} [\href{https://arxiv.org/abs/2204.13125}{{\ttfamily 2204.13125}}].

\bibitem{Yamada:2022imq}
M.~Yamada and K.~Yonekura, \emph{{Cosmic strings from pure Yang\textendash{}Mills theory}}, \href{https://doi.org/10.1103/PhysRevD.106.123515}{\emph{Phys. Rev. D} {\bfseries 106} (2022) 123515} [\href{https://arxiv.org/abs/2204.13123}{{\ttfamily 2204.13123}}].

\bibitem{Kitajima:2023cek}
N.~Kitajima, J.~Lee, K.~Murai, F.~Takahashi and W.~Yin, \emph{{Nanohertz Gravitational Waves from Axion Domain Walls Coupled to QCD}},  \href{https://arxiv.org/abs/2306.17146}{{\ttfamily 2306.17146}}.

\bibitem{Matarrese:1992rp}
S.~Matarrese, O.~Pantano and D.~Saez, \emph{{A General relativistic approach to the nonlinear evolution of collisionless matter}}, \href{https://doi.org/10.1103/PhysRevD.47.1311}{\emph{Phys. Rev. D} {\bfseries 47} (1993) 1311}.

\bibitem{Matarrese:1993zf}
S.~Matarrese, O.~Pantano and D.~Saez, \emph{{General relativistic dynamics of irrotational dust: Cosmological implications}}, \href{https://doi.org/10.1103/PhysRevLett.72.320}{\emph{Phys. Rev. Lett.} {\bfseries 72} (1994) 320} [\href{https://arxiv.org/abs/astro-ph/9310036}{{\ttfamily astro-ph/9310036}}].

\bibitem{Matarrese:1997ay}
S.~Matarrese, S.~Mollerach and M.~Bruni, \emph{{Second order perturbations of the Einstein-de Sitter universe}}, \href{https://doi.org/10.1103/PhysRevD.58.043504}{\emph{Phys. Rev. D} {\bfseries 58} (1998) 043504} [\href{https://arxiv.org/abs/astro-ph/9707278}{{\ttfamily astro-ph/9707278}}].

\bibitem{Nakamura:2004rm}
K.~Nakamura, \emph{{Second-order gauge invariant cosmological perturbation theory: Einstein equations in terms of gauge invariant variables}}, \href{https://doi.org/10.1143/PTP.117.17}{\emph{Prog. Theor. Phys.} {\bfseries 117} (2007) 17} [\href{https://arxiv.org/abs/gr-qc/0605108}{{\ttfamily gr-qc/0605108}}].

\bibitem{Ananda:2006af}
K.N.~Ananda, C.~Clarkson and D.~Wands, \emph{{The Cosmological gravitational wave background from primordial density perturbations}}, \href{https://doi.org/10.1103/PhysRevD.75.123518}{\emph{Phys. Rev. D} {\bfseries 75} (2007) 123518} [\href{https://arxiv.org/abs/gr-qc/0612013}{{\ttfamily gr-qc/0612013}}].

\bibitem{Baumann:2007zm}
D.~Baumann, P.J.~Steinhardt, K.~Takahashi and K.~Ichiki, \emph{{Gravitational Wave Spectrum Induced by Primordial Scalar Perturbations}}, \href{https://doi.org/10.1103/PhysRevD.76.084019}{\emph{Phys. Rev. D} {\bfseries 76} (2007) 084019} [\href{https://arxiv.org/abs/hep-th/0703290}{{\ttfamily hep-th/0703290}}].

\bibitem{Domenech:2021ztg}
G.~Dom\`enech, \emph{{Scalar Induced Gravitational Waves Review}}, \href{https://doi.org/10.3390/universe7110398}{\emph{Universe} {\bfseries 7} (2021) 398} [\href{https://arxiv.org/abs/2109.01398}{{\ttfamily 2109.01398}}].

\bibitem{Dandoy:2023jot}
V.~Dandoy, V.~Domcke and F.~Rompineve, \emph{{Search for scalar induced gravitational waves in the International Pulsar Timing Array Data Release 2 and NANOgrav 12.5 years datasets}},  \href{https://arxiv.org/abs/2302.07901}{{\ttfamily 2302.07901}}.

\bibitem{Franciolini:2023pbf}
G.~Franciolini, A.~Iovino, Junior., V.~Vaskonen and H.~Veermae, \emph{{The recent gravitational wave observation by pulsar timing arrays and primordial black holes: the importance of non-gaussianities}},  \href{https://arxiv.org/abs/2306.17149}{{\ttfamily 2306.17149}}.

\bibitem{Vagnozzi:2023lwo}
S.~Vagnozzi, \emph{{Inflationary interpretation of the stochastic gravitational wave background signal detected by pulsar timing array experiments}}, \href{https://doi.org/10.1016/j.jheap.2023.07.001}{\emph{JHEAp} {\bfseries 39} (2023) 81} [\href{https://arxiv.org/abs/2306.16912}{{\ttfamily 2306.16912}}].

\bibitem{Guo:2023hyp}
S.-Y.~Guo, M.~Khlopov, X.~Liu, L.~Wu, Y.~Wu and B.~Zhu, \emph{{Footprints of Axion-Like Particle in Pulsar Timing Array Data and JWST Observations}},  \href{https://arxiv.org/abs/2306.17022}{{\ttfamily 2306.17022}}.

\bibitem{Ellis:2023tsl}
J.~Ellis, M.~Lewicki, C.~Lin and V.~Vaskonen, \emph{{Cosmic Superstrings Revisited in Light of NANOGrav 15-Year Data}},  \href{https://arxiv.org/abs/2306.17147}{{\ttfamily 2306.17147}}.

\bibitem{Cai:2023dls}
Y.-F.~Cai, X.-C.~He, X.~Ma, S.-F.~Yan and G.-W.~Yuan, \emph{{Limits on scalar-induced gravitational waves from the stochastic background by pulsar timing array observations}},  \href{https://arxiv.org/abs/2306.17822}{{\ttfamily 2306.17822}}.

\bibitem{Madge:2023cak}
E.~Madge, E.~Morgante, C.~Puchades-Ib\'a\~nez, N.~Ramberg, W.~Ratzinger, S.~Schenk et~al., \emph{{Primordial gravitational waves in the nano-Hertz regime and PTA data -- towards solving the GW inverse problem}},  \href{https://arxiv.org/abs/2306.14856}{{\ttfamily 2306.14856}}.

\bibitem{Bai:2023cqj}
Y.~Bai, T.-K.~Chen and M.~Korwar, \emph{{QCD-Collapsed Domain Walls: QCD Phase Transition and Gravitational Wave Spectroscopy}},  \href{https://arxiv.org/abs/2306.17160}{{\ttfamily 2306.17160}}.

\bibitem{Liu:2023ymk}
L.~Liu, Z.-C.~Chen and Q.-G.~Huang, \emph{{Implications for the non-Gaussianity of curvature perturbation from pulsar timing arrays}},  \href{https://arxiv.org/abs/2307.01102}{{\ttfamily 2307.01102}}.

\bibitem{Unal:2023srk}
C.~Unal, A.~Papageorgiou and I.~Obata, \emph{{Axion-Gauge Dynamics During Inflation as the Origin of Pulsar Timing Array Signals and Primordial Black Holes}},  \href{https://arxiv.org/abs/2307.02322}{{\ttfamily 2307.02322}}.

\bibitem{Servant:2023mwt}
G.~Servant and P.~Simakachorn, \emph{{Constraining Post-Inflationary Axions with Pulsar Timing Arrays}},  \href{https://arxiv.org/abs/2307.03121}{{\ttfamily 2307.03121}}.

\bibitem{Ellis:2023oxs}
J.~Ellis, M.~Fairbairn, G.~Franciolini, G.~H\"utsi, A.~Iovino, M.~Lewicki et~al., \emph{{What is the source of the PTA GW signal?}},  \href{https://arxiv.org/abs/2308.08546}{{\ttfamily 2308.08546}}.

\bibitem{Wang:2023ost}
S.~Wang, Z.-C.~Zhao, J.-P.~Li and Q.-H.~Zhu, \emph{{Implications of Pulsar Timing Array Data for Scalar-Induced Gravitational Waves and Primordial Black Holes: Primordial Non-Gaussianity $f_{\mathrm{NL}}$ Considered}},  \href{https://arxiv.org/abs/2307.00572}{{\ttfamily 2307.00572}}.

\bibitem{NANOGrav:2020bcs}
{\scshape NANOGrav} collaboration, \emph{{The NANOGrav 12.5 yr Data Set: Search for an Isotropic Stochastic Gravitational-wave Background}}, \href{https://doi.org/10.3847/2041-8213/abd401}{\emph{Astrophys. J. Lett.} {\bfseries 905} (2020) L34} [\href{https://arxiv.org/abs/2009.04496}{{\ttfamily 2009.04496}}].

\bibitem{EPTA:2023fyk}
{\scshape EPTA} collaboration, \emph{{The second data release from the European Pulsar Timing Array III. Search for gravitational wave signals}},  \href{https://arxiv.org/abs/2306.16214}{{\ttfamily 2306.16214}}.

\bibitem{Janssen:2014dka}
G.~Janssen et~al., \emph{{Gravitational wave astronomy with the SKA}}, \href{https://doi.org/10.22323/1.215.0037}{\emph{PoS} {\bfseries AASKA14} (2015) 037} [\href{https://arxiv.org/abs/1501.00127}{{\ttfamily 1501.00127}}].

\bibitem{Abazajian:2016yjj}
{\scshape CMB-S4} collaboration, \emph{{CMB-S4 Science Book, First Edition}},  \href{https://arxiv.org/abs/1610.02743}{{\ttfamily 1610.02743}}.

\bibitem{Sesana:2019vho}
A.~Sesana et~al., \emph{{Unveiling the gravitational universe at $\mu$-Hz frequencies}}, \href{https://doi.org/10.1007/s10686-021-09709-9}{\emph{Exper. Astron.} {\bfseries 51} (2021) 1333} [\href{https://arxiv.org/abs/1908.11391}{{\ttfamily 1908.11391}}].

\bibitem{LISA:2017pwj}
{\scshape LISA} collaboration, \emph{{Laser Interferometer Space Antenna}},  \href{https://arxiv.org/abs/1702.00786}{{\ttfamily 1702.00786}}.

\bibitem{Ruan:2020smc}
W.-H.~Ruan, C.~Liu, Z.-K.~Guo, Y.-L.~Wu and R.-G.~Cai, \emph{{The LISA-Taiji network}}, \href{https://doi.org/10.1038/s41550-019-1008-4}{\emph{Nature Astron.} {\bfseries 4} (2020) 108} [\href{https://arxiv.org/abs/2002.03603}{{\ttfamily 2002.03603}}].

\bibitem{Baker:2019pnp}
J.~Baker et~al., \emph{{Space Based Gravitational Wave Astronomy Beyond LISA}}, {\emph{Bull. Am. Astron. Soc.} {\bfseries 51} (2019) 243} [\href{https://arxiv.org/abs/1907.11305}{{\ttfamily 1907.11305}}].

\bibitem{Kawamura:2011zz}
S.~Kawamura et~al., \emph{{The Japanese space gravitational wave antenna: DECIGO}}, \href{https://doi.org/10.1088/0264-9381/28/9/094011}{\emph{Class. Quant. Grav.} {\bfseries 28} (2011) 094011}.

\bibitem{Mei:2020lrl}
{\scshape TianQin} collaboration, \emph{{The TianQin project: current progress on science and technology}}, \href{https://doi.org/10.1093/ptep/ptaa114}{\emph{PTEP} {\bfseries 2021} (2021) 05A107} [\href{https://arxiv.org/abs/2008.10332}{{\ttfamily 2008.10332}}].

\bibitem{Kuns:2019upi}
K.A.~Kuns, H.~Yu, Y.~Chen and R.X.~Adhikari, \emph{{Astrophysics and cosmology with a decihertz gravitational-wave detector: TianGO}}, \href{https://doi.org/10.1103/PhysRevD.102.043001}{\emph{Phys. Rev. D} {\bfseries 102} (2020) 043001} [\href{https://arxiv.org/abs/1908.06004}{{\ttfamily 1908.06004}}].

\bibitem{Reitze:2019iox}
D.~Reitze et~al., \emph{{Cosmic Explorer: The U.S. Contribution to Gravitational-Wave Astronomy beyond LIGO}}, {\emph{Bull. Am. Astron. Soc.} {\bfseries 51} (2019) 035} [\href{https://arxiv.org/abs/1907.04833}{{\ttfamily 1907.04833}}].

\bibitem{Punturo:2010zz}
M.~Punturo et~al., \emph{{The Einstein Telescope: A third-generation gravitational wave observatory}}, \href{https://doi.org/10.1088/0264-9381/27/19/194002}{\emph{Class. Quant. Grav.} {\bfseries 27} (2010) 194002}.

\bibitem{Graham:2017pmn}
{\scshape MAGIS} collaboration, \emph{{Mid-band gravitational wave detection with precision atomic sensors}},  \href{https://arxiv.org/abs/1711.02225}{{\ttfamily 1711.02225}}.

\bibitem{Bertoldi:2019tck}
{\scshape AEDGE} collaboration, \emph{{AEDGE: Atomic Experiment for Dark Matter and Gravity Exploration in Space}}, \href{https://doi.org/10.1140/epjqt/s40507-020-0080-0}{\emph{EPJ Quant. Technol.} {\bfseries 7} (2020) 6} [\href{https://arxiv.org/abs/1908.00802}{{\ttfamily 1908.00802}}].

\bibitem{Aggarwal:2020olq}
N.~Aggarwal et~al., \emph{{Challenges and opportunities of gravitational-wave searches at MHz to GHz frequencies}}, \href{https://doi.org/10.1007/s41114-021-00032-5}{\emph{Living Rev. Rel.} {\bfseries 24} (2021) 4} [\href{https://arxiv.org/abs/2011.12414}{{\ttfamily 2011.12414}}].

\bibitem{Cranmer_2020}
K.~Cranmer, J.~Brehmer and G.~Louppe, \emph{The frontier of simulation-based inference}, \href{https://doi.org/10.1073/pnas.1912789117}{\emph{Proceedings of the National Academy of Sciences} {\bfseries 117} (2020) 30055}.

\bibitem{brehmer2020}
J.~Brehmer and K.~Cranmer, \emph{Simulation-based inference methods for particle physics},  \href{https://arxiv.org/abs/2010.06439}{{\ttfamily 2010.06439}}.

\bibitem{lueckmann2021}
J.-M.~Lueckmann, J.~Boelts, D.S.~Greenberg, P.J.~Gonçalves and J.H.~Macke, \emph{Benchmarking simulation-based inference},  \href{https://arxiv.org/abs/2101.04653}{{\ttfamily 2101.04653}}.

\bibitem{Miller2022}
B.K.~Miller, A.~Cole, C.~Weniger, F.~Nattino, O.~Ku and M.W.~Grootes, \emph{swyft: Truncated marginal neural ratio estimation in python}, \href{https://doi.org/10.21105/joss.04205}{\emph{Journal of Open Source Software} {\bfseries 7} (2022) 4205}.

\bibitem{Tejero-Cantero2020}
A.~Tejero-Cantero, J.~Boelts, M.~Deistler, J.-M.~Lueckmann, C.~Durkan, P.J.~Gonçalves et~al., \emph{sbi: A toolkit for simulation-based inference}, \href{https://doi.org/10.21105/joss.02505}{\emph{Journal of Open Source Software} {\bfseries 5} (2020) 2505}.

\bibitem{Cole:2021gwr}
A.~Cole, B.K.~Miller, S.J.~Witte, M.X.~Cai, M.W.~Grootes, F.~Nattino et~al., \emph{{Fast and credible likelihood-free cosmology with truncated marginal neural ratio estimation}}, \href{https://doi.org/10.1088/1475-7516/2022/09/004}{\emph{JCAP} {\bfseries 09} (2022) 004} [\href{https://arxiv.org/abs/2111.08030}{{\ttfamily 2111.08030}}].

\bibitem{Montel:2022fhv}
N.A.~Montel, A.~Coogan, C.~Correa, K.~Karchev and C.~Weniger, \emph{{Estimating the warm dark matter mass from strong lensing images with truncated marginal neural ratio estimation}}, \href{https://doi.org/10.1093/mnras/stac3215}{\emph{Mon. Not. Roy. Astron. Soc.} {\bfseries 518} (2022) 2746} [\href{https://arxiv.org/abs/2205.09126}{{\ttfamily 2205.09126}}].

\bibitem{Makinen:2021nly}
T.L.~Makinen, T.~Charnock, J.~Alsing and B.D.~Wandelt, \emph{{Lossless, scalable implicit likelihood inference for cosmological fields}}, \href{https://doi.org/10.1088/1475-7516/2021/11/049}{\emph{JCAP} {\bfseries 11} (2021) 049} [\href{https://arxiv.org/abs/2107.07405}{{\ttfamily 2107.07405}}].

\bibitem{Dimitriou:2022cvc}
A.~Dimitriou, C.~Weniger and C.A.~Correa, \emph{{Towards reconstructing the halo clustering and halo mass function of N-body simulations using neural ratio estimation}},  \href{https://arxiv.org/abs/2206.11312}{{\ttfamily 2206.11312}}.

\bibitem{Gagnon-Hartman:2023soa}
S.~Gagnon-Hartman, J.~Ruan and D.~Haggard, \emph{{Debiasing standard siren inference of the Hubble constant with marginal neural ratio estimation}}, \href{https://doi.org/10.1093/mnras/stad069}{\emph{Mon. Not. Roy. Astron. Soc.} {\bfseries 520} (2023) 1} [\href{https://arxiv.org/abs/2301.05241}{{\ttfamily 2301.05241}}].

\bibitem{Delaunoy:2020zcu}
A.~Delaunoy, A.~Wehenkel, T.~Hinderer, S.~Nissanke, C.~Weniger, A.R.~Williamson et~al., \emph{{Lightning-Fast Gravitational Wave Parameter Inference through Neural Amortization}},  \href{https://arxiv.org/abs/2010.12931}{{\ttfamily 2010.12931}}.

\bibitem{Karchev:2022xyn}
K.~Karchev, R.~Trotta and C.~Weniger, \emph{{SICRET: Supernova Ia Cosmology with truncated marginal neural Ratio EsTimation}},  \href{https://arxiv.org/abs/2209.06733}{{\ttfamily 2209.06733}}.

\bibitem{Lin:2022ayr}
K.~Lin, M.~von Wietersheim-Kramsta, B.~Joachimi and S.~Feeney, \emph{{A simulation-based inference pipeline for cosmic shear with the Kilo-Degree Survey}}, \href{https://doi.org/10.1093/mnras/stad2262}{\emph{Mon. Not. Roy. Astron. Soc.} {\bfseries 524} (2023) 6167} [\href{https://arxiv.org/abs/2212.04521}{{\ttfamily 2212.04521}}].

\bibitem{Alvey:2023naa}
J.~Alvey, U.~Bhardwaj, S.~Nissanke and C.~Weniger, \emph{{What to do when things get crowded? Scalable joint analysis of overlapping gravitational wave signals}},  \href{https://arxiv.org/abs/2308.06318}{{\ttfamily 2308.06318}}.

\bibitem{Bhardwaj:2023xph}
U.~Bhardwaj, J.~Alvey, B.K.~Miller, S.~Nissanke and C.~Weniger, \emph{{Sequential simulation-based inference for gravitational wave signals}}, \href{https://doi.org/10.1103/PhysRevD.108.042004}{\emph{Phys. Rev. D} {\bfseries 108} (2023) 042004} [\href{https://arxiv.org/abs/2304.02035}{{\ttfamily 2304.02035}}].

\bibitem{rezende2016}
D.J.~Rezende and S.~Mohamed, \emph{Variational inference with normalizing flows},  2016.

\bibitem{Karnesis:2019mph}
N.~Karnesis, M.~Lilley and A.~Petiteau, \emph{{Assessing the detectability of a Stochastic Gravitational Wave Background with LISA, using an excess of power approach}}, \href{https://doi.org/10.1088/1361-6382/abb637}{\emph{Class. Quant. Grav.} {\bfseries 37} (2020) 215017} [\href{https://arxiv.org/abs/1906.09027}{{\ttfamily 1906.09027}}].

\bibitem{Caprini:2019pxz}
C.~Caprini, D.G.~Figueroa, R.~Flauger, G.~Nardini, M.~Peloso, M.~Pieroni et~al., \emph{{Reconstructing the spectral shape of a stochastic gravitational wave background with LISA}}, \href{https://doi.org/10.1088/1475-7516/2019/11/017}{\emph{JCAP} {\bfseries 11} (2019) 017} [\href{https://arxiv.org/abs/1906.09244}{{\ttfamily 1906.09244}}].

\bibitem{Flauger:2020qyi}
R.~Flauger, N.~Karnesis, G.~Nardini, M.~Pieroni, A.~Ricciardone and J.~Torrado, \emph{{Improved reconstruction of a stochastic gravitational wave background with LISA}}, \href{https://doi.org/10.1088/1475-7516/2021/01/059}{\emph{JCAP} {\bfseries 01} (2021) 059} [\href{https://arxiv.org/abs/2009.11845}{{\ttfamily 2009.11845}}].

\bibitem{Baghi:2023qnq}
Q.~Baghi, N.~Karnesis, J.-B.~Bayle, M.~Besan\c{c}on and H.~Inchausp\'e, \emph{{Uncovering gravitational-wave backgrounds from noises of unknown shape with LISA}}, \href{https://doi.org/10.1088/1475-7516/2023/04/066}{\emph{JCAP} {\bfseries 04} (2023) 066} [\href{https://arxiv.org/abs/2302.12573}{{\ttfamily 2302.12573}}].

\bibitem{Pozzoli:2023lgz}
F.~Pozzoli, R.~Buscicchio, C.J.~Moore, F.~Haardt and A.~Sesana, \emph{{Weakly parametric approach to stochastic background inference in LISA}}, \href{https://doi.org/10.1103/PhysRevD.109.083029}{\emph{Phys. Rev. D} {\bfseries 109} (2024) 083029} [\href{https://arxiv.org/abs/2311.12111}{{\ttfamily 2311.12111}}].

\bibitem{Audley:2017drz}
{\scshape LISA} collaboration, \emph{{Laser Interferometer Space Antenna}},  \href{https://arxiv.org/abs/1702.00786}{{\ttfamily 1702.00786}}.

\bibitem{Domcke:2019zls}
V.~Domcke, J.~Garcia-Bellido, M.~Peloso, M.~Pieroni, A.~Ricciardone, L.~Sorbo et~al., \emph{{Measuring the net circular polarization of the stochastic gravitational wave background with interferometers}}, \href{https://doi.org/10.1088/1475-7516/2020/05/028}{\emph{JCAP} {\bfseries 05} (2020) 028} [\href{https://arxiv.org/abs/1910.08052}{{\ttfamily 1910.08052}}].

\bibitem{Colpi:2024xhw}
M.~Colpi et~al., \emph{{LISA Definition Study Report}},  \href{https://arxiv.org/abs/2402.07571}{{\ttfamily 2402.07571}}.

\bibitem{LISAnoise}
S.~Babak and A.~Petiteau, \emph{Lisa data challenge manual}, {\emph{Tech. Rep. LISA-LCST-SGS-MAN-002, APC Paris, 7, 2020} }.

\bibitem{Hogan:2001jn}
C.J.~Hogan and P.L.~Bender, \emph{{Estimating stochastic gravitational wave backgrounds with Sagnac calibration}}, \href{https://doi.org/10.1103/PhysRevD.64.062002}{\emph{Phys. Rev. D} {\bfseries 64} (2001) 062002} [\href{https://arxiv.org/abs/astro-ph/0104266}{{\ttfamily astro-ph/0104266}}].

\bibitem{Adams:2010vc}
M.R.~Adams and N.J.~Cornish, \emph{{Discriminating between a Stochastic Gravitational Wave Background and Instrument Noise}}, \href{https://doi.org/10.1103/PhysRevD.82.022002}{\emph{Phys. Rev. D} {\bfseries 82} (2010) 022002} [\href{https://arxiv.org/abs/1002.1291}{{\ttfamily 1002.1291}}].

\bibitem{Armano:2016bkm}
M.~Armano et~al., \emph{{Sub-Femto- g Free Fall for Space-Based Gravitational Wave Observatories: LISA Pathfinder Results}}, \href{https://doi.org/10.1103/PhysRevLett.116.231101}{\emph{Phys. Rev. Lett.} {\bfseries 116} (2016) 231101}.

\bibitem{Muratore:2023gxh}
M.~Muratore, J.~Gair and L.~Speri, \emph{{Impact of the noise knowledge uncertainty for the science exploitation of cosmological and astrophysical stochastic gravitational wave background with LISA}}, \href{https://doi.org/10.1103/PhysRevD.109.042001}{\emph{Phys. Rev. D} {\bfseries 109} (2024) 042001} [\href{https://arxiv.org/abs/2308.01056}{{\ttfamily 2308.01056}}].

\bibitem{Cornish:2018dyw}
T.~Robson, N.J.~Cornish and C.~Liu, \emph{{The construction and use of LISA sensitivity curves}}, \href{https://doi.org/10.1088/1361-6382/ab1101}{\emph{Class. Quant. Grav.} {\bfseries 36} (2019) 105011} [\href{https://arxiv.org/abs/1803.01944}{{\ttfamily 1803.01944}}].

\bibitem{Giese:2021dnw}
F.~Giese, T.~Konstandin and J.~van~de Vis, \emph{{Finding sound shells in LISA mock data using likelihood sampling}}, \href{https://doi.org/10.1088/1475-7516/2021/11/002}{\emph{JCAP} {\bfseries 11} (2021) 002} [\href{https://arxiv.org/abs/2107.06275}{{\ttfamily 2107.06275}}].

\bibitem{durkan2019}
C.~Durkan, A.~Bekasov, I.~Murray and G.~Papamakarios, \emph{Neural spline flows},  2019.

\bibitem{greenberg2019}
D.S.~Greenberg, M.~Nonnenmacher and J.H.~Macke, \emph{Automatic posterior transformation for likelihood-free inference},  2019.

\bibitem{Foreman_Mackey_2013}
D.~Foreman-Mackey, D.W.~Hogg, D.~Lang and J.~Goodman, \emph{$ttemcee/tt$: The {MCMC} hammer}, \href{https://doi.org/10.1086/670067}{\emph{Publications of the Astronomical Society of the Pacific} {\bfseries 125} (2013) 306}.

\bibitem{neal:2012}
R.M.~Neal, \emph{{MCMC using Hamiltonian dynamics}},  \href{https://arxiv.org/abs/1206.1901}{{\ttfamily 1206.1901}}.

\bibitem{LIGOScientific:2019vic}
{\scshape LIGO Scientific, Virgo} collaboration, \emph{{Search for the isotropic stochastic background using data from Advanced LIGO\textquoteright{}s second observing run}}, \href{https://doi.org/10.1103/PhysRevD.100.061101}{\emph{Phys. Rev. D} {\bfseries 100} (2019) 061101} [\href{https://arxiv.org/abs/1903.02886}{{\ttfamily 1903.02886}}].

\bibitem{Fenu:2009qf}
E.~Fenu, D.G.~Figueroa, R.~Durrer and J.~Garcia-Bellido, \emph{{Gravitational waves from self-ordering scalar fields}}, \href{https://doi.org/10.1088/1475-7516/2009/10/005}{\emph{JCAP} {\bfseries 10} (2009) 005} [\href{https://arxiv.org/abs/0908.0425}{{\ttfamily 0908.0425}}].

\bibitem{Jones-Smith:2007hib}
K.~Jones-Smith, L.M.~Krauss and H.~Mathur, \emph{{A Nearly Scale Invariant Spectrum of Gravitational Radiation from Global Phase Transitions}}, \href{https://doi.org/10.1103/PhysRevLett.100.131302}{\emph{Phys. Rev. Lett.} {\bfseries 100} (2008) 131302} [\href{https://arxiv.org/abs/0712.0778}{{\ttfamily 0712.0778}}].

\bibitem{Enqvist:2012im}
K.~Enqvist, D.G.~Figueroa and T.~Meriniemi, \emph{{Stochastic Background of Gravitational Waves from Fermions}}, \href{https://doi.org/10.1103/PhysRevD.86.061301}{\emph{Phys. Rev. D} {\bfseries 86} (2012) 061301} [\href{https://arxiv.org/abs/1203.4943}{{\ttfamily 1203.4943}}].

\bibitem{Figueroa:2013vif}
D.G.~Figueroa and T.~Meriniemi, \emph{{Stochastic Background of Gravitational Waves from Fermions -- Theory and Applications}}, \href{https://doi.org/10.1007/JHEP10(2013)101}{\emph{JHEP} {\bfseries 10} (2013) 101} [\href{https://arxiv.org/abs/1306.6911}{{\ttfamily 1306.6911}}].

\bibitem{Figueroa:2014aya}
D.G.~Figueroa, \emph{{A gravitational wave background from the decay of the standard model Higgs after inflation}}, \href{https://doi.org/10.1007/JHEP11(2014)145}{\emph{JHEP} {\bfseries 11} (2014) 145} [\href{https://arxiv.org/abs/1402.1345}{{\ttfamily 1402.1345}}].

\bibitem{Figueroa:2016ojl}
D.G.~Figueroa, J.~Garc\'\i{}a-Bellido and F.~Torrent\'\i{}, \emph{{Gravitational wave production from the decay of the standard model Higgs field after inflation}}, \href{https://doi.org/10.1103/PhysRevD.93.103521}{\emph{Phys. Rev. D} {\bfseries 93} (2016) 103521} [\href{https://arxiv.org/abs/1602.03085}{{\ttfamily 1602.03085}}].

\bibitem{Figueroa:2022iho}
D.G.~Figueroa, A.~Florio, N.~Loayza and M.~Pieroni, \emph{{Spectroscopy of particle couplings with gravitational waves}}, \href{https://doi.org/10.1103/PhysRevD.106.063522}{\emph{Phys. Rev. D} {\bfseries 106} (2022) 063522} [\href{https://arxiv.org/abs/2202.05805}{{\ttfamily 2202.05805}}].

\bibitem{Cosme:2022htl}
C.~Cosme, D.G.~Figueroa and N.~Loayza, \emph{{Gravitational wave production from preheating with trilinear interactions}}, \href{https://doi.org/10.1088/1475-7516/2023/05/023}{\emph{JCAP} {\bfseries 05} (2023) 023} [\href{https://arxiv.org/abs/2206.14721}{{\ttfamily 2206.14721}}].

\bibitem{Machado:2019xuc}
C.S.~Machado, W.~Ratzinger, P.~Schwaller and B.A.~Stefanek, \emph{{Gravitational wave probes of axionlike particles}}, \href{https://doi.org/10.1103/PhysRevD.102.075033}{\emph{Phys. Rev. D} {\bfseries 102} (2020) 075033} [\href{https://arxiv.org/abs/1912.01007}{{\ttfamily 1912.01007}}].

\bibitem{Ratzinger:2020oct}
W.~Ratzinger, P.~Schwaller and B.A.~Stefanek, \emph{{Gravitational Waves from an Axion-Dark Photon System: A Lattice Study}}, \href{https://doi.org/10.21468/SciPostPhys.11.1.001}{\emph{SciPost Phys.} {\bfseries 11} (2021) 001} [\href{https://arxiv.org/abs/2012.11584}{{\ttfamily 2012.11584}}].

\bibitem{Schwaller:2015tja}
P.~Schwaller, \emph{{Gravitational Waves from a Dark Phase Transition}}, \href{https://doi.org/10.1103/PhysRevLett.115.181101}{\emph{Phys. Rev. Lett.} {\bfseries 115} (2015) 181101} [\href{https://arxiv.org/abs/1504.07263}{{\ttfamily 1504.07263}}].

\bibitem{Greljo:2019xan}
A.~Greljo, T.~Opferkuch and B.A.~Stefanek, \emph{{Gravitational Imprints of Flavor Hierarchies}}, \href{https://doi.org/10.1103/PhysRevLett.124.171802}{\emph{Phys. Rev. Lett.} {\bfseries 124} (2020) 171802} [\href{https://arxiv.org/abs/1910.02014}{{\ttfamily 1910.02014}}].

\bibitem{Phinney:2001di}
E.S.~Phinney, \emph{{A Practical theorem on gravitational wave backgrounds}},  \href{https://arxiv.org/abs/astro-ph/0108028}{{\ttfamily astro-ph/0108028}}.

\bibitem{Adams:2013qma}
M.R.~Adams and N.J.~Cornish, \emph{{Detecting a Stochastic Gravitational Wave Background in the presence of a Galactic Foreground and Instrument Noise}}, \href{https://doi.org/10.1103/PhysRevD.89.022001}{\emph{Phys. Rev. D} {\bfseries 89} (2014) 022001} [\href{https://arxiv.org/abs/1307.4116}{{\ttfamily 1307.4116}}].

\bibitem{Cornish:2017vip}
N.~Cornish and T.~Robson, \emph{{Galactic binary science with the new LISA design}}, \href{https://doi.org/10.1088/1742-6596/840/1/012024}{\emph{J. Phys. Conf. Ser.} {\bfseries 840} (2017) 012024} [\href{https://arxiv.org/abs/1703.09858}{{\ttfamily 1703.09858}}].

\bibitem{Schmitz:2020rag}
K.~Schmitz, \emph{{LISA Sensitivity to Gravitational Waves from Sound Waves}}, \href{https://doi.org/10.3390/sym12091477}{\emph{Symmetry} {\bfseries 12} (2020) 1477} [\href{https://arxiv.org/abs/2005.10789}{{\ttfamily 2005.10789}}].

\bibitem{Cornish:2005qw}
N.J.~Cornish and J.~Crowder, \emph{{LISA data analysis using MCMC methods}}, \href{https://doi.org/10.1103/PhysRevD.72.043005}{\emph{Phys. Rev. D} {\bfseries 72} (2005) 043005} [\href{https://arxiv.org/abs/gr-qc/0506059}{{\ttfamily gr-qc/0506059}}].

\bibitem{Vallisneri:2008ye}
M.~Vallisneri, \emph{{A LISA Data-Analysis Primer}}, \href{https://doi.org/10.1088/0264-9381/26/9/094024}{\emph{Class. Quant. Grav.} {\bfseries 26} (2009) 094024} [\href{https://arxiv.org/abs/0812.0751}{{\ttfamily 0812.0751}}].

\bibitem{MockLISADataChallengeTaskForce:2009wir}
{\scshape Mock LISA Data Challenge Task Force} collaboration, \emph{{The Mock LISA Data Challenges: From Challenge 3 to Challenge 4}}, \href{https://doi.org/10.1088/0264-9381/27/8/084009}{\emph{Class. Quant. Grav.} {\bfseries 27} (2010) 084009} [\href{https://arxiv.org/abs/0912.0548}{{\ttfamily 0912.0548}}].

\bibitem{Valerie}
J.~Alvey, U.~Bhardwaj, V.~Domcke, M.~Pieroni and C.~Weniger, \emph{Simulation-based inference for stochastic gravitational wave background data analysis},  \href{https://arxiv.org/abs/2309.xxxxx}{{\ttfamily 2309.xxxxx}}.

\bibitem{2021arXiv211006581H}
J.~{Hermans}, A.~{Delaunoy}, F.~{Rozet}, A.~{Wehenkel}, V.~{Begy} and G.~{Louppe}, \emph{{A Trust Crisis In Simulation-Based Inference? Your Posterior Approximations Can Be Unfaithful}}, \href{https://doi.org/10.48550/arXiv.2110.06581}{\emph{arXiv e-prints} (2021) arXiv:2110.06581} [\href{https://arxiv.org/abs/2110.06581}{{\ttfamily 2110.06581}}].

\bibitem{Cook2006}
S.R.~Cook, A.~Gelman and D.B.~Rubin, \emph{Validation of software for bayesian models using posterior quantiles}, {\emph{Journal of Computational and Graphical Statistics} {\bfseries 15} (2006) 675}.

\bibitem{2022JCAP...09..004C}
A.~{Cole}, B.K.~{Miller}, S.J.~{Witte}, M.X.~{Cai}, M.W.~{Grootes}, F.~{Nattino} et~al., \emph{{Fast and credible likelihood-free cosmology with truncated marginal neural ratio estimation}}, \href{https://doi.org/10.1088/1475-7516/2022/09/004}{\emph{jcap} {\bfseries 2022} (2022) 004} [\href{https://arxiv.org/abs/2111.08030}{{\ttfamily 2111.08030}}].

\bibitem{2019arXiv191202762P}
G.~{Papamakarios}, E.~{Nalisnick}, D.~{Jimenez Rezende}, S.~{Mohamed} and B.~{Lakshminarayanan}, \emph{{Normalizing Flows for Probabilistic Modeling and Inference}}, \href{https://doi.org/10.48550/arXiv.1912.02762}{\emph{arXiv e-prints} (2019) arXiv:1912.02762} [\href{https://arxiv.org/abs/1912.02762}{{\ttfamily 1912.02762}}].

\bibitem{2019arXiv190604032D}
C.~{Durkan}, A.~{Bekasov}, I.~{Murray} and G.~{Papamakarios}, \emph{{Neural Spline Flows}}, \href{https://doi.org/10.48550/arXiv.1906.04032}{\emph{arXiv e-prints} (2019) arXiv:1906.04032} [\href{https://arxiv.org/abs/1906.04032}{{\ttfamily 1906.04032}}].

\end{thebibliography}\endgroup

\end{document}